\documentclass[aps,prd,superscriptaddress,amsfonts,amssymb,
amsmath,showpacs,floatfix,fleqn,preprintnumbers]{revtex4-1}

\usepackage{macros}
\usepackage{dcolumn}
\usepackage{bm}
\usepackage{multirow}
\usepackage{verbatim}
\usepackage{graphicx}
\usepackage[normalem]{ulem}
\usepackage{color}
\usepackage{bbm}
\usepackage{simplewick}
\usepackage{slashed}
\usepackage{tikz}
\usetikzlibrary{shapes.misc}
\usepackage{subfig}
\usepackage{fancyhdr}

\newcommand{\widesim}[2][1.5]{
  \mathrel{\overset{#2}{\scalebox{#1}[1]{$\sim$}}}
}
\DeclareRobustCommand{\CP}{$\mathcal{CP}$}

\graphicspath{{Graphics/}}

\begin{document}

\preprint{JLAB-THY-20-3190}

\title{Short flow-time coefficients of CP-violating operators}



\author{Matthew D.~Rizik}
\email[e-mail: ]{rizik@nscl.msu.edu}
\affiliation{Facility for Rare Isotope Beams,
  Physics Department, Michigan State University, East
Lansing, Michigan 48824, USA}

\author{Christopher J.~Monahan}
\email[e-mail: ]{cjm373@uw.edu}
\affiliation{Department of Physics, The College of William \& Mary, Williamsburg, VA 23187, USA}
\affiliation{Theory Center, Thomas Jefferson National Accelerator Facility, Newport News, VA 23606, USA}

\author{Andrea Shindler}
\email[e-mail: ]{shindler@frib.msu.edu}
\affiliation{Facility for Rare Isotope Beams,
  Physics Department, Michigan State University, East
Lansing, Michigan 48824, USA}

\collaboration{SymLat Collaboration}

\noaffiliation
\vspace{-1.0cm}
\begin{figure}[h!]
\flushleft{  \includegraphics[scale=0.1]{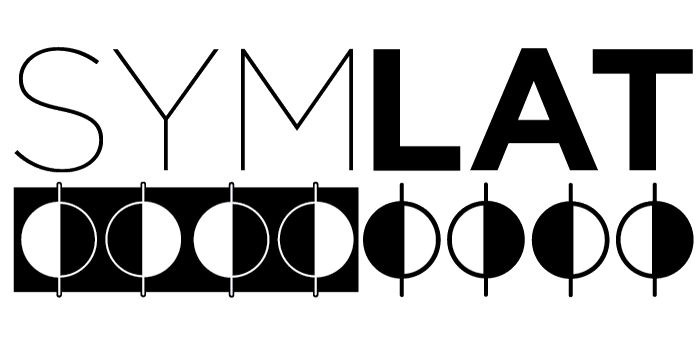}}
\end{figure}

\begin{abstract}
Measurements of a permanent neutron electric dipole moment (EDM) potentially probe Beyond-the-Standard Model (BSM) sources of CP-violation. At low energy the CP-violating BSM
interactions are parametrized by flavor-conserving CP-violating operators of dimension higher than four. QCD calculations of the nucleon matrix elements of these operators are required to fully reconstruct the sources and magnitudes of the different CP-violating contributions to the nucleon EDM. Herein we study the quark-chromo electric dipole moment (qCEDM) operator and the three-gluon Weinberg operator. The non-perturbative determination, using lattice QCD, of the nucleon matrix elements of these CP-violating operators is hampered by their short-distance behavior. Under renormalization these operators mix with lower dimensional operators, which induces power divergences in the lattice spacing, as the continuum limit is approached. We study the short-distance behavior of the qCEDM and the Weinberg operators using the gradient flow. We perform a short flow time expansion and determine, in perturbation theory, the expansion coefficients of the linearly-divergent terms stemming from the mixing with the pseudoscalar density and the topological charge, confirming the expectations of the operator product expansion.
We introduce a new method to perform calculations at non-zero flow-time for arbitrary values of the external momenta. This method allows us to work in four dimensions
for most of the calculations described in this paper, avoiding the complications associated with defining $\gamma_5$ in generic $d$ dimensions. We show that leading contributions in the external momenta can be reproduced by defining $\gamma_5$ using the ’t Hooft-Veltman-Breitenlohner-Maison scheme.
\end{abstract}

\maketitle

\section{Introduction}
\label{sec:intro}

The nucleon electric dipole moment (EDM) is a physical quantity that, once measured, will
provide a unique opportunity to detect and investigate beyond-the-standard model (BSM)
sources of charge and parity (CP) violation.
In principle, there are multiple sources for a non-vanishing nucleon EDM,
including the Cabibbo-Kobayashi-Maskawa (CKM) quark-mixing matrix,
the quantum chromodyamics (QCD) $\theta$ term,
higher-dimensional CP-violating operators, or any combination of these.
The current experimental limit for the neutron EDM~\cite{Abel:2020gbr,Afach:2015sja},
$|d_n| \le 1.8 \times 10^{-26}~e$ cm at $90\%$ confidence level,
leaves open the possibility of a dominant BSM source of
CP-violation, which could be several orders of magnitude larger than Standard Model sources
(see Ref.~\cite{Chupp:2017rkp} for a recent review).

In addition to the Standard Model contributions
to the nucleon EDM from the CKM matrix~\cite{Seng:2014lea}
and from the $\theta$ term~\cite{Dragos:2019oxn},
BSM theories that contain complex CP-violating couplings can induce
a non-vanishing EDM at the one loop level.
At low energies the BSM degrees of freedom are heavy enough
that one can parametrize their effects through effective,
higher-dimension CP-violating operators.
In this paper we consider two such operators,
the quark-color EDM (qCEDM) operator and the CP-violating three-gluon operator,
i.e.~the Weinberg operator.
To constrain couplings in BSM theories at high energies, one needs to determine the QCD
contribution to the EDM at low energy. Broadly speaking, there are three approaches to determining the relevant matrix elements: QCD sum rules \cite{Demir:2002gg,Haisch:2019bml}; chiral perturbation theory \cite{deVries:2010ah,Mereghetti:2010kp}; and lattice QCD.

Lattice QCD provides the most systematic method to calculate
individual contributions from different CP-violating sources
to the nucleon EDM in terms of the QCD fundamental degrees of freedom,
quark and gluons.
There is a long history of attempts to determine the nucleon EDM
from lattice
QCD~\cite{Shintani:2005xg,Berruto:2005hg,Guo:2015tla,Shindler:2015aqa,Alexandrou:2015spa,Shintani:2015vsx,Abramczyk:2017oxr,Yoon:2017tag,Izubuchi:2020ngl},
and several technical difficulties have been encountered.

The first difficulty arises from the fact that in Euclidean space the $\theta$
term renders the QCD action complex, which prevents the use of stochastic methods. The current experimental bound on the neutron EDM implies a very small value
for $\theta \sim 10^{-10}$, justifying a perturbative expansion in $\theta$.
Correlators that include an insertion of the $\theta$ term,
once the topological charge has been properly renormalized,
are theoretically well-defined. Despite the very poor signal-to-noise ratio
it is possible to determine the nucleon EDM induced by the $\theta$ term
using signal-to-noise improved ratios~\cite{Dragos:2018uzd,Dragos:2019oxn}. 

The second difficulty arises from the renormalization of the relevant composite operators. In Ref.~\cite{Shindler:2014oha} we proposed using the
gradient flow~\cite{Narayanan:2006rf,Luscher:2010iy,Luscher:2011bx,Luscher:2013cpa}
to renormalize the $\theta$ term and the BSM CP-violating operators.
We are currently pursuing this program and in
Refs.~\cite{Shindler:2015aqa,Dragos:2017wms,Dragos:2018uzd,Dragos:2019oxn}
we investigated and calculated the nucleon EDM from the $\theta$ term.

The properties of the gradient flow have led to a wide variety of applications
in lattice gauge theories. These applications include determining
the fundamental parameters of QCD,
such as the running coupling
constant~\cite{Fodor:2012td,Fodor:2014cpa,Fritzsch:2013je,Ramos:2014kla,Ramos:2015baa,DallaBrida:2016kgh,DallaBrida:2019wur,Ishikawa:2017xam}
and the equation of state at finite
temperature~\cite{Asakawa:2013laa,Kitazawa:2016dsl,Taniguchi:2016ofw,Kitazawa:2017qab,Iritani:2018idk,Eller:2018yje},
extracted from a non-perturbative definition of the
energy-momentum tensor at finite lattice
spacing~\cite{Suzuki:2013gza,DelDebbio:2013zaa,Makino:2014taa,Harlander:2018zpi}.
The gradient flow has also provided an important tool for relative scale-setting in
lattice calculations~\cite{Borsanyi:2012zs,Hollwieser:2020qri}.
Many of these techniques have been applied in other
theories~\cite{Lin:2015zpa,Hasenfratz:2016dou,DeGrand:2017gbi,Bribian:2019ybc,Hirakida:2018uoy,Carosso:2018bmz,Hasenfratz:2019dpr,Bennett:2019jzz,DeGrand:2020utq}.

Renormalization schemes based on the gradient flow include
non-perturbative step-scaling approaches~\cite{Luscher:2014kea,Monahan:2013lwa},
removing power divergences in nonlocal operators relevant
to hadron structure \cite{Monahan:2016bvm,Monahan:2017hpu},
and defining regularization-independent quark-bilinear
currents \cite{Endo:2015iea,Hieda:2016lly}.
Perturbative calculations of the gradient flow have been
carried out to three loops for certain quantities using
automated perturbation theory routines~\cite{Harlander:2016vzb,Harlander:2018zpi,Artz:2019bpr}
and to two loops via numerical stochastic perturbation theory~\cite{DallaBrida:2017pex,DallaBrida:2017tru}.

Analytic loop-order calculations with the gradient flow often
introduce some difficulties related to dimensional regularization.
One method to avoid these complications employs an expansion in
the external momentum $p$ to some desired order.
This can induce extraneous, non-physical infrared poles at fairly
low orders in the external momentum.
(In the calculation of the Wilson coefficient $c_{CP}$ below,
for example, these appear as early as $\mcO(p^2)$.)
We have used a novel combinatorial scheme to track the
external momentum at all orders,
which maintains finiteness at positive flow
time throughout all of the calculations
in this paper with the exception of those related
to the renormalization of the flowed fermion propagator in App.~\ref{app:Zfermion}.

In this paper we focus on the renormalization
of the higher dimensional CP-violating operators using the gradient flow. 
First results appeared in~\cite{Rizik:2018lrz,Kim:2018rce,Reyes:2018ucu} and in this paper
we determine the leading contribution to the short flow-time expansion (SFTE)
coefficients of the CP-violating operators defined using the gradient flow. The renormalization and mixing of these operators was first studied in the $\overline{\mathrm{MS}}$ scheme in \cite{Weinberg:1989dx,Braaten:1990zt,Braaten:1990gq}, the determined at two loops for the qCEDM in \cite{Degrassi:2005zd}, and at two and three loops for the Weinberg in \cite{deVries:2019nsu}. After describing our perturbative strategy for determining these coefficients, we focus
on the leading linearly-divergent expansion coefficients and some logarithmic terms.

The paper is organized as follows.  
We first introduce the gradient flow and some technical details relevant 
for our perturbative expansion in Sec.~\ref{sec:flow}. We calculate the expansion coefficients of the quark-chromo operator,
parameterizing the mixing with the pseudoscalar density and the topological charge density,
in Sec.~\ref{sec:qcedm}, and the corresponding coefficient of the Weinberg operator,
induced by the mixing with the topological charge density, in Sec.~\ref{sec:weinbergResults}.
We summarize our results and our conclusions in Sec.~\ref{sec:summary}.

In Appendix~\ref{app:conventions} we 
detail our notations and conventions including the $d-$dimensional Dirac
gamma matrices.
In Appendix~\ref{app:rules} we list Feynman rules for the flowed vertices and for the relevant operators. In Appendix~\ref{app:Zfermion} we use the calculation of the quark propagator as an example
to elucidate the computational techniques for finite flow time.

\section{\label{sec:flow}The gradient flow}

In this section we give a brief introduction to the gradient flow, emphasizing
the technical details needed for our perturbative expansion.
The gradient flow equations define the evolution of the bulk gauge and fermion fields,
$B_\mu(x;t)$ and $\chi(x;t)$ respectively,
as a function of the flow time, $t$ \cite{Luscher:2010iy,Luscher:2013cpa}:
\begin{align}
\partial_t B_\mu= {} & D_\nu G_{\nu\mu} +\alpha_0D_\mu \partial_\nu B_\nu,\label{eq:ymflow} \\ 
\partial_t\chi = {} & D_\mu D_\mu\chi-\alpha_0 \partial_\nu B_\nu \chi,\label{eq:fermflow} \\ 
\partial_t\bar{\chi} = {} & \bar{\chi}\overleftarrow{D}_\mu \overleftarrow{D}_\mu +  \alpha_0 \bar{\chi} \partial_\nu B_\nu ,\label{eq:antifermflow}
\end{align}
where
\begin{equation}
G_{\mu\nu} = \partial_\mu B_\nu - \partial_\nu 
B_\mu +[B_\mu,B_\nu], 
\end{equation}
and the covariant derivatives are
\begin{align}
D_\mu G_{\nu\sigma} = {} & \partial_\mu G_{\nu\sigma} +
[B_\mu,G_{\nu\sigma}], \\
D_\mu \chi  = {} & \big(\partial_\mu + B_\mu\big)\chi\,, \quad \chibar\overleftarrow{D}_\mu  =
{}  \chibar \big(\overleftarrow{\partial}_\mu - B_\mu\big)\,.
\end{align}
The bulk fields are related via Dirichlet boundary conditions 
to the boundary fields, that is, the integration variables of the functional integral
defining the theory, through
\begin{align}\label{eq:ymbc}
B_\mu(x;t=0) = {} & A_\mu(x), \\
\chi(x;t=0) = {} & \psi(x), \\
\bar{\chi}(x;t=0) = {} & \bar{\psi}(x).
\end{align}
The generalized gauge-fixing terms proportional to  $\alpha_0$
remove some technical complications associated with perturbation theory~\cite{Luscher:2010iy,Luscher:2011bx,Luscher:2013cpa}.
The solutions of the flow equations for $\alpha_0 >0$ are related to the solutions
at $\alpha_0 = 0$ by a flow-time dependent gauge transformation.
We work in Feynman gauge and take $\alpha_0 = 1$ throughout this work.

We solve the flow equations
~\eqref{eq:ymflow} and~\eqref{eq:fermflow} in $d$-dimensions
by casting them into the integral forms
\begin{align}
B_\mu(x;t) = {} & \int \mathrm{d}^dy\bigg[K_{\mu\nu}(x-y;t)A_\nu(y) + \int_0^t \mathrm{d}s 
K_{\mu\nu}(x-y;t - s)R_\nu(y;s)\bigg], \label{eq:Bint}\\
\chi(x,t) = {} & \int \mathrm{d}^dy\, 
\bigg[J(x-y; t)\psi(y) + \int_0^t \mathrm{d}s\, J(x-y;t-s)\Delta'\chi(y;s)\bigg]\,, \label{eq:chiint}\\
\chibar(x,t) = {} & \int \mathrm{d}^dy\, \bigg[\psibar(y)\bar{J}(x-y; t) + \int_0^t \mathrm{d}s\,
\chibar(y;s)\overleftarrow{\Delta}'\bar{J}(x-y;t-s)\bigg]\,.
\label{eq:chibarint}
\end{align}
Here $K_{\mu\nu}(x;t)$ and $J(x;t)$ are the heat kernels
\begin{align}
K_{\mu\nu}(x;t) = {} & \int_p 
\frac{e^{ipx}}{p^2}
\left\{\Big(\delta_{\mu\nu}p^2 - p_\mu p_\nu\Big) e^{-t p^2} + p_\mu p_\nu {\rm e}^{-\alpha_0 t p^2}\right\},
\label{eq:kdef}\\
J(x;t) = {} & \bar{J}(x;t) = \int_p e^{ipx} e^{-t p^2}, \label{eq:jdef}\\
\end{align}
and the interaction terms are
\begin{align}
  R_\mu = {} & 2[B_\nu,\partial_\nu B_\mu]   - [B_\nu, \partial_\mu B_\nu] +
  (\alpha_0 -1) [B_\mu,\partial_\nu B_\nu] + [B_\nu, [B_\nu,B_\mu]], \\
\Delta' = {} & (1-\alpha_0)\partial_\nu B_\nu + 2B_\nu \partial_\nu + B_\nu B_\nu\,, \\
\overleftarrow{\Delta}' = {} & -(1-\alpha_0)\partial_\nu B_\nu - 2\overleftarrow{\partial}_\nu B_\nu + B_\nu B_\nu\,.
\label{eq:remainder}
\end{align}

We can solve the integral form of the flow 
equations, Eqs.~\eqref{eq:Bint} and \eqref{eq:chiint}, by iteration, generating a tree expansion
of the bulk fields in powers of the boundary fields.
Bulk vertices are then connected by ``flow lines'',
which are flow-time ordered and 
governed by the heat kernel.
We give explicit expressions for the relevant
Feynman rules in Appendix~\ref{app:rules}.

In pure Yang-Mills theory, all correlation functions are finite at finite flow time,
provided the boundary theory is renormalized \cite{Luscher:2011bx}.
Fermions, however, require an additional wave-function renormalization at
finite flow time, generally denoted by $Z_\chi$ \cite{Luscher:2013cpa}.
The pole contribution to this additional fermionic wave-function renormalization first appeared in \cite{Luscher:2013cpa} and was reproduced in \cite{Makino:2014taa},
through a next-to-leading-order perturbative calculation of
$\langle \bar{\chi}(x;t)\gamma_\mu \overleftrightarrow{D_\mu}\chi(x;t))\rangle$,
and in \cite{Monahan:2017hpu}, in the context of nonlocal Wilson-line operators.
In Appendix~\ref{app:Zfermion} we calculate the finite contributions
to this extra wave-function renormalization that, to our knowledge, have not appeared in the literature.
The calculation in Appendix~~\ref{app:Zfermion} also serves as a sample calculation with flowed
fermions fields.

Once the fermions have been renormalized, composite operators composed
of fields at finite flow time are therefore finite and
all scale dependence carried by these operators can be related to the flow time. In particular, any potential power divergence in the cutoff of the theory is removed. 
At small flow times, a short flow-time expansion (SFTE)
can be used to relate these composite operators to linear combinations
of local renormalized operators at vanishing flow time.
The SFTE is an operator product expansion in the neighborhood
of vanishing flow time, with coefficients, calculable in perturbation theory,
that carry the flow time dependence~\cite{Luscher:2013vga}.  The SFTE provides a perturbative understanding of the way in which power divergences are removed and the form of the flow-time dependence for which the power divergences are traded.

On the lattice, correlation functions involving higher-dimension operators
can be plagued by power-divergent mixings with lower-dimension operators.
In large volume calculations, the only accessible energy scale
is the inverse lattice spacing $\sim 1/a$, so the regularization and
renormalization of correlation functions may depend only on the lattice spacing. Disentangling the dual roles of the lattice spacing, as cutoff and as energy renormalization scale, can be arduous,
particularly in the presence of power divergences,
which must be removed non-perturbatively.
	
The gradient flow provides a workaround:
the flow renders all operators finite, and, in the continuum limit,
the scale of all flowed correlators is parameterized by the flow time, $\mu^2\propto1/t$.
The SFTE then provides a method to extract renormalized  operators evaluated at $t=0$
from finite operators calculated on the lattice at finite flow time, $t>0$.
In other words, we calculate correlation functions of local operators
at non-vanishing flow time and relate them to physical correlation functions of boundary operators via a SFTE.
The challenge associated with the renormalization of the correlators at $t=0$ is traded for the difficulty
of determining the expansion coefficients in the SFTE.
One advantage of the SFTE, however, is that we can perform
the analysis in the continuum, thus avoiding spurious chiral-symmetry breaking effects. In addition, the SFTE connects operators at several
values of the flow time in a gauge-invariant way. This is a significant advantage compared to standard techniques, based for example on
RI-MOM schemes, where determining
the coefficients of the power divergent terms requires a non-perturbative gauge-fixing procedure~\cite{Martinelli:1993dq,Martinelli:1994ty,Donini:1999sf,Bhattacharya:2015rsa,Cirigliano:2020msr}. An alternative gauge-invariant way to study power divergences is to use
coordinate space renormalization methods~\cite{Gimenez:2004me,Chetyrkin:2010dx,Tomii:2018zix,Izubuchi:2020ngl}, although this does not provide a continuous probe of the fields, in practice.

We consider our theory in continuum Euclidean $4$-dimensional space-time.
For some gauge-invariant and local operator $\mathcal{O}_i(t)$
in an associative operator algebra with basis $B$, defined at flow time $t$,
the SFTE is~\cite{Luscher:2013vga}
\be
\left(\mcO_i\right)_R(t)\stackrel{t\rightarrow0}{\widesim{}}\sum_{\mathcal{O}_j\in B}c_{ij}(t)\left(\mcO_j\right)_R(0)\,,
\label{eq:sfte}
\ee
where the label $R$ denotes a renormalized operator. 
Here, the Wilson, or expansion, coefficients
$c_{ij}(t)$ have absorbed all flow time dependence,
and the SFTE connects renormalized operators
in the bulk and on the physical boundary.
The SFTE is valid only if all fields are renormalized and
all operators appearing in the SFTE are evaluated in correlation functions
at non-zero physical distances to avoid spurious and additional contact terms.

If the renormalized operators at vanishing flow time
do not share the symmetries of the flowed operator,
their expansion coefficients vanish.
More specifically, the form of the SFTE and the operators contributing to the SFTE
are dictated by the symmetries of the regulated theory. Thus, if our regulator breaks
certain symmetries, those symmetries cannot be used to classify all the operators
$\left(\mcO_j\right)_R(0)$ contributing to the right-hand side of the SFTE in Eq.~\eqref{eq:sfte}.
The leading contributions in the SFTE stem from the
lowest dimension operators and 
the renormalization group equation satisfied by the expansion coefficients
dictates their asymptotic behavior at short flow time. 
In general, OPE's are linear and gauge-independent, so
we are free to study the expansion in an arbitrary correlation function. 
Hence we are able, with the appropriate choice of external probes, 
to study the SFTE termwise, order-by-order. Moreover, the Wilson coefficients of the SFTE are universal,
that is, the coefficients are insensitive to our choice of external states.
This universality ensures that, once the Wilson coefficients
are determined using one particular choice of external state,
the resulting coefficients can be used with any other
choice of external state.

\section{Quark-chromo electric dipole moment}
\label{sec:qcedm}

The effects of BSM physics at high energies can generate a set of effective, dimension-six, CP-violating operators
at the electroweak scale. The five-dimensional qCEDM operator, which induces the nEDM at low energies, arises from the effects of electroweak symmetry breaking on the CP-violating Gluon-Higgs-Fermion operator~\cite{Engel:2013lsa}.
We define the bare qCEDM to be
\be
\mcO_C=k_C\psibar\widetilde{\sigma}_{\mu\nu}G_{\mu\nu}\psi,
\ee
where
\be
\widetilde{\sigma}_{\mu\nu}=\frac{1}{2}\left\{\sigma_{\mu\nu},\gamma_5\right\}
\ee
is a generalization of $\sigma_{\mu\nu}\gamma_5$ that preserves
Hermiticity in $d$ dimensions \cite{Bhattacharya:2015rsa}.
All operators in this paper carry an arbitrary normalization factor,
to simplify comparison to other results; in this case, $k_C$ is a
complex number normalizing $\mathcal{O}_C$.

The calculation of a renormalized qCEDM matrix element on the lattice is plagued by the presence of mixing with the other
CP-violating operators~\cite{Bhattacharya:2015rsa}.
In particular the mixing with the lower-dimensional pseudoscalar density
\be
P=k_P\psibar\gamma_5\psi\,,
\ee
generates power divergences in the lattice spacing $a$.
A second lower-dimensional operator that mixes with the qCEDM is the topological
charge density (TCD)
\be
q=k_q\text{Tr}\left[G_{\mu\nu}\widetilde{G}_{\mu\nu}\right]\,, \qquad
\widetilde{G}_{\mu\nu} = \frac{1}{2}\varepsilon_{\mu\nu\alpha\beta}G_{\alpha\beta}\,. 
\ee
The chirality of the TCD, opposite to the qCEDM, ensures the mixing is proportional to the quark mass.
Our perturbative results confirm our expectations for the form of the power divergence
and the mass dependence of the pseudoscalar density and the TCD, respectively.
We remark that, if the lattice QCD calculation is performed with chiral symmetry breaking terms in the lattice action,
chirality no longer protects the mixing of the qCEDM and the TCD and therefore a linearly divergent term in the inverse lattice spacing $1/a$ can arise.
Although other operators of the same dimension mix with the qCEDM, in this work we focus on the calculation of the SFTE coefficients
of the lower dimensional pseudoscalar density and TCD operators.

The SFTE for the qCEDM reads
\be
\label{eq:SFTEqCEDM}
\left(\mcO_C\right)_R(t)\stackrel{t\rightarrow0}{\widesim{}}c_{CP}(t)P_R(0)+c_{Cq}(t)q_R(0)+ \cdots,
\ee
where we have retained only the lowest-dimension operators in the expansion. For the remainder of this section we will not include contributions from higher dimensional operators, such as the renormalized qCEDM itself. The study of this logarithmic mixing will be considered in future work. In perturbation theory it is possible to extract the lowest dimensional operator
contributions by selecting appropriate external sources~\cite{Panagopoulos:1989zn}. Working at ${\cal O}(g^2)$, we can extract the expansion coefficients of the pseudoscalar and TCD by selecting two-fermion and two-gluon sources, respectively. With these choices of external states, and at ${\cal O}(g^2)$ in perturbation theory, higher dimensional operators do not contribute to the expansion coefficients.

\subsection{Mixing with the pseudoscalar density}
\label{ssec:pseudo}

We start by extracting the coefficient for the pseudoscalar density.
Choosing a two-fermion external state, we define, for any operator $\mathcal{O}$,
the connected correlation functions~\footnote{We will not consider contributions
  from totally disconnected diagrams.}
\be
\label{eq:GammafCf}
\Gamma_{\psi\mathcal{O}\bar{\psi}}(x,y;t)=\left\langle\psi(x)\mathcal{O}(t)\bar{\psi}(y)\right\rangle.
\ee
We may then distribute over Eq.~\eqref{eq:SFTEqCEDM}, so that
\be
\Gamma^R_{\psi\mathcal{O}_C\bar{\psi}}(x,y;t)=c_{CP}(t)\Gamma^R_{\psi P \bar{\psi}}(x,y;0)+
c_{Cq}(t)\Gamma^R_{\psi q \bar{\psi}}(x,y;0)+\cdots
\label{eq:SFTEGamma}
\ee
where $R$ once again denotes a renormalized quantity.
Expanding both the correlation functions and the Wilson
coefficients in powers of the renormalized coupling $g$, we find
\be
\begin{aligned}
  g^2\Gamma^{(1)R}_{\psi\mathcal{O}_C\bar{\psi}}(x,y;t)
  =&
  \left[c_{CP}^{(0)}(t)
    +
    g^2c_{CP}^{(1)}(t)\right]
  \left[\Gamma^{(0)R}_{\psi\mathcal{O}_P\bar{\psi}}(x,y;0)
    +
    g^2\Gamma^{(1)R}_{\psi\mathcal{O}_P\bar{\psi}}(x,y;0)\right]\\
  &+
  \left[c_{Cq}^{(0)}(t)
    +
    g^2c_{Cq}^{(1)}(t)\right]
  \left[\Gamma^{(0)R}_{\psi\mathcal{O}_q\bar{\psi}}(x,y;0)
    +
    g^2\Gamma^{(1)R}_{\psi\mathcal{O}_q\bar{\psi}}(x,y;0)\right]\\
  &+
  \mathcal{O}(g^4) + \cdots\,,
\end{aligned}
\ee
where the first term in the expansion of the left hand side of Eq.~\eqref{eq:SFTEGamma}
vanishes because the correlator $\Gamma^{R}_{\psi\mathcal{O}_C\bar{\psi}}(x,y;t)$ has no tree-level contributions, that is, the first term in the expansion of this correlator is the one-loop contribution proportional to $g^2$.

Equating terms order-by-order and neglecting higher dimensional operators we obtain, up to ${\cal O}(g^4)$,
\begin{subequations}
	\begin{align}
	  0&=c_{CP}^{(0)}(t)\Gamma^{(0)}_{\psi\mathcal{O}_P\bar{\psi}}(x,y;0)+
          c_{Cq}^{(0)}(t)\Gamma^{(0)}_{\psi\mathcal{O}_q\bar{\psi}}(x,y;0)\,,\\
		\Gamma^{(1)}_{\psi\mathcal{O}_C\bar{\psi}}(x,y;t)&=
			\begin{aligned}[t]
			  &c_{CP}^{(0)}(t)\Gamma^{(1)}_{\psi\mathcal{O}_P\bar{\psi}}(x,y;0)+
                          c_{CP}^{(1)}(t)\Gamma^{(0)}_{\psi\mathcal{O}_P\bar{\psi}}(x,y;0)+
                          c_{Cq}^{(0)}(t)\Gamma^{(1)}_{\psi\mathcal{O}_q\bar{\psi}}(x,y;0)\\
			  &+c_{Cq}^{(1)}(t)\Gamma^{(0)}_{\psi\mathcal{O}_q\bar{\psi}}(x,y;0)\,.
			\end{aligned}
		\label{eq:fCf1}
	\end{align}
\end{subequations}
The TCD vanishes at tree-level with two external quarks,
$\Gamma^{(0)}_{\psi\mathcal{O}_q\bar{\psi}}(x,y;0) = 0$, and we obtain
\be
c_{CP}^{(0)}(t)\Gamma^{(0)}_{\psi\mathcal{O}_P\bar{\psi}}(x,y;0)=0\,.
\ee
The tree-level of the pseudoscalar density with two external quarks
does not vanish $\Gamma^{(0)}_{\psi\mathcal{O}_P\bar{\psi}}(x,y;0) \neq 0$,
implying that the expansion coefficient $c_{CP}(t)$
vanishes at leading order, $c_{CP}^{(0)}(t) = 0$.
Applying this to Eq.~\eqref{eq:fCf1}, we have
\be
\Gamma^{(1)}_{\psi\mathcal{O}_C\bar{\psi}}(x,y;t)=c_{CP}^{(1)}(t)\Gamma^{(0)}_{\psi\mathcal{O}_P\bar{\psi}}(x,y;0)+c_{Cq}^{(0)}(t)\Gamma^{(1)R}_{\psi\mathcal{O}_q\bar{\psi}}(x,y;0)\,.
\ee	
To extract $c_{Cq}(t)$ at leading order, we choose an external state with two gluons and define
\be
\label{eq:SFTEGammaACA}
\Gamma_{A\mathcal{O}A}(x,y;t)_{\alpha\beta}^{ab}=\left\langle A_\alpha^a(x)\mathcal{O}(t)A_\beta^b(y)\right\rangle,
\ee
in analogy to Eq.~\eqref{eq:GammafCf}. Applying the methods and results from above, 
\begin{subequations}
	\begin{align}
        0&  = c_ {Cq}^{(0)}(t)\Gamma^{(0)}_{A\mathcal{O}_qA}(x,y;0)\,, \\
	\Gamma^{(1)}_{A\mathcal{O}_C A}(x,y;t)&=c_{Cq}^{(0)}(t)\Gamma^{(1)}_{A\mathcal{O}_qA}(x,y;0)+
        c_{Cq}^{(1)}(t)\Gamma^{(0)}_{A\mathcal{O}_qA}(x,y;0)\,,
	\end{align}
\end{subequations}
because the tree-level of the qCEDM with $2$ external gluons vanishes,
$\Gamma^{(0)}_{A\mathcal{O}_C A}(x,y;t)=0$,
and $c_{CP}^{(0)}(t) = 0$.
The tree-level contribution to the TCD does not vanish, $\Gamma^{(0)}_{A\mathcal{O}_qA}(x,y;0) \neq 0$, from which we deduce that the leading order of the expansion coefficient
$c_{Cq}^{(0)}(t)$ vanishes, $c_{Cq}^{(0)}(t)=0$. 

\begin{figure}[h]
	\begin{minipage}[b]{.5\linewidth}
		\centering
		\subfloat[][$g^2\Gamma^{(1)a}_{\psi\mathcal{O}_C\bar{\psi}}$]{\label{fig:fCfa}\includegraphics[width=142pt,height=29pt]{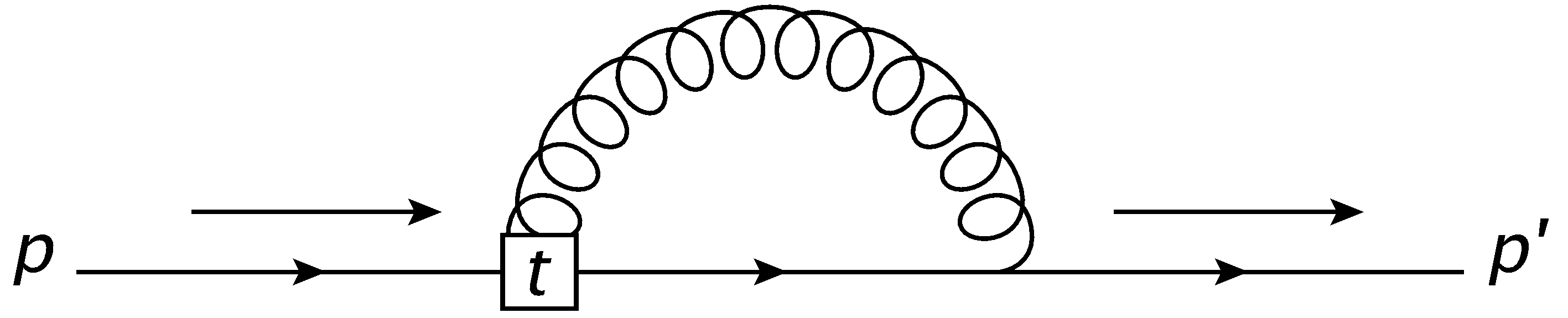}}
	\end{minipage}%
	\begin{minipage}[b]{.5\linewidth}
		\centering
		\subfloat[][$g^2\Gamma^{(1)b}_{\psi\mathcal{O}_C\bar{\psi}}$]{\label{fig:fCfb}\includegraphics[width=142pt,height=29pt]{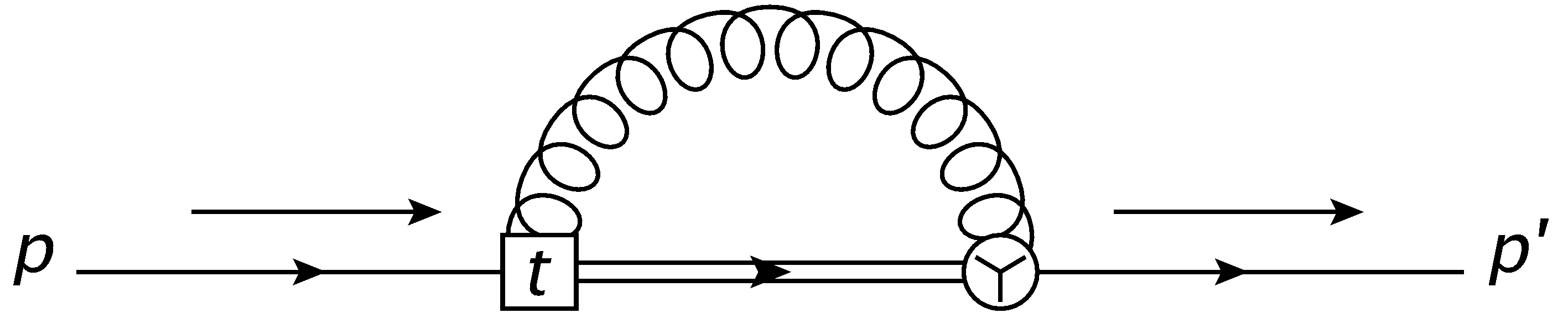}}
	\end{minipage}\par\medskip
	\centering
	\subfloat[][$g^2\Gamma^{(1)c}_{\psi\mathcal{O}_C\bar{\psi}}$]{\label{fig:fCfc}\includegraphics[width=142pt,height=50pt]{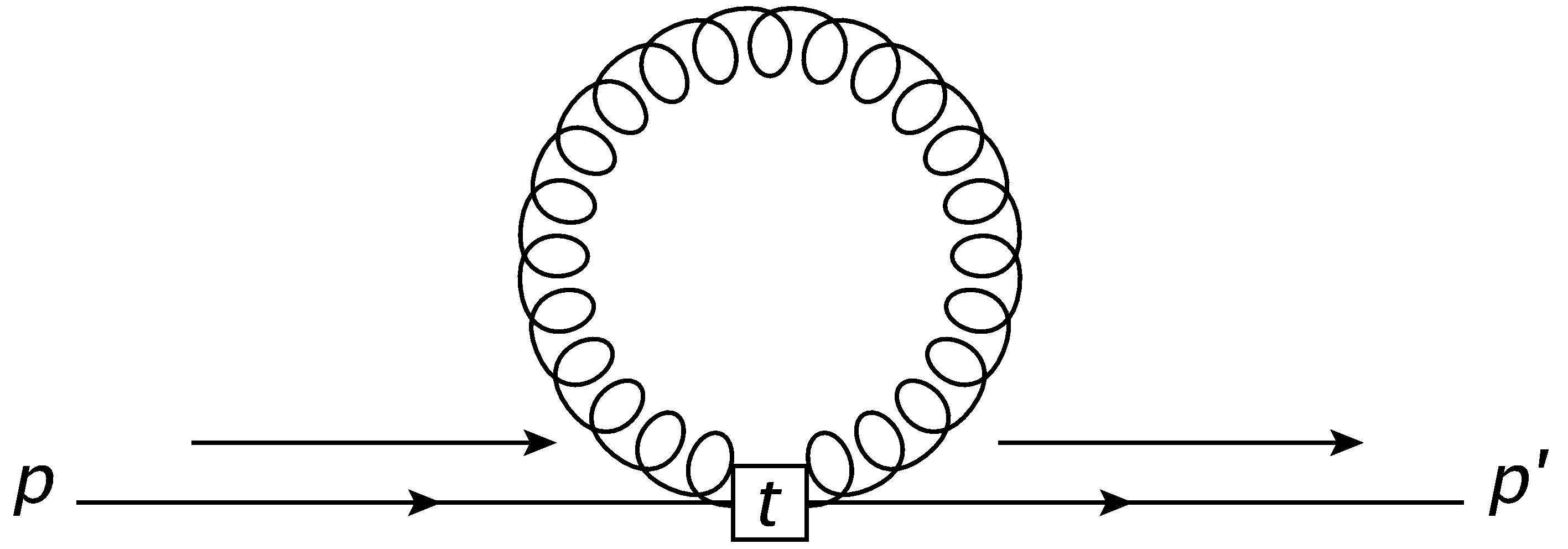}}
	\caption{Leading order contributions to the mixing of the pseudoscalar density with the qCEDM}
	\label{fig:fCfGraphs}
\end{figure}

To summarize, at ${\cal O}(g^2)$ we obtain
\begin{subequations}
	\begin{align}
		&\label{eq:c_CP}
		\Gamma^{(1)}_{\psi\mathcal{O}_C\bar{\psi}}(x,y;t)=c_{CP}^{(1)}(t)\Gamma^{(0)}_{\psi\mathcal{O}_P\bar{\psi}}(x,y;0)\,,\\
		&\label{eq:c_Cq}
		\Gamma^{(1)}_{A\mathcal{O}_CA}(x,y;t)=c_{Cq}^{(1)}(t)\Gamma^{(0)}_{A\mathcal{O}_qA}(x,y;0)\,.
	\end{align}
\end{subequations}
We are now in a position to extract $c_{CP}^{(1)}(t)$.
There are three one-loop graphs that contribute to the left-hand-side of Eq.~\eqref{eq:c_CP}, which we show in Fig.~\ref{fig:fCfGraphs}),
and the correlator on the right-hand side is simply the tree-level for the pseudoscalar density. 

We calculate these graphs to all orders in the external momenta and flow time. The inclusion of the mass results in a particularly cumbersome asymptotic analysis that lies outside the scope of this paper; a nonzero external momentum is sufficient to regulate all infrared divergences. The Feynman rules and mathematical details can be found in
Apps.~(\ref{app:conventions}, \ref{app:rules}, and \ref{app:Zfermion}).
Additional mathematical details can be found in ref.~\cite{Rizik:new}.
We expand in powers of the quark mass and flow time to obtain
\begin{subequations}
	\label{fCfResults}
	\begin{align}
		&\label{eq:fCfa}
		\widetilde{\Gamma}^{(1)a}_{\psi\mathcal{O}_C\bar{\psi}}(p,p';t)=3i\frac{k_C}{k_P}\frac{C_2(F)}{(4\pi)^2}\left\{\frac{1}{t}+p^2\left[\log\left(2p^2t\right)+\gamma_E-\frac{11}{4}\right]\right\}\cdot\gamma_5+\mathcal{O}(m,p^2t),\\
		&\widetilde{\Gamma}^{(1)b}_{\psi\mathcal{O}_C\bar{\psi}}(p,p';t)=0,\\
		&\widetilde{\Gamma}^{(1)c}_{\psi\mathcal{O}_C\bar{\psi}}(p,p';t)=0.
	\end{align}
\end{subequations}

There are symmetric counterparts for diagrams (a) and (b), so the sum of these contributions is
\be
\begin{aligned}
  \Gamma^{(1)}_{\psi\mathcal{O}_C\bar{\psi}}(x,y;t)
  &=\int d^4z\int_{p,p'}\frac{e^{ip(x-z)}}{i\slashed{p}+m} \left[2\widetilde{\Gamma}^a_{\psi\mathcal{O}_C\bar{\psi}}(p,p';t)+2\widetilde{\Gamma}^b_{\psi\mathcal{O}_C\bar{\psi}}(p,p';t)+\widetilde{\Gamma}^c_{\psi\mathcal{O}_C\bar{\psi}}(p,p';t)\right]\frac{e^{ip'(y-z)}}{i\slashed{p'}+m}\\
  &=6i\frac{k_C}{k_P}\frac{C_2(F)}{(4\pi)^2}\int_{p,p'}\left[\frac{1}{t}+p^2\left(\log\left(2p^2t\right)+\gamma_E-\frac{11}{4}\right)\right]\int d^4z\frac{e^{ip(x-z)}}{i\slashed{p}+m}\gamma_5\frac{e^{ip'(y-z)}}{i\slashed{p'}+m}\\
  &=6i\frac{k_C}{k_P}\frac{C_2(F)}{(4\pi)^2}\left\{\frac{1}{t}+p^2\left[\log\left(2p^2t\right)+\gamma_E-\frac{11}{4}\right]\right\}\Gamma_{\psi\mathcal{O}_P\bar{\psi}}^{(0)}(x,y;0)\,,
\end{aligned}
\ee
where we have omitted higher order corrections in flow time and quark mass.
The final expression for the expansion coefficient reads 
\be
c_{CP}(t)=6ig^2\frac{k_C}{k_P}\frac{C_2(F)}{(4\pi)^2}\left\{\frac{1}{t}+p^2\left[\log\left(2p^2t\right)+\gamma_E-\frac{11}{4}\right]\right\}+\mathcal{O}(m,p^2t,g^4).
\ee
We confirm the general expectation, based on symmetry and dimensional
considerations, that the dominant contribution to the SFTE of the qCEDM is the pseudoscalar density, which has a corresponding expansion coefficient that diverges linearly 
in flow time.

\subsection{Mixing with the topological charge density}

\begin{figure}[h]
	\begin{minipage}[b]{.5\linewidth}
		\centering
		\subfloat[][$g^2\Gamma^{(1)a}_{A\mathcal{O}_CA}$]{\label{fig:ACAa}\includegraphics[width=140pt,height=26pt]{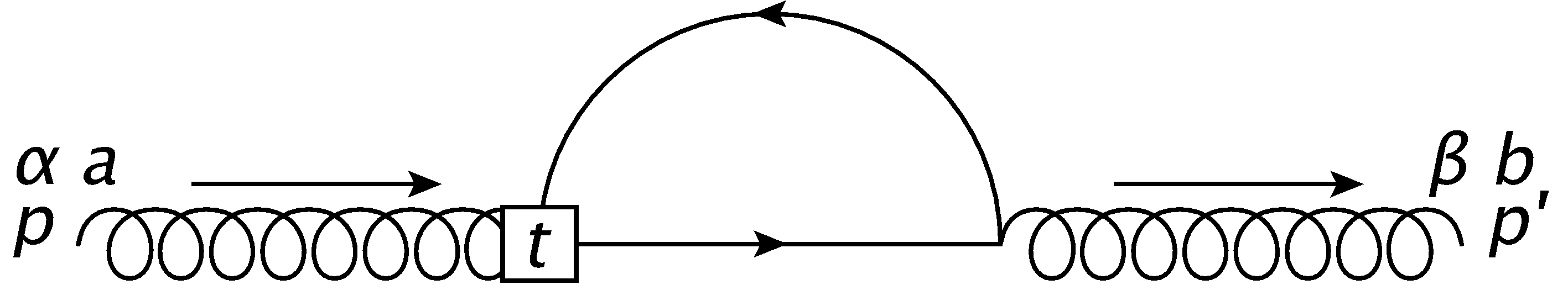}}
	\end{minipage}%
	\begin{minipage}[b]{.5\linewidth}
		\centering
		\subfloat[][$g^2\Gamma^{(1)b}_{A\mathcal{O}_CA}$]{\label{fig:ACAb}\includegraphics[width=140pt,height=26pt]{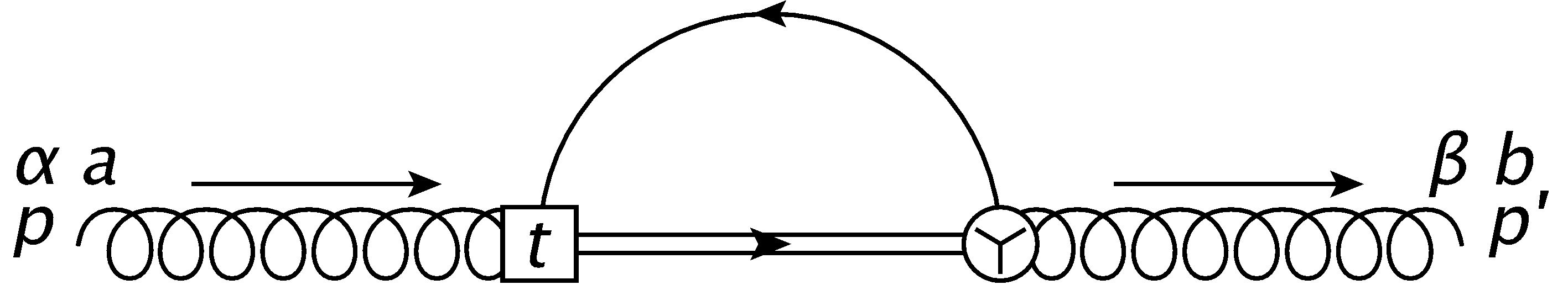}}
	\end{minipage}\par\medskip
	\centering
	\subfloat[][$g^2\Gamma^{(1)c}_{A\mathcal{O}_CA}$]{\label{fig:ACAc}\includegraphics[width=140pt,height=47pt]{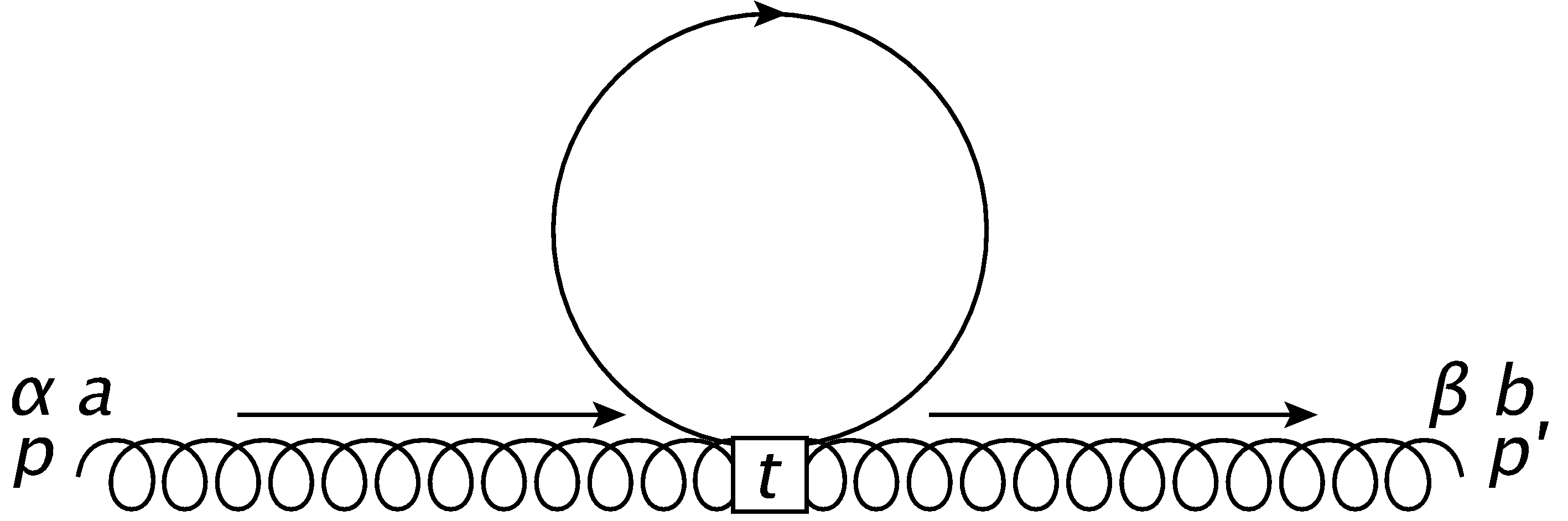}}
	\caption{Leading order contributions to the mixing of the TCD with the qCEDM.}
	\label{fig:ACAGraphs}
\end{figure}

To calculate the expansion coefficient $c_{Cq}(t)$, following Eq.~\eqref{eq:c_Cq},
we need to calculate the one-loop contribution $\Gamma^{(1)}_{A\mathcal{O}_CA}(x,y;t)$, stemming from 
the three Feynman diagrams shown in Fig.~\ref{fig:ACAGraphs}.
The graphs displayed in both~\ref{fig:ACAb} and~\ref{fig:ACAc}
vanish under the traces of the fermion loops, so we are again left to calculate a single Feynman graph.
To calculate the $d$-dimensional traces over fermion loops
one could employ the 't Hooft-Veltman-Breitenlohner-Maison (HVBM)
scheme~\cite{tHooft:1972tcz,Breitenlohner:1977hr,Buras:1989xd}.
Our conventions and details on the HVBM scheme can be found in
Appendix~\ref{app:conventions}. This is, however, only necessary when these calculations are performed by expanding near $p^2=0$. Starting at $\mcO(p^2)$, this removes an essential IR regulator, the momentum, and introduces spurious divergences. The correlators listed below have been calculated by applying
a new method that includes all orders in momenta, so our results are IR safe.
The flow further removes all UV divergences, and the diagrams are finite in four dimensions.

Following the methods outlined for the pseudoscalar density, we obtain
\begin{subequations}
	\label{ACAResults}
	\begin{align}
		&\widetilde{\Gamma}^{(1)a}_{A\mathcal{O}_CA}(p,p';t)=2i\frac{k_C}{k_q}\frac{m}{(4\pi)^2}\left[\log(2p^2t)+\gamma_E-1\right]\cdot(-2k_q)\delta^{ab}\epsilon_{\alpha\beta\mu\nu}p_\mu p'_\nu+\mathcal{O}(m_0,p^2t),\\
		&\widetilde{\Gamma}^{(1)b}_{A\mathcal{O}_CA}(p,p';t)=0,\\
		&\widetilde{\Gamma}^{(1)c}_{A\mathcal{O}_CA}(p,p';t)=0.
	\end{align}
\end{subequations}
We therefore find
\be
	\begin{aligned}
		\Gamma^{(1)}_{A\mathcal{O}_CA}(x,y;t)
		&=\int d^4z\int_{p,p'}\frac{e^{ip(x-z)}}{p^2}\left[2\widetilde{\Gamma}^{(1)a}_{A\mathcal{O}_CA}(p,p';t)\right]\frac{e^{ip'(y-z)}}{p'^2}\\
		&=4i\frac{k_C}{k_q}\frac{m}{(4\pi)^2}\left[\log(2p^2t)+\gamma_E-1\right]\Gamma_{A\mathcal{O}_qA}^{(0)R}(x,y;0)+\mathcal{O}(m^2,t),
	\end{aligned}
\ee
and
\be
c_{Cq}(t)=4ig^2\frac{k_C}{k_q}\frac{m}{(4\pi)^2}\left[\log(2p^2t)+\gamma_E-1\right]+\mathcal{O}(m^2,t,g^4)\,.
\ee
We again confirm, following general chiral symmetry considerations, that
the expansion coefficient for the TCD has a logarithmic dependence on the flow time.
Chiral symmetry enforces the presence of a quark mass factor multiplying the TCD and this factor arises naturally in our calculation.

Then, at non-zero mass, the qCEDM behaves, to leading-order, as
\be
	\begin{aligned}
		\mathcal{O}_C^R(t)\stackrel{t\rightarrow0}{\widesim{}}\ &6ig^2\frac{k_C}{k_P}\frac{C_2(F)}{(4\pi)^2}\left\{\frac{1}{t}+p^2\left[\log\left(2p^2t\right)+\gamma_E-\frac{11}{4}\right]\right\}\mathcal{O}_P^R(0)\\
		&+4ig^2\frac{k_C}{k_q}\frac{m}{(4\pi)^2}\left[\log(2p^2t)+\gamma_E-1\right]\mathcal{O}_q^R(0)+\cdots\,,
	\end{aligned}
\ee
where the dots indicate contributions from renormalized higher-dimensional operators.

\section{Weinberg operator}
\label{sec:weinbergResults}

Among the higher dimensional CP-violating operators obtained by integrating out
heavy quarks and Higgs bosons, there is a dimension six gluonic operator, Weinberg's three-gluon operator~\cite{Weinberg:1989dx},
\be
\mcO_W=k_W\text{Tr}\left\{\left[G_{\mu\rho},G_{\nu\rho}\right]\tilde{G}_{\mu\nu}\right\}\,
\ee
The Weinberg operator could potentially generate a large contribution
to the nucleon EDM because it is purely gluonic and therefore not supressed by any small quark mass factor or by a small CKM phase.

To determine the SFTE of the Weinberg operator we need to isolate the lower dimensional
CP-violating operators with the same symmetry properties.
In principle, the pseudoscalar density, multiplied by a mass factor,
could contribute to the SFTE of the Weinberg,
but its leading contribution is ${\cal O}(g^4)$, because the first non-vanishing term of the correlator with
the Weinberg operator and $2$ external fermions arises at this order. 

As with the qCEDM operator, we do not consider the contributions of operators with the same dimension as     the
Weinberg operator. The operators that could potentially contribute to the SFTE of the Weinberg operator 
originate from terms proportional to $m \mcO_C$ and the Weinberg operator itself. By choosing external states of two quarks or two gluons, we can ensure that the leading contributions appear only at higher
order in the external scales, such as momentum and flow-time, or at higher order in the coupling.

Expanding the Weinberg operator at short flow time, in a manner similar to the qCEDM,
we obtain
\be
\mathcal{O}_W^R(t)\stackrel{t\rightarrow0}{\widesim{}}c_{Wq}(t)q_R(0)+ \cdots\,,
\ee
where we have considered only operators contributing to the expansion coeffcient $c_{Wq}(t)$.
These considerations confirm that the expansion coeffcient contribution
from the qCEDM to the SFTE of the Weinberg operator starts at ${\cal O}(g^2)$.

We choose two gauge bosons as the external state and expand in powers of the coupling, leading to
\be
\begin{aligned}
  g^2\Gamma_{A\mathcal{O}_WA}^{(1)R}(t)
  =&
  \left[c_{Wq}^{(0)}(t)
    +
    g^2c_{Wq}^{(1)}(t)\right]
  \left[\Gamma^{(0)}_{A\mathcal{O}_qA}(x,y;0)
    +
    g^2\Gamma^{(1)R}_{A\mathcal{O}_qA}(x,y;0)\right]\\
  &+
  \mathcal{O}(g^4).
\end{aligned}
\ee
Equating order-by-order in the coupling, we obtain
\begin{subequations}
  \begin{align}
    0&=c_{Wq}^{(0)}(t)\Gamma^{(0)}_{A\mathcal{O}_qA}(x,y;0)\,,\\
    \Gamma^{(1)}_{A\mathcal{O}_WA}(x,y;t)&=
    \begin{aligned}[t]
      &c_{Wq}^{(0)}(t)\Gamma^{(1)}_{A\mathcal{O}_qA}(x,y;0)+
      c_{Wq}^{(1)}(t)\Gamma^{(0)}_{A\mathcal{O}_qA}(x,y;0)\,.
    \end{aligned}
    \label{eq:AWA1}
  \end{align}
\end{subequations}
Thus the leading contribution to the expansion coefficient $c_{Wq}$ vanishes, $c_{Wq}^{(0)}=0$.
The next order in the coupling expansion reads
\be
\Gamma^{(1)}_{A\mathcal{O}_WA}(x,y;t)=c_{Wq}^{(1)}(t)\Gamma^{(0)}_{A\mathcal{O}_qA}(x,y;0)\,,
\ee
which allows us to determine $c_{Wq}^{(1)}(t)$ once we have determined the one-loop
contribution $\Gamma^{(1)}_{A\mathcal{O}_WA}(x,y;t)$.
\begin{figure}
	\begin{minipage}[b]{.5\linewidth}
		\centering
		\subfloat[][$g^2\Gamma^a_{A\mathcal{O}_WA}$]{\label{fig:AWAa}\includegraphics[width=142pt,height=29pt]{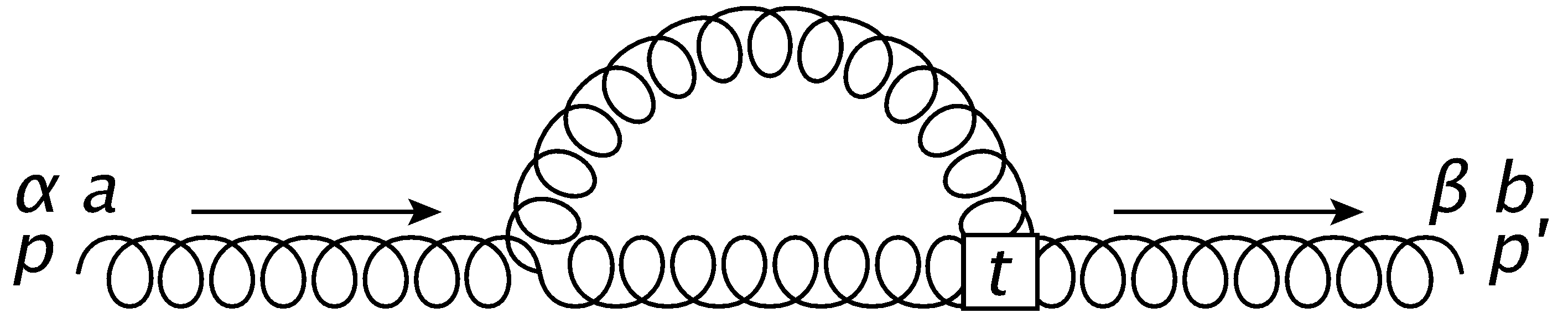}}
	\end{minipage}%
	\begin{minipage}[b]{.5\linewidth}
		\centering
		\subfloat[][$g^2\Gamma^b_{A\mathcal{O}_WA}$]{\label{fig:AWAb}\includegraphics[width=142pt,height=30pt]{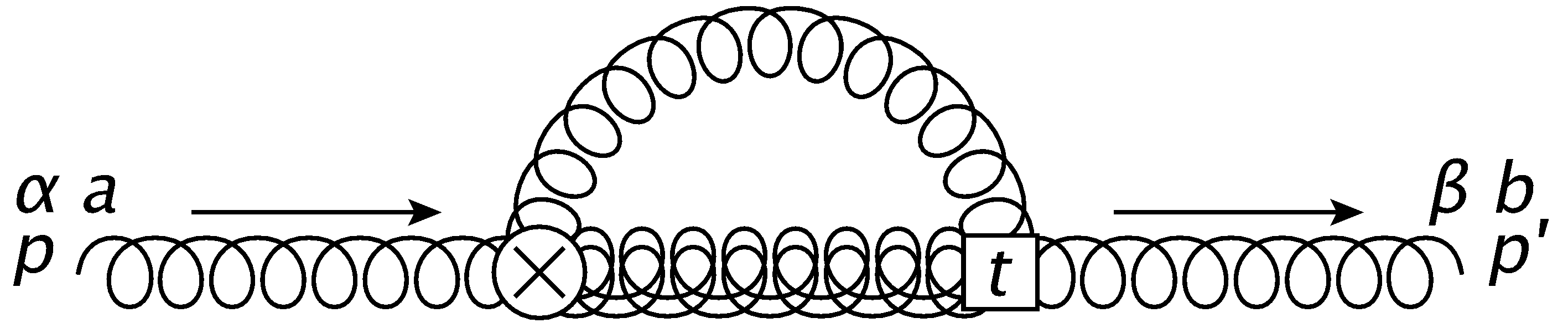}}
	\end{minipage}\par\medskip
	\centering
	\subfloat[][$g^2\Gamma^c_{A\mathcal{O}_WA}$]{\label{fig:AWAc}\includegraphics[width=142pt,height=50pt]{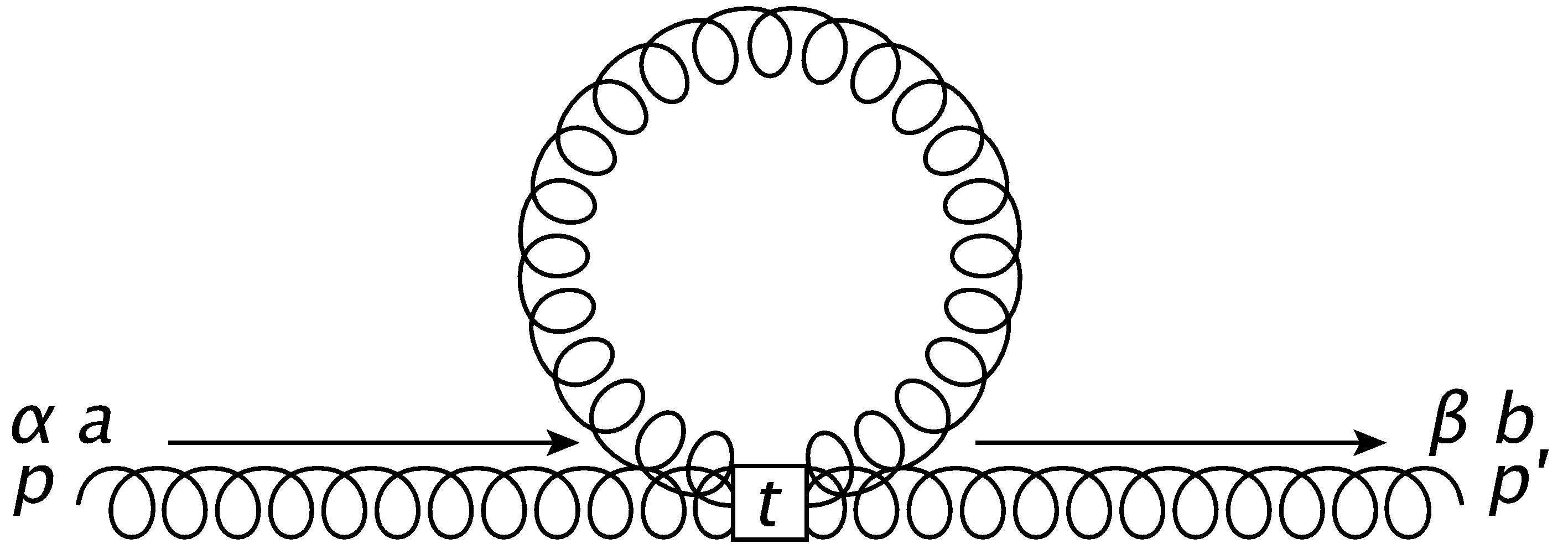}}
	\caption{Leading order contributions to the mixing of the TCD with the Weinberg operator}
	\label{fig:WeinbergGraphs}
\end{figure}
There are, once again, three Feynman graphs that contribute, which we show in Fig.~\ref{fig:WeinbergGraphs}.
There are a large number of equivalent permutations of the
fields of the Weinberg operator, so to simplify our calculations
we employ a relation valid for any alternating $2$-tensor
\be
A_{\mu\tau}A_{\nu\tau}A_{\rho\sigma}\epsilon_{\mu\nu\rho\sigma}=\frac{1}{16i}\text{Tr}\left[\sigma_{\alpha\beta}\sigma_{\gamma\delta}\sigma_{\epsilon\eta}\gamma_5\right]A_{\alpha\beta}A_{\gamma\delta}A_{\epsilon\eta},
\ee
which slightly generalizes the corresponding relation with Minkowski metric \cite{Braaten:1990zt,Braaten:1990gq}. This relation decouples the indices of $A$,
so that the permutations of any fields that may be contained in $A$
become well-defined permutations on the indices within the trace.
It should be noted that this formula is available in $d$-dimensions,
but upon evaluation we reproduce exactly the four-dimensional trace in the HVBM scheme, so it may only contract nontrivially with other four-dimensional structures.
This leaves only those pieces of a dimensionally-regularized
integral that take values in the four-dimensional subalgebra.

In the calculation of the correlators involving the Weinberg operator, the flow automatically regulates the UV modes of the 
the bulk gauge field, and the external momentum controls infrared divergences. Thus all integrals are finite in four dimensions. Inserting the field tensor $G$ in place of $A$,
we find a simple expression for the Weinberg operator conducive to perturbative calculations: 
\be
\mathcal{O}_W=k_W\text{Tr}\left\{\left[G_{\mu\rho},G_{\nu\rho}\right]\tilde{G}_{\mu\nu}\right\}=\frac{1}{64}ik_Wf^{abc}\text{Tr}\left[\sigma_{\alpha\beta}\sigma_{\gamma\delta}\sigma_{\epsilon\eta}\right]G^a_{\alpha\beta}G^b_{\gamma\delta}G^c_{\epsilon\eta}.
\ee
The Feynman rules for this operator are derived in Appendix~\ref{app:rules}.
The calculation of the Feynman diagrams in Fig.~\ref{fig:WeinbergGraphs} leads to
\begin{subequations}
  \label{AWAResults}
  \begin{align}
    &\tilde{\Gamma}^{(1)a}_{A\mathcal{O}_WA}(p,p';t)=-\frac{9}{4}\frac{k_W}{k_q}\frac{C_2(A)}{(4\pi)^2}\left\{\frac{1}{t}+\frac{2}{3}p^2\left[\log(2p^2t)+\gamma_E-\frac{25}{12}\right]\right\}\cdot(-2k_q)\delta^{ab}\epsilon_{\alpha\beta\mu\nu}p_\mu p'_\nu+\mathcal{O}\left(p^2t\right),\\
    &\tilde{\Gamma}^{(1)b}_{A\mathcal{O}_WA}(p,p';t)=-\frac{9}{16}\frac{k_W}{k_q}\frac{C_2(A)}{(4\pi)^2}\left\{\frac{1}{t}-\frac{5}{18}p^2\right\}\cdot(-2k_q)\delta^{ab}\epsilon_{\alpha\beta\mu\nu}p_\mu p'_\nu+\mathcal{O}\left(p^2t\right),\\
    &\tilde{\Gamma}^{(1)c}_{A\mathcal{O}_WA}(p,p';t)=0.
  \end{align}
\end{subequations}
The second diagram has no logarithmic divergence in the flow time; a kernel line appears in place of the gauge boson propagator, which generates two additional powers of the loop momentum.
The third diagram vanishes, because two of the legs
on the Weinberg operator are contracted, and the Weinberg operator
is antisymmetric with respect to its fields.
Summing these contributions and factoring out the tree-level
structure for the TCD,
we isolate the Weinberg operator's leading-order divergent behavior:
\be
\mathcal{O}_{W}^R(t)\stackrel{t\rightarrow0}{\widesim{}}-
\frac{45}{8} g^2 \frac{k_W}{k_q}\frac{C_2(A)}{(4\pi)^2}\left\{\frac{1}{t}+
\frac{8}{15}p^2\left[\log(2p^2t)+\gamma_E-\frac{35}{16}\right]\right\}
\mathcal{O}_{q}^R(0)+\cdots.
\ee
Our calculation again confirms the expectation that the leading contribution to the
SFTE of the Weinberg operator stems from the lowest dimensional operator with
the same symmetry properties; the TCD generates the linear
divergence of the Weinberg operator at short flow time.

\section{Summary and conclusions}
\label{sec:summary}

The nucleon electric dipole moment (EDM) provides a unique opportunity to probe of sources of charge and parity (CP) violation in the Standard Model and beyond (BSM). BSM theories that contain complex CP-violating couplings can induce a non-vanishing EDM, and at low energies one can parametrize the effects of the BSM degrees of freedom through effective, higher-dimensional CP-violating operators.

We have calculated, at one loop in perturbation theory, selected Wilson coefficients of the
short flow time expansion (SFTE) for two CP-violating operators: the quark-chromo EDM (qCEDM) and
the Weinberg operator.
We have studied the leading contributions generated by the
pseudoscalar density and the topological charge density, and confirmed the
general expectation that the lowest-dimensional operators generate the dominant contributions at short flow time.

For the qCEDM, the Wilson coefficient of the pseudoscalar density is proportional
to the inverse of the flow time, $1/t$, and we have calculated the corresponding coefficient. In addition, we have calculated the logarithmic contribution
to the qCEDM proportional to the topological charge density. Our calculation confirms the general expectation that chiral symmetry forces the
contribution of the topological charge density to be proportional to the quark mass.

For the Weinberg operator, the leading contribution, which is proportional to the inverse of the flow time,
stems from the topological charge density. We have determined both the coefficient of this $1/t$ term
and additional logarithmic terms.

Further, we have introduced a method of evaluation for flowed loop-integrals,
which permits, in many applications, the calculation of correlation functions
in a natural four-dimensional setting.
We fully avoid artificial divergences related to the zero-momentum
or zero-mass calculations,
while latently allowing for the study of these correlation functions
at any or all positive values of momentum or mass.
This also sidesteps the various problems that arise in continuing
the spacetime algebra to any arbitrary dimension.
This is particularly useful for our considerations,
since the source of potential technical difficulties, $\gamma_5$,
is pervasive in CP-odd calculations yet well-defined only in four dimensions.

Our calculation is intended to provide a new framework to study the ultraviolet behavior of CP-violating operators contributing to the electric dipole moment.
Ideally, the Wilson coefficients should be determined non-perturbatively and work in this direction is in progress~\cite{SymLat:2020}. Alternative strategies to pursue the same goals have been recently proposed based on coordinate space methods~\cite{Izubuchi:2020ngl} and the RI-MOM scheme~\cite{Bhattacharya:2015rsa,Cirigliano:2020msr}.
The one-loop calculation of the linearly divergent coefficients is also of practical importance
for the non-perturbative determination of the Wilson coefficient, by constraining the perturbative
behavior at small values of the gauge coupling.

We consider this calculation a first step toward the non-perturbative renormalization
of all CP-violating operators contributing to the EDM.
The next steps in our program are the non-perturbative determination of the 
linear divergence in the Wilson coefficients and a
perturbative analysis that includes higher dimensional operators
and their corresponding Wilson coefficients.

\begin{acknowledgments}
  We thank the members of the SymLat collaboration, Jack Dragos, Jangho Kim, Thomas Luu,
  Giovanni Pederiva, and Jordy de Vries for very useful discussions
  and a most enjoyable collaboration.
  In particular, we thank Jordy for his valuable
  insight in discussions regarding the renormalization properties of CP-violating operators
  and a careful reading of this manuscript.
  C.J.M. is supported in part by the U.S. Department of Energy,
  Office of Science,
  Office of Nuclear Physics under contract No.~DE-AC05-06OR23177.
  M.D.R. and A.S. acknowledge funding support under the
  National Science Foundation grant PHY-1913287.
\end{acknowledgments}

\appendix
\section{Conventions}
\label{app:conventions}

\subsection{$SU(N)$ Conventions}
First, we define the set of generators for the gauge group, $SU(N)$, to be traceless and skew-Hermitian, so that the algebra is defined by
\begin{equation}
	\left[t^a,t^b\right]=f^{abc}t^c,
	\label{eq:SUNCommutator}
\end{equation}
for the $N^2-1$ generators $t^a\in\mathfrak{su}(N)$, and for structure constants $f^{abc}$. For any representation $\rho:\text{SU}(N)\rightarrow GL(\mathbb{C})$, the trace over any two generators provides a natural Killing form for $\mathfrak{su}(N)$, normalized by the Dynkin index, $T_\rho=-\frac{\dim(\rho)}{\dim(\mathfrak{su}(N))}C_2(\rho)$, where $C_2(\rho)\mathbbm{1}_{\dim(\rho)}=-t^a_\rho t^a_\rho$ is the quadratic Casimir invariant. Thus, we have
\begin{equation}
	\text{Tr}\left\{t^a_\rho t^b_\rho\right\}=T_\rho\delta^{ab}.
\end{equation}

We now turn our attention to two particular representations, the fundamental ($F$) and the adjoint ($A$) representations, which have dimensions $N$ and $N^2-1$, respectively. In these cases, our Casimir elements are $C_2(F)=(N^2-1)/(2N)$ and $C_2(A)=N$, so the Dynkin indices become $T_F=-1/2$ and $T_A=-N$. Further, we can obtain an explicit set of generators for the adjoint representation by defining
\begin{equation}
	(t^a_A)_{bc}=-f^{abc}.
\end{equation}
Clearly this definition is traceless and skew-symmetric, and it is trivial to prove that $f^{abc}$ must be real. Moreover, the Jacobi identity for $f^{abc}$ implicitly satisfies \eqref{eq:SUNCommutator}, so that the $N^2-1$ matrices defined above indeed generate $SU(N)$. This allows for quick computations of objects such as
\begin{equation}
	f^{acd}f^{bcd}=C_2(A)\delta^{ab}.
\end{equation}

\subsection{Quantum Chromodynamics}
We work in $d$ dimensions with a Euclidean metric, taking the $d\rightarrow4$ limit at the end. 
For all momentum integrals, we adopt the shorthand notation
\begin{equation}
	\int_p=\mu^{4-d}\int_{\mathbb{R}^d}\frac{d^dp}{(2\pi)^d},
\end{equation}
where $\mu$ is the energy scale introduced in dimensional regularization. We also define Fourier transforms so that the factor of $(2\pi)^d$ appears only in the momentum space measure:
\begin{equation}
	\tilde{f}(p)=\int_{\mathbb{R}^d}d^dxf(x)e^{-ipx},\hfill f(x)=\int_{\mathbb{R}^d}\frac{d^dp}{(2\pi)^d}\tilde{f}(p)e^{ipx}.\hfill
\end{equation}
All calculations are performed on a QCD background, so that for any local operator $\mathcal{O}$, correlation functions are given by
\begin{equation}
	\langle\mathcal{O}\rangle=Z_0^{-1}\int\mathcal{D}\left[\bar{\psi},\psi,A,\cdots\right]\mathcal{O}e^{-\int d^dx\mathcal{L}\left[\bar{\psi},\psi,A,\cdots\right](x)},
\end{equation}
with the gauge-fixed Lagrangian
\begin{equation}
	\mathcal{L}=\bar{\psi}\left(\slashed{D}+m\right)\psi+\frac{1}{4}G_{\mu\nu}^aG_{\mu\nu}^a+\frac{1}{2\xi}\left(\partial_\mu A_\mu^a\right)\left(\partial_\nu A_\nu^a\right)+\left(\partial_\mu c^a\right)\left(\delta^{ab}\partial_\mu-f^{abc}A_\mu^c\right)\tilde{c}^b\,.
\end{equation}
The generators of $SU(N)$ were chosen to be skew-Hermitian, so the covariant derivative is simply
\begin{equation}
	D_\mu=\partial_\mu+A_\mu\,, \qquad A_\mu = A_\mu^at^a
\end{equation}
when acting on objects in the fundamental representation, where the coupling has been absorbed in to the fields, $A_\mu^a$. When acting on objects in the adjoint representation, it assumes the form
\begin{equation}
	D_\mu=\partial_\mu+[A_\mu,\cdot]\,.
\end{equation}
Then the field strength-tensor is
\begin{equation}
	G_{\mu\nu}=\partial_\mu A_\nu-\partial_\nu A_\mu+[A_\mu,A_\nu].
\end{equation}

\subsection{Higher-Dimensional Gamma Matrices}
This paper deals primarily with \CP-odd operators, so the parity-violating $\gamma_5$ is ubiquitous. To maintain algebraic consistency in generic $d$ dimensions, we follow 't Hooft, Veltman, Breitenlohner, and Maison \cite{tHooft:1972tcz,Breitenlohner:1977hr,Buras:1989xd} and split the geometric algebra into two disjoint, orthogonal subalgebras, one containing the four-dimensional gamma matrices, $\tilde{\gamma}_\mu$, and a second containing their $(d-4)$-dimensional ``evanescent'' extensions, $\hat{\gamma}_\mu$, where $\mu=1,2,\dots,d$. The $d$-dimensional algebra itself is a direct sum of the subalgebras, defined by the anticommutator
\begin{equation}
	\left\{\gamma_\mu,\gamma_\nu\right\}=2\delta_{\mu\nu},
\end{equation}
where the $d$-dimensional generalizations of the gamma matrices and metric tensor are given by
\begin{equation}
	\gamma_\mu=\tilde{\gamma}_\mu+\hat{\gamma}_\mu
\end{equation}
and
\begin{equation}
	\delta_{\mu\nu}=\tilde{\delta}_{\mu\nu}+\hat{\delta}_{\mu\nu}.
\end{equation}
By definition, inner products between the two subalgebras vanish:
\begin{equation}
	\tilde{\gamma}_\mu\hat{\gamma}_\mu=0,\hfill
\end{equation}
and the metric tensors have a trace equal to the dimension of the subspace to which they belong:
\begin{equation}
	\tilde{\delta}_{\mu\mu}=4,\hfill\hat{\delta}_{\mu\mu}=d-4.\hfill
\end{equation}
In the absence of $\gamma_5$, this simply reduces to the natural $d$-dimensional generalization of the Dirac algebra. With $\gamma_5$, however, there are some complications. In four dimensions, $\gamma_5$ is completely characterized by three properties:
\begin{subequations}\label{G5:properties}
  \begin{align}
	&\{\gamma_5,\gamma_\mu\}=0,\label{G5:Anticommutation}\\
	&\text{Tr}[AB]=\text{Tr}[BA],\label{G5:TraceCyclitity}\\
	&\text{Tr}[\gamma_\mu\gamma_\nu\gamma_\rho\gamma_\sigma\gamma_5]=4\epsilon_{\mu\nu\rho\sigma},\label{G5:Trace12345}	
  \end{align}
\end{subequations}
from which we find that, in $d$-dimensions,
\begin{equation}
	(d-2)(d-4)\text{Tr}[\gamma_\mu\gamma_\nu\gamma_\rho\gamma_\sigma\gamma_5]=0.
\end{equation}
Since this prohibits a smooth limit for $d\rightarrow4$, we conclude that one of the above properties must be sacrificed to continue analytically to an arbitrary dimension. Our choice, introduced by 't Hooft and Veltman and systematized by Breitenlohner and Maison, relaxes the first condition \eqref{G5:Anticommutation}, so that $\gamma_5$ anticommutes with the four-dimensional subspace and commutes with the $(d-4)$-dimensional subspace. Thus
\begin{equation}
	\{\gamma_5,\tilde{\gamma}_\mu\}=[\gamma_5,\hat{\gamma}_\mu]=0.
\end{equation}
Furthermore, the trace in \eqref{G5:Trace12345} is taken to be
fundamental, and the Levi-Civita symbol
$\epsilon_{\mu\nu\rho\sigma}$ is strictly four-dimensional, containing no evanescent components.
As such, it is best to algebraically reduce expressions containing
$\epsilon_{\mu\nu\rho\sigma}$ after the $d\rightarrow4$ limit is taken. As a form of dimensional regularization, this scheme is manifestly Lorentz invariant, so that the reduction of tensor integrals is fairly straightforward. Moreover, the HVBM scheme maintains algebraic consistency in our applications; we have at most one instance of $\gamma_5$ in any correlation function.
Finally, to maintain Hermiticity in all dimensions,
we generalize the ``pseudo-tensor'' 
 $\sigma_{\mu\nu}\gamma_5=\frac{i}{2}[\gamma_\mu,\gamma_\nu]\gamma_5$ to \cite{Bhattacharya:2015rsa,Cirigliano:2020msr}
\begin{equation}
	\tilde{\sigma}_{\mu\nu}=\frac{1}{2}\left\{\sigma_{\mu\nu},\gamma_5\right\}.
\end{equation}
Note that the tilde here does not signify
a four-dimensional object as in the HVBM scheme;
rather it is an unfortunate artifact of the literature.
This modified version is central to the calculation
of any correlation functions including the quark chromo-electric dipole moment operator.

\section{Feynman Rules}
\label{app:rules}
We adopt the standard Feynman rules for QCD in $d$ Euclidean dimensions.
Below we describe in more detail the Feynman rules
for gauge bosons and fermions at non-vanishing flow time.
Some Feynman rules for flowed fields,
and similar details relevant to perturbative calculations,
have appeared already in the literature~\cite{Luscher:2010iy,Luscher:2011bx,Luscher:2013cpa,Suzuki:2013gza,
  Makino:2014taa,Endo:2015iea,Hieda:2016lly,Harlander:2016vzb,Monahan:2017hpu,Harlander:2018zpi,Artz:2019bpr}.
To keep this paper self-contained and provide a future reference, we list all the Feynman rules
for flowed fields that we have used in these calculations, along with the relevant vertices arising from our operators. We note that all vertices with $n$-interacting fields are defined with inward-directed momenta $p_1,\dots,p_n$ and that, unless stated otherwise (see Sec.~\ref{sec:ops}), there is an implicit factor of $(2\pi)^d\delta^{(d)}(p_1+\cdots+p_n)$ that ensures momentum conservation.

\subsection{Gradient Flow}
The nonlinearity of the flow equations produces extra vertices,
which must be included in perturbation theory. For bosons, the vertices $X^{(n,0)}$
appear in the solutions of the flow equation, where $n$ is the number of gluon fields involved.
These flow vertices must always be connected to a kernel line.
Kernels, called so for their role as the integral kernel of the solution to the
flow equation, appropriately carry the information within a bulk field to its higher-order corrections.
Diagrammatically, a kernel line may initiated at any vertex at positive flow time,
replacing a bulk field leg, and terminating at a flow vertex.
Thus for any interaction involving bulk fields with some functional form $\Delta(t)$,
we will have corrections starting at $O(g_0)$ attached with a kernel line.
Let $\Gamma(s)$ represent the associated flow vertex and all relevant subsidiary interactions
involving all attached bulk fields.
Then, representing a bosonic kernel as a double curly line,
we define the Feynman rule:
\begin{equation}
	\vcenter{\hbox{\includegraphics[width=71pt,height=26pt]{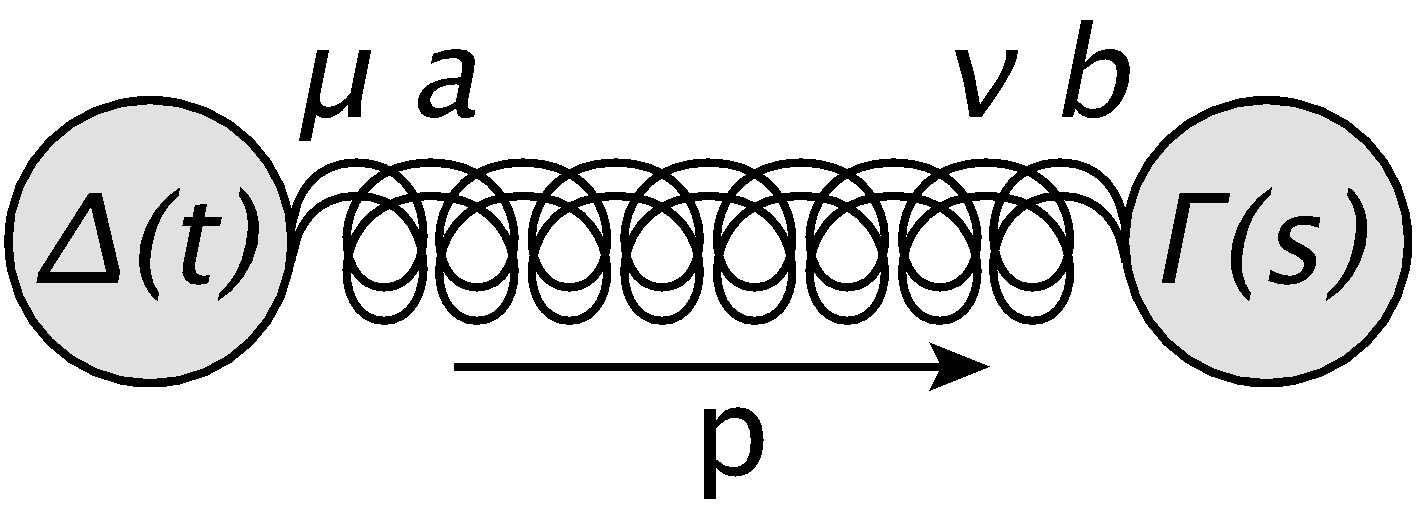}}}
	=\int_0^\infty ds\ \theta(t-s)\Gamma_\nu(s)\tilde{K}_{t-s}(p)^{ab}_{\nu\mu}\Delta_\mu(t),
	\label{eq:KernelK}
\end{equation}
where
\begin{equation}
	\tilde{K}_t(p)_{\mu\nu}^{ab}=\frac{\delta^{ab}}{p^2}\left[(\delta_{\mu\nu}p^2 - p_\mu p_\nu)e^{-p^2t}-p_\mu p_\nu e^{-\alpha_0p^2t}\right]
\end{equation}
is the bosonic kernel.
Observe that it collapses to a simple Gaussian in the ``generalized Feynman gauge,'' $\alpha_0=1$. For clarity, note also that the ordering of the structures $\Gamma$ and $\Delta$ above is only restricted by the ordering of the fermionic fields contained within them.
Turning our attention to the vertices, we have $\frac{1}{2}X^{(2,0)}$ at first order:
\begin{equation}
	\vcenter{\hbox{\includegraphics[width=88pt,height=85pt]{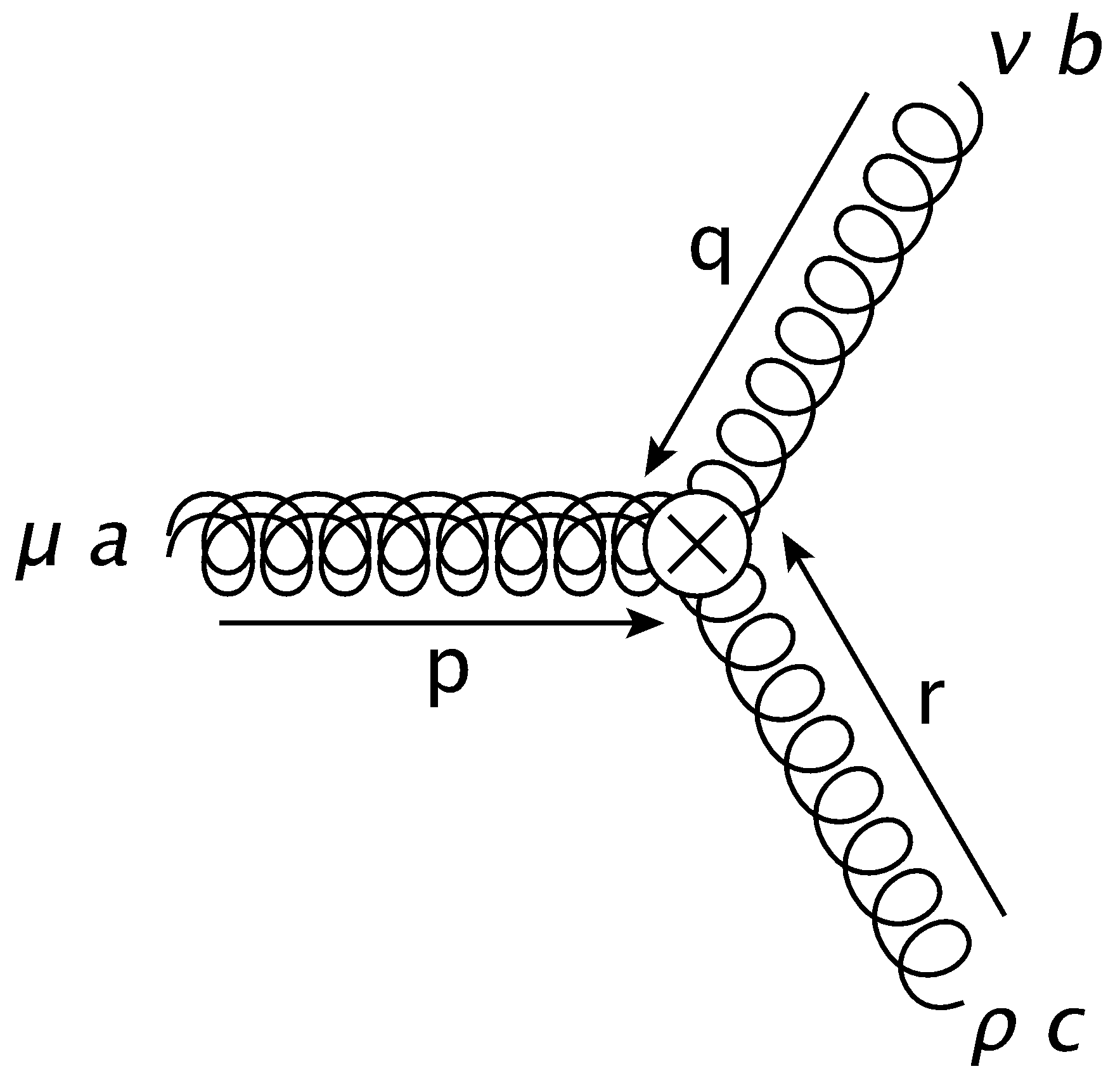}}}
	=\frac{1}{2}if^{abc}\left\{(r-q)_\mu\delta_{\nu\rho}+2q_\rho\delta_{\mu\nu}-2r_\nu\delta_{\rho\mu}+\left(\alpha_0-1\right)\left(q_\nu\delta_{\rho\mu}-r_\rho\delta_{\mu\nu}\right)\right\}.
\end{equation}
The fields radiating out of this and all other flow vertices are bulk fields at some positive flow time, which in Eq.~\ref{eq:KernelK} we denote as $s$, whereas the kernel is generated by a bulk field at a flow time that, in Eq.~\ref{eq:KernelK}, we denote $t$. The second-order vertex is $\frac{1}{6}X^{(3,0)}$:
\begin{equation}
	\vcenter{\hbox{\includegraphics[width=110pt,height=99pt]{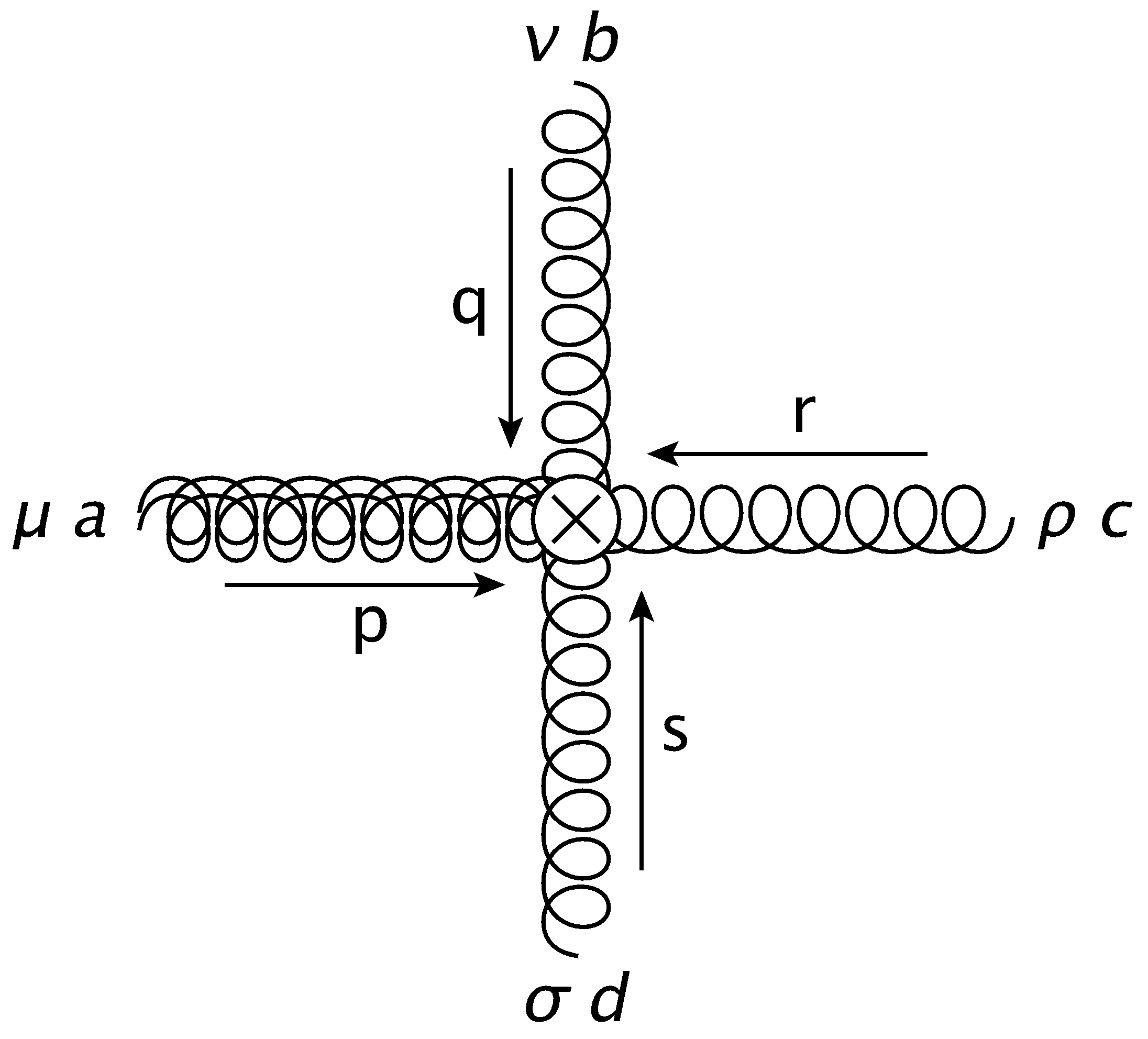}}}
	=
	\begin{aligned}[t]
		\frac{1}{6}\big\{&f^{abe}f^{cde}\left(\delta_{\mu\sigma}\delta_{\rho\nu}-\delta_{\mu\rho}\delta_{\sigma\nu}\right)\\
		+&f^{ade}f^{bce}\left(\delta_{\mu\rho}\delta_{\sigma\nu}-\delta_{\mu\nu}\delta_{\rho\sigma}\right)\\
		+&f^{ace}f^{dbe}\left(\delta_{\mu\nu}\delta_{\rho\sigma}-\delta_{\mu\sigma}\delta_{\rho\nu}\right)\big\}.
	\end{aligned}
\end{equation}
The factors of $1/n!$ are placed within the vertex rules above
so that the kernel line has the same Feynman rule regardless of the
flow vertex to which it is attached. There are no intrinsic higher-order vertices,
but these vertices may be nested to the desired order,
ensuring that proper symmetry factors are included.
For example, in the calculation a two-point Green's function at positive flow time and at one-loop order,
we must account for all combinations up to $\mathcal{O}(g_0^2)$.
Both vertices will contribute, along with the (at least) second-order
structure:
\begin{equation}
	\begin{aligned}
		2\times&\int_0^tds\ \tilde{K}_{t-s}(p)^{aa'}_{\mu\mu'}\frac{1}{2}X^{(2,0)}(p,q,-p-q)_{\mu'\nu\rho}^{a'bc}\tilde{B}_\nu^b(-q;s)\\
		&\times\int_0^sdu\ \tilde{K}_{s-u}(p+q)^{cc'}_{\rho\rho'}\frac{1}{2}X^{(2,0)}(p+q,k,-p-q-k)_{\rho'\sigma\tau}^{c'de}\tilde{B}_\sigma^d(-k;u)\tilde{B}_\tau^e(p+q+k;u),
	\end{aligned}
	\label{eq:NestedX20}
\end{equation}
or, pictorially:
\begin{equation}
	2\times\left(\vcenter{\hbox{\includegraphics[width=151pt,height=64pt]{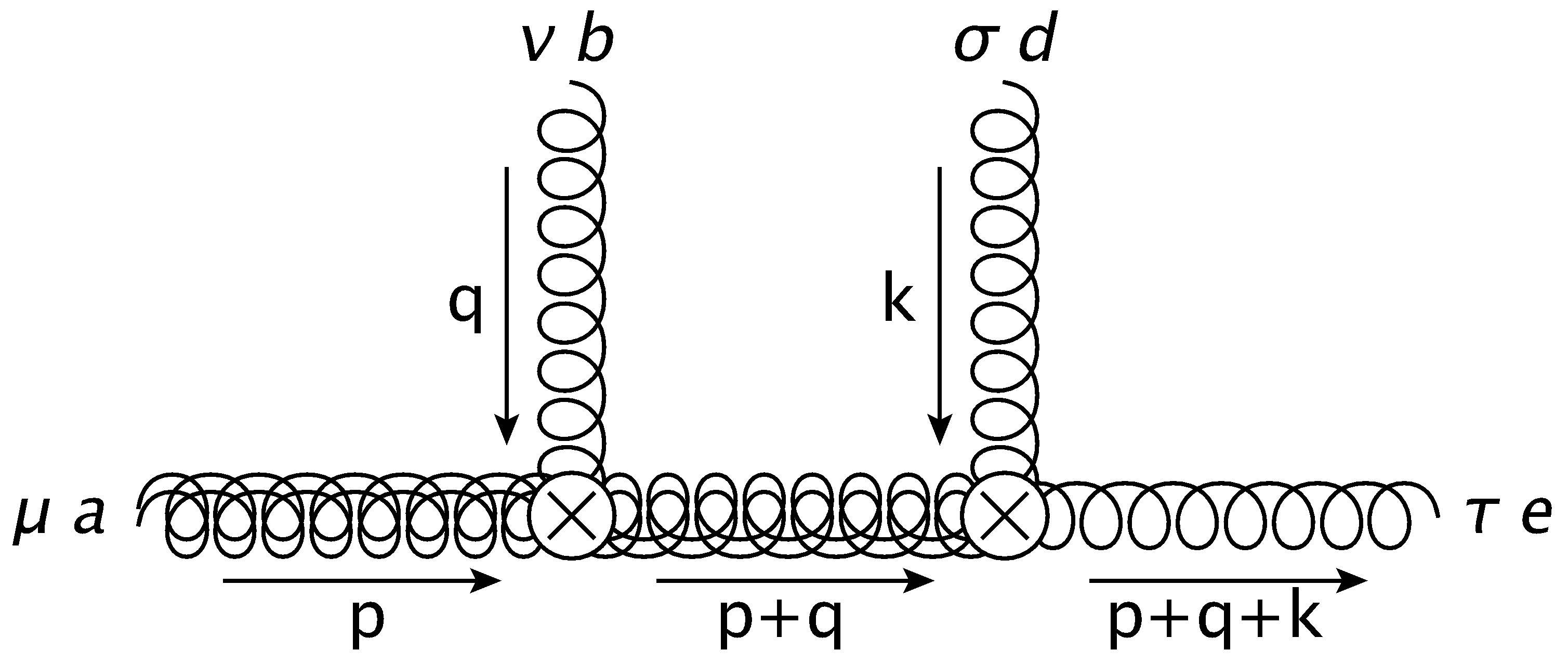}}}\right).
\end{equation}
The second line in \eqref{eq:NestedX20} is simply the NLO contribution
to either of the two fields attached to the vertex $X^{(2,0)}(p,q,-p-q)_{\mu\nu\rho}^{abc}$.
The initial factor of $2$ accounts for the symmetry in choosing which
of the $B$ fields to expand. Since both fields include
the same nonlinear corrections, either may be expanded,
so long as the result is summed over all of these redundancies.

Fermions have similar rules. The fermionic kernels,
\begin{equation}
\begin{aligned}
	J(x-y;t) &= \int_p \e^{ip(x-y)}\tilde{J}_t(p)\,, \qquad \tilde{J}_t(p)=e^{-p^2t}\,,\\
	\bar{J}(x-y;t) &= \int_p \e^{ip(x-y)}\tilde{\bar{J}}_t(p)\,, \qquad \tilde{\bar{J}}_t(p)=e^{-p^2t}\,,
\end{aligned}
\end{equation}
produce Feynman rules analogous to the bosonic kernel.
Letting $\Delta$ and $\Gamma$ be defined as before,
and representing the fermionic kernel line by a double straight line, we have
\begin{subequations}
	\begin{align}
		&\vcenter{\hbox{\includegraphics[width=71pt,height=22pt]{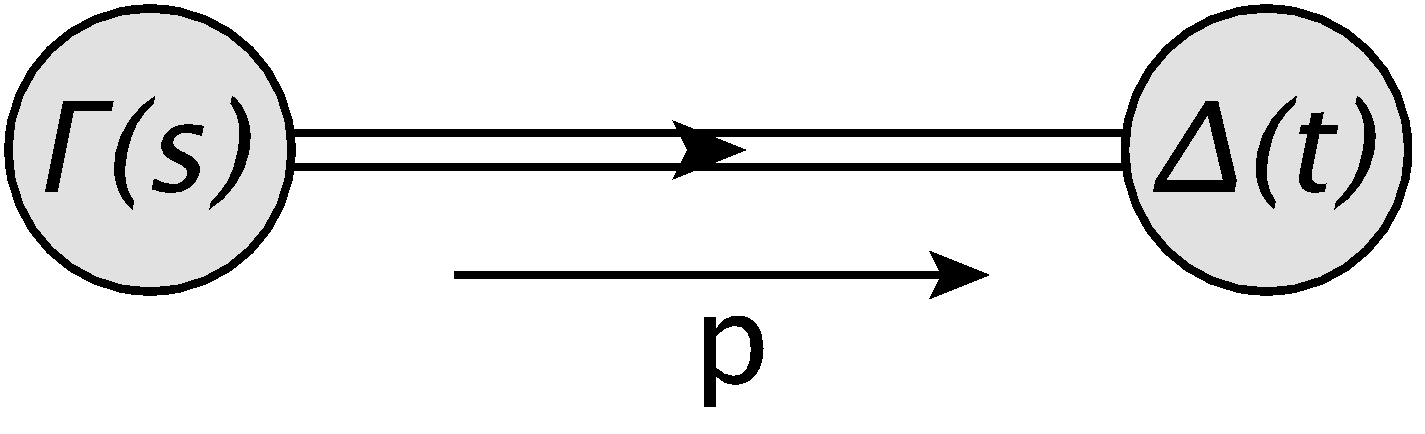}}}
	  =\int_0^\infty ds\ \theta(t-s)\Delta(t)\tilde{J}_{t-s}(p)\Gamma(s)\,, \\
		&\vcenter{\hbox{\includegraphics[width=71pt,height=22pt]{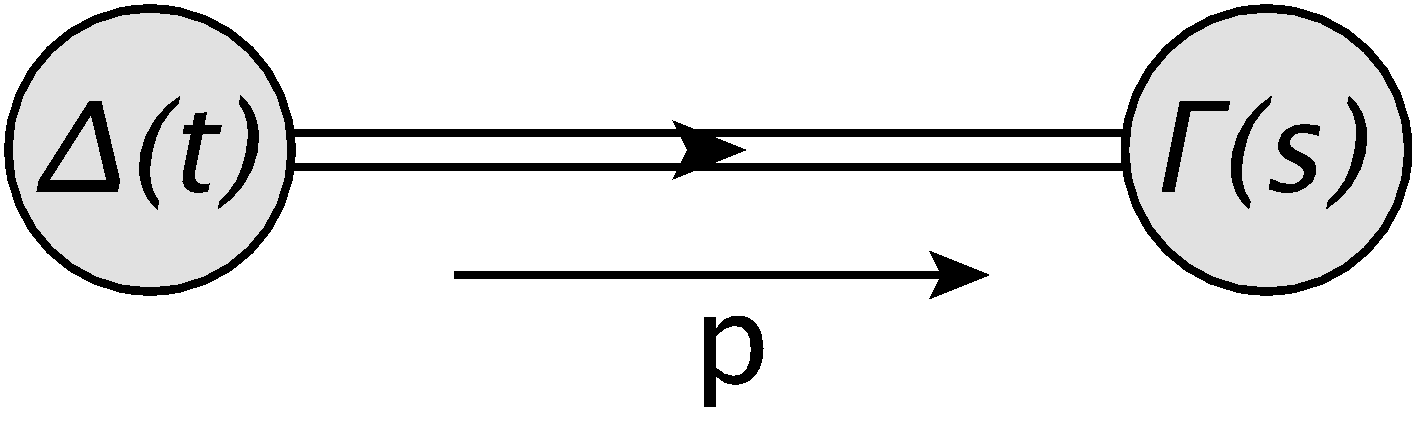}}}
		=\int_0^\infty ds\ \theta(t-s)\Gamma(s)\tilde{\bar{J}}_{t-s}(p)\Delta(t)\,,
	\end{align}
\end{subequations}
where the first rule applies to the flow-time evolution of the
$\chi$ field while the second rule to the $\chibar$ field. The distinction between $J$ and $\bar{J}$ is purely formal; $J$ acts from the left on $\chi$, while and $\bar{J}$ acts from the right on $\chibar$.
In the same manner as the fermion propagator, the direction of the arrow indicates the flow of fermion number from $\chibar$ to $\chi$.
Analogously to what happens for the gauge bosons, the flow equations for the
fermion fields~(\ref{eq:chibarint},\ref{eq:remainder}) can be solved in an iterative
manner, generating higher-order vertices containing one fermion field and $n$ gauge fields, $Y^{(1,n)}$.
The term linear in $B$ in the fermion flow equation produces $Y^{(1,1)}$:
\begin{equation}
	\vcenter{\hbox{\includegraphics[width=68pt,height=80pt]{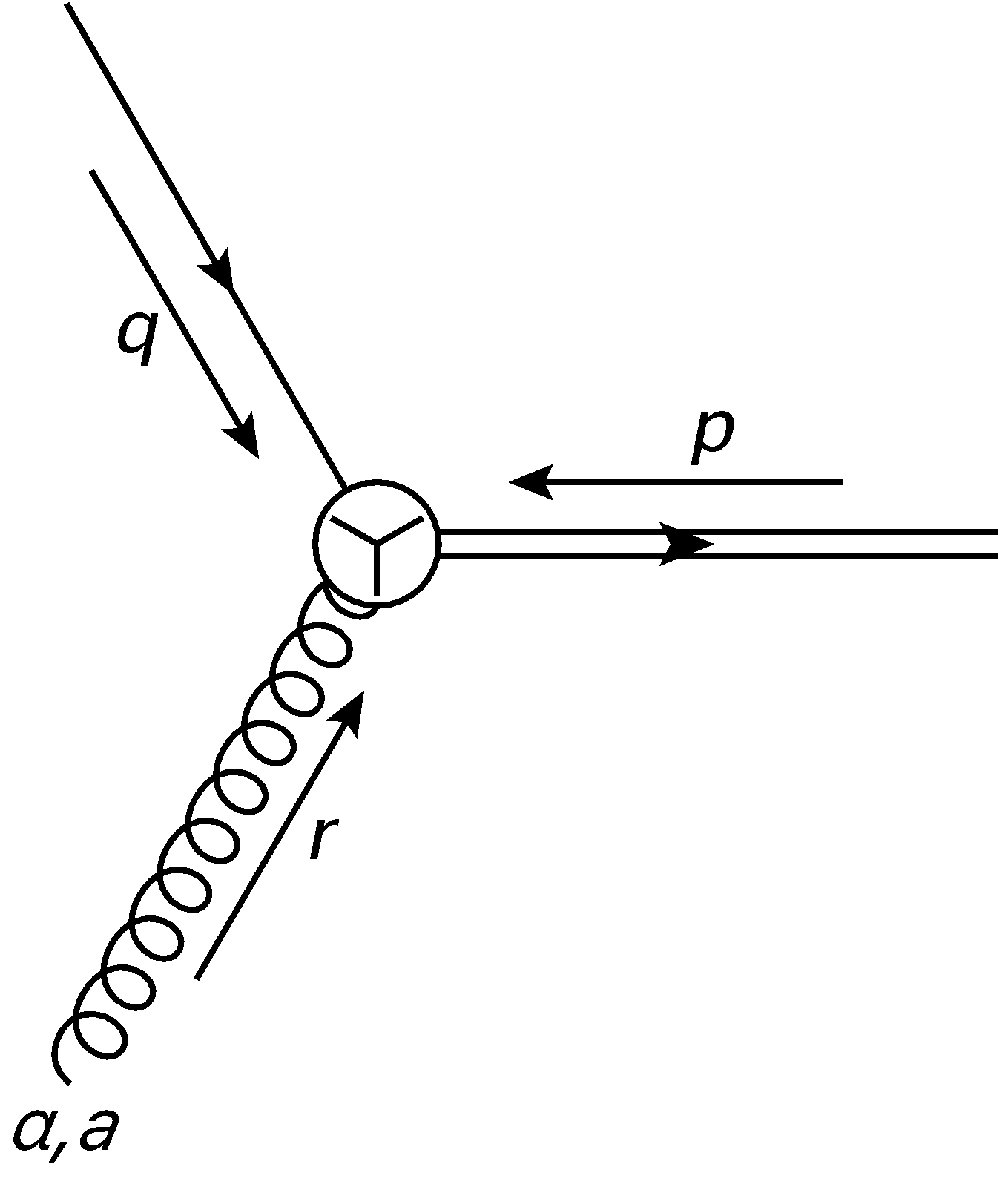}}}
	=it^a\left\{2q_\mu+\left(1-\alpha_0\right)r_\mu\right\}\,,
\end{equation}
while the analogous term in the adjoint fermion flow equation produces $\bar{Y}^{(1,1)}$:
\begin{equation}
	\vcenter{\hbox{\includegraphics[width=68pt,height=80pt]{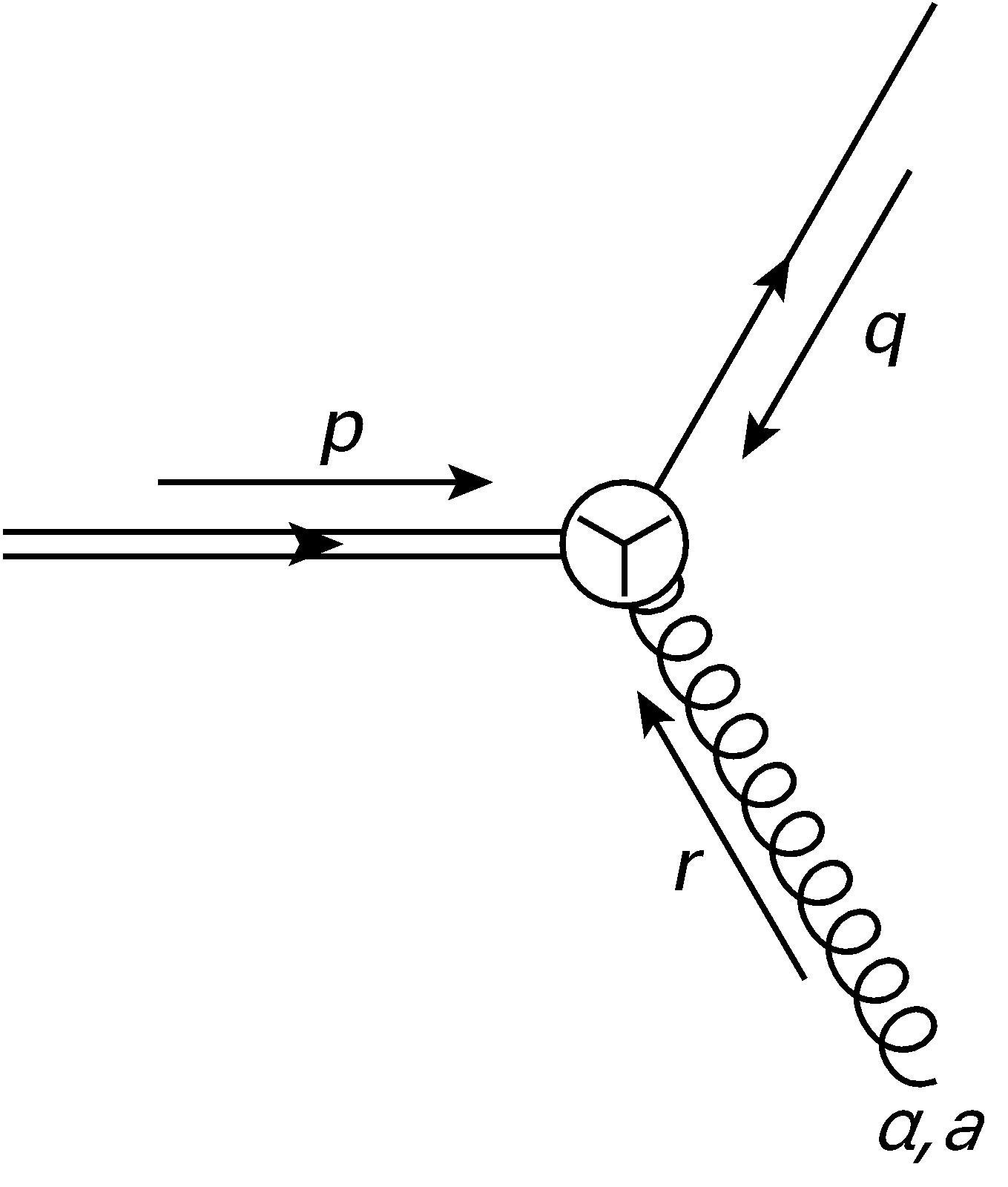}}}
	=-it^a\left\{2q_\mu-\left(1-\alpha_0\right)r_\mu\right\}\,,
\end{equation}
where the first diagram refers to the perturbative expansion of the $\chi$ field and
the second to the expansion of the $\chibar$ field. 
The vertex $Y^{(1,2)}$ is thoroughly simpler:
\begin{equation}
	\vcenter{\hbox{\includegraphics[width=85pt,height=92pt]{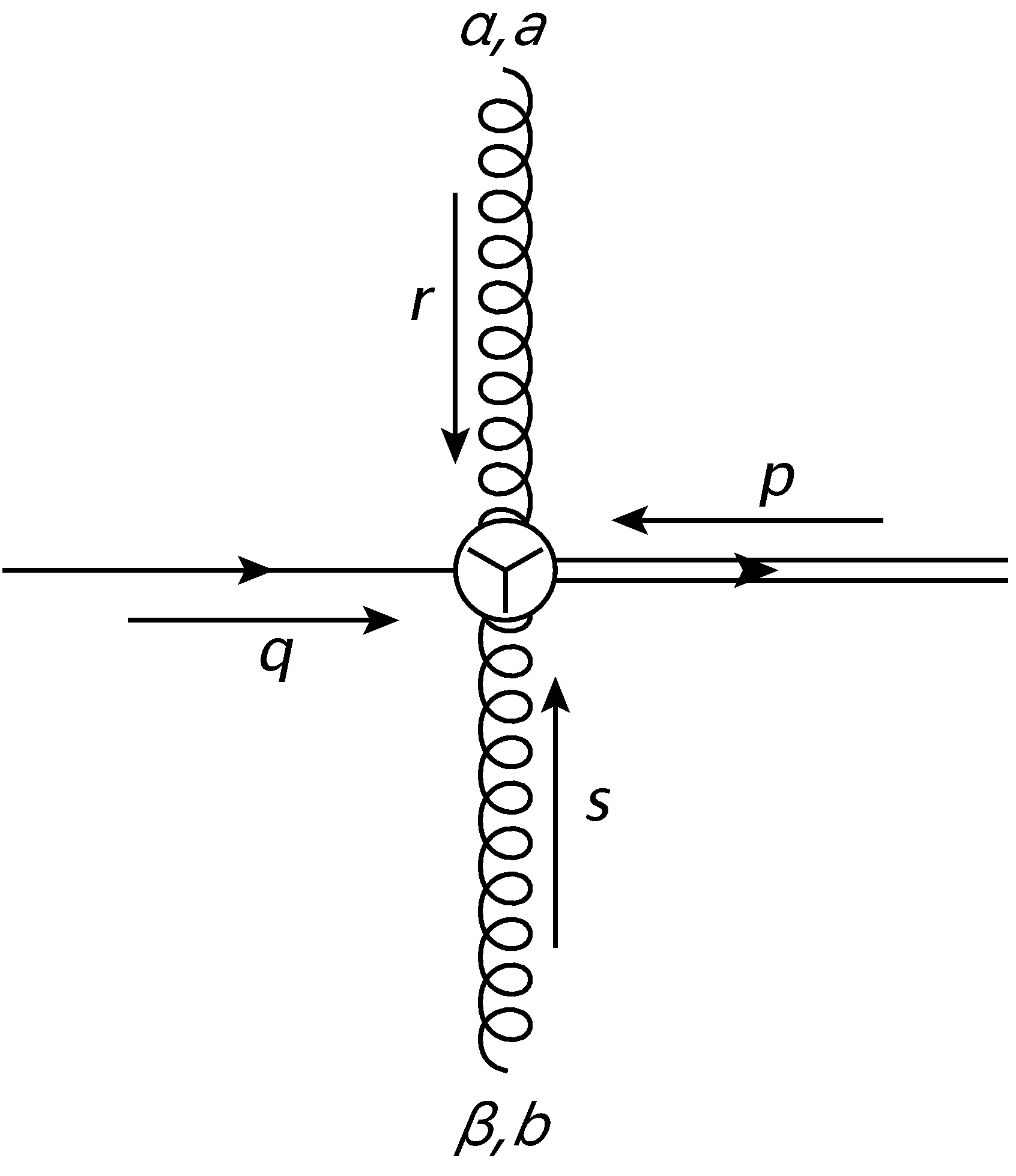}}}
	=\frac{1}{2}\delta_{\mu\nu}\left\{t^a,t^b\right\}.
\end{equation}
Since this term is quadratic in $B$,
there is no sign change with respect to the direction of fermion flow, and $\bar{Y}^{(1,2)}$ is identical to $Y^{(1,2)}$
\begin{equation}
	\vcenter{\hbox{\includegraphics[width=85pt,height=92pt]{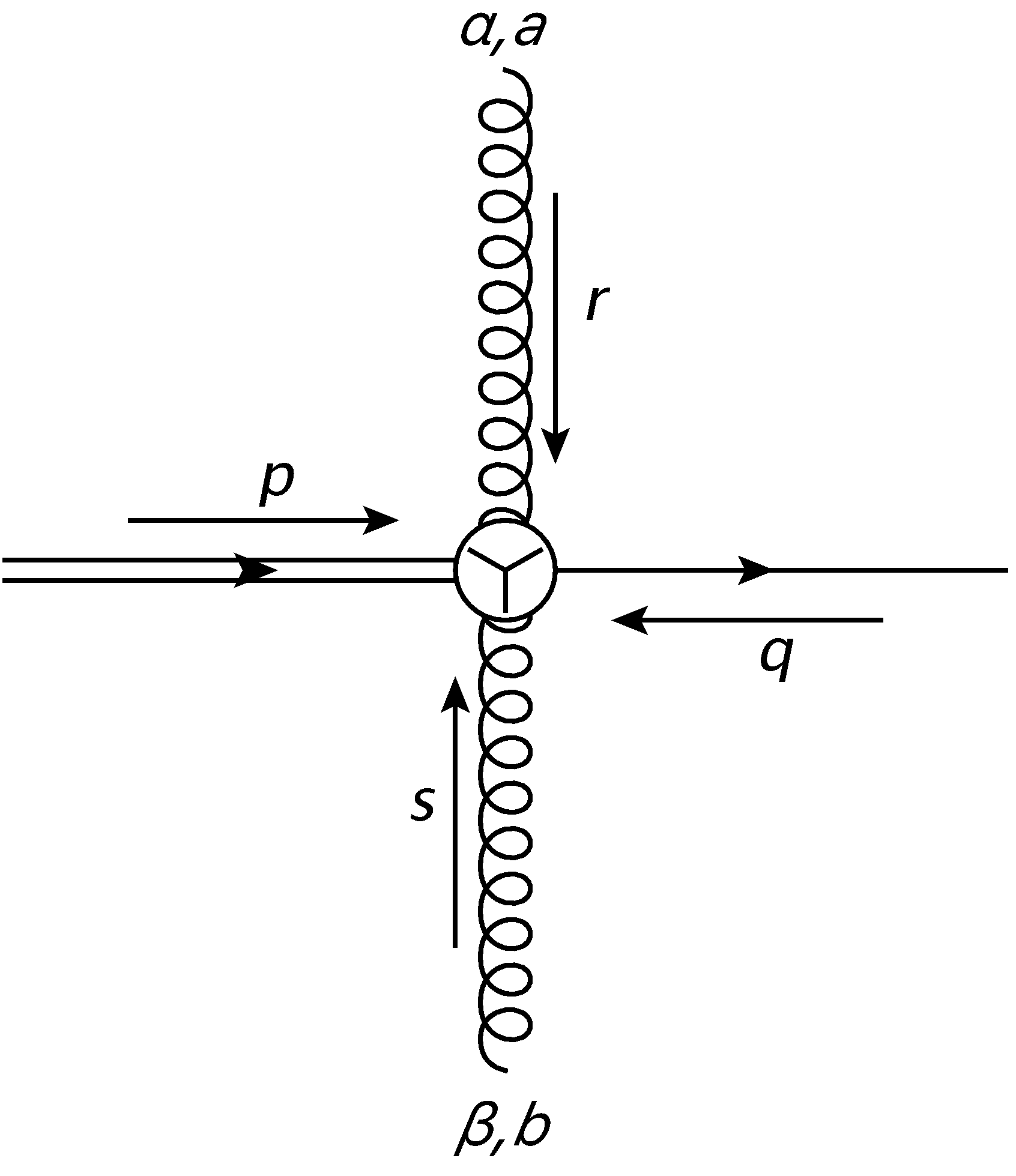}}}
	=\frac{1}{2}\delta_{\mu\nu}\left\{t^a,t^b\right\}.
\end{equation}
\subsection{Operators}\label{sec:ops}
In this section we list the Feynman rules for the CP-violating operators.
The Feynman rules are flow-time independent, but the fields connected to these vertices
may be flowed. The Feynman rules arising specifically from the perturbative
expansion of the flowed fields are described in the previous subsection; only the tree-level fields enter our operator Feynman rules.

There is some subtlety in the implementation of these operators in perturbative QCD. A na\"ive calculation of any correlator with an odd number of \CP-violating operators will always vanish. This should actually be expected; all correlation functions are calculated within a QCD background, so there may be no expectation values that violate \CP. We circumvent this problem by temporarily ignoring momentum conservation; equivalently, we calculate all such correlations functions pointwise in coordinate space, integrating the point of interaction for our \CP-violating operators over all spacetime only after we subtract off the desired quantities 
\cite{Espriu:1982bw,Braaten:1990zt,Braaten:1990gq}. If momentum were to be conserved throughout these calculations, all operators would project to zero momentum at the onset, and structures like $\epsilon_{\alpha\beta\mu\nu}p_\mu p'_\nu$ would contract to zero identically, trivializing the entire calculation. This trick allows us to break translational symmetry,
giving the $in$ and $out$ states different total momenta and
subsequently different transformations under the Lorentz group.
After identifying the Wilson coefficients, we dynamically restore
the conservation of momentum by integrating over all spacetime.
In so doing, we also restore the appropriate discrete symmetries. We are simply keeping track of the various structures that vanish perturbatively.

\subsubsection{Topological Charge Density}
\be
\begin{aligned}
  \mathcal{O}_q=&k_q\text{Tr}\left\{G_{\mu\nu}\tilde{G}_{\mu\nu}\right\}\\
  \rightarrow&-\frac{1}{4}k_q\epsilon_{\mu\nu\rho\sigma}G^a_{\mu\nu}G^a_{\rho\sigma}
\end{aligned}
\ee
\be
\vcenter{\hbox{\includegraphics[width=110pt,height=17pt]{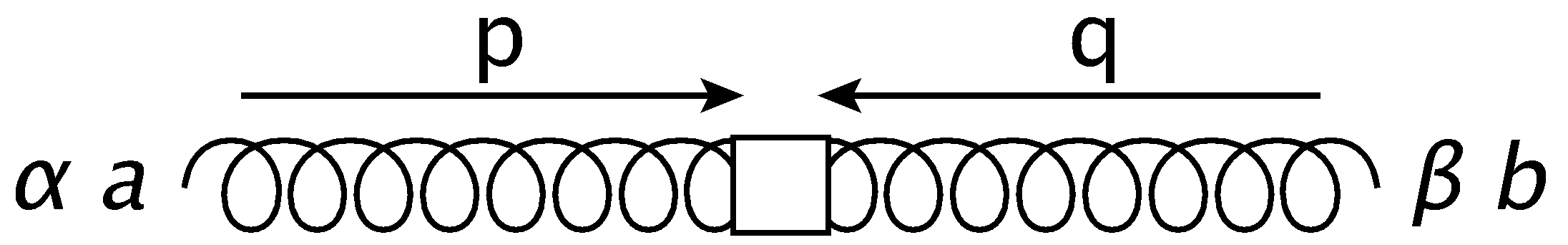}}}=-2k_q\delta^{ab}\epsilon_{\alpha\beta\mu\nu}p_\mu q_\nu
		\ee
		\be
		\vcenter{\hbox{\includegraphics[width=88pt,height=85pt]{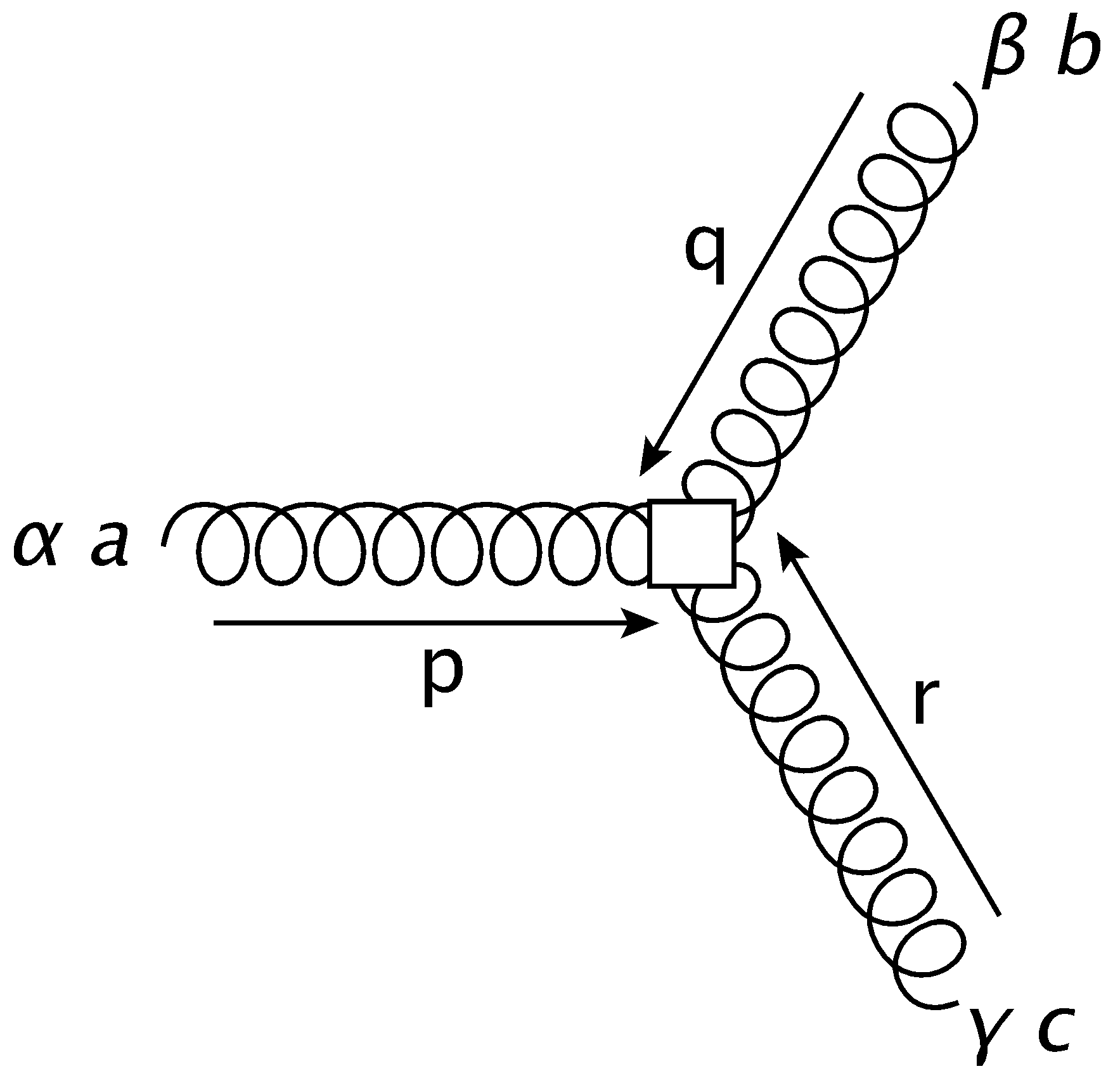}}}=-2ik_qf^{abc}\epsilon_{\alpha\beta\gamma\mu}(p+q+r)_\mu
		\ee
	        \subsubsection{qCEDM}
		\be
		\mathcal{O}_C=k_C\bar{\psi}G_{\mu\nu}\tilde{\sigma}_{\mu\nu}\psi
		\ee
		\be
		\vcenter{\hbox{\includegraphics[width=76pt,height=80pt]{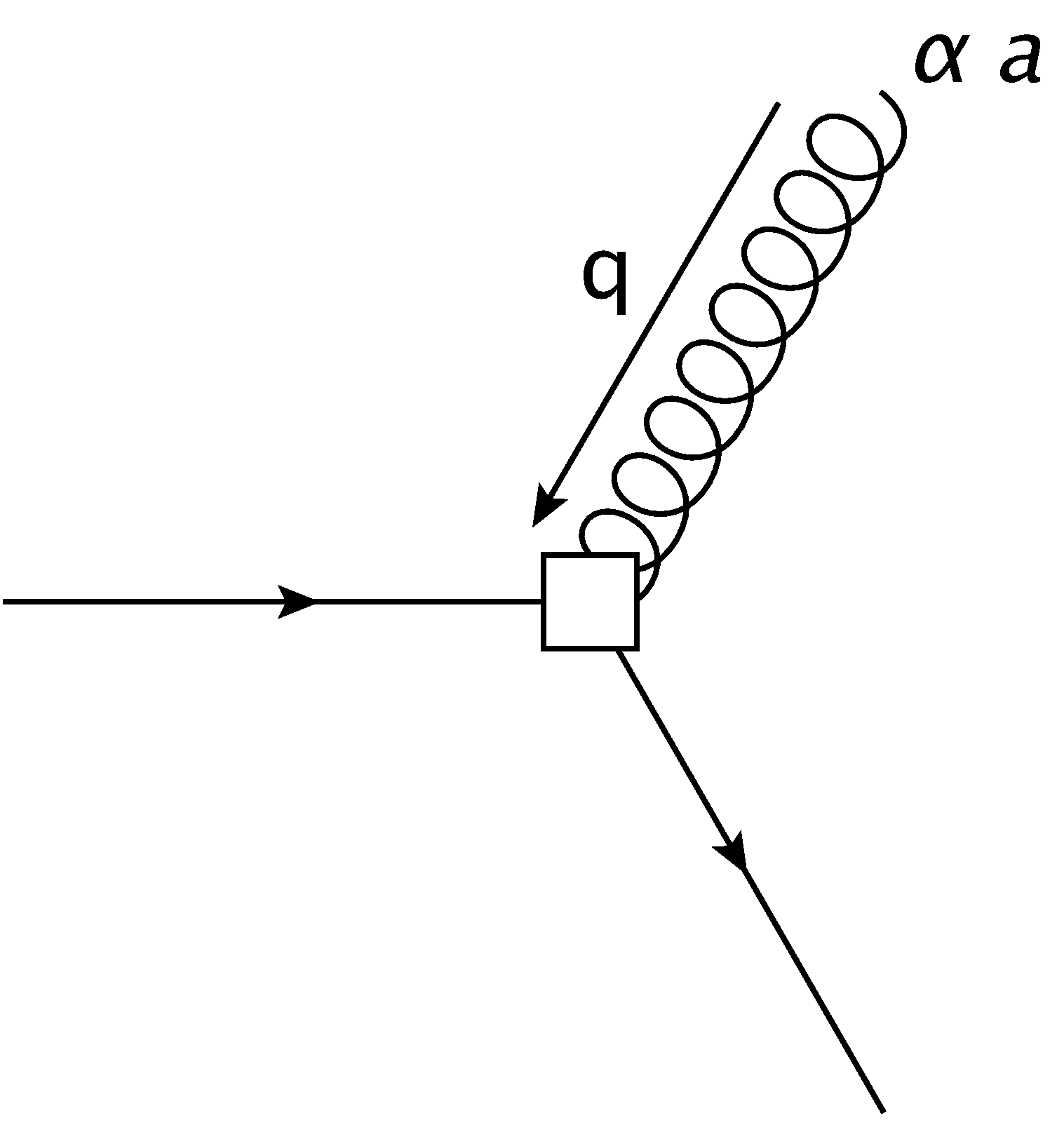}}}=-2ik_ct^a\sigma_{\mu\nu}\gamma_5q_\nu
		\ee
		\be
		\vcenter{\hbox{\includegraphics[width=85pt,height=100pt]{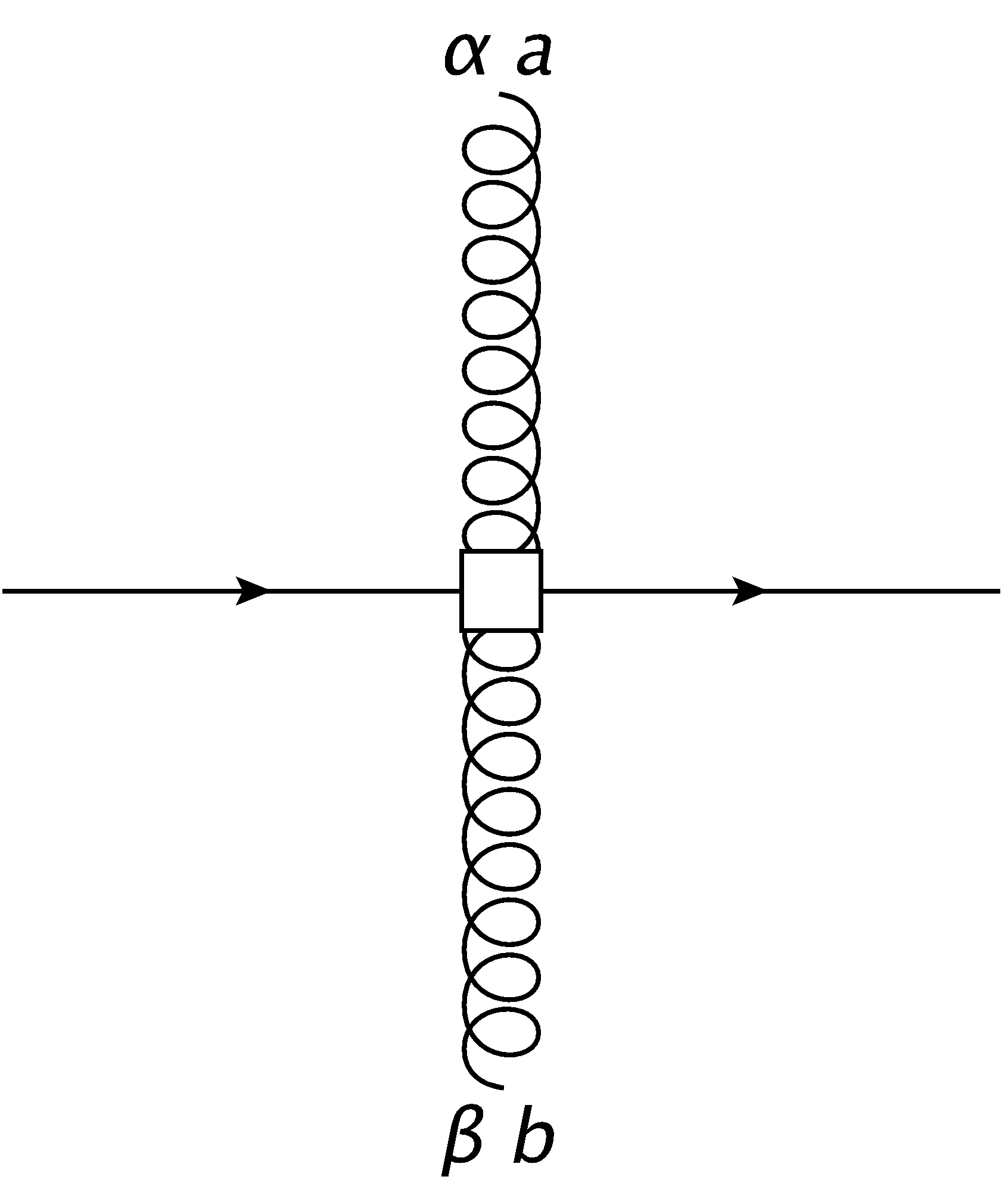}}}=2k_cf^{abc}t^c\sigma_{\mu\nu}\gamma_5
		\ee
	        \subsubsection{Weinberg Operator}
		\be
		\begin{aligned}
		  \mathcal{O}_W=&k_W\text{Tr}\left\{\left[G_{\mu\rho},G_{\nu\rho}\right]\tilde{G}_{\mu\nu}\right\}\\
		  \rightarrow&-\frac{1}{4}k_Wf^{abc}\epsilon_{\mu\nu\rho\sigma}G^a_{\mu\nu}G^b_{\mu\tau}G^c_{\nu\tau}\\
		  \xrightarrow{d\rightarrow4}&\frac{1}{64}ik_Wf^{abc}\text{Tr}\left\{\sigma_{\mu\nu}\sigma_{\rho\sigma}\sigma_{\tau\chi}\gamma_5\right\}G_{\mu\nu}^aG_{\rho\sigma}^bG_{\tau\chi}^c
		\end{aligned}
		\ee
		\be
		\vcenter{\hbox{\includegraphics[width=88pt,height=85pt]{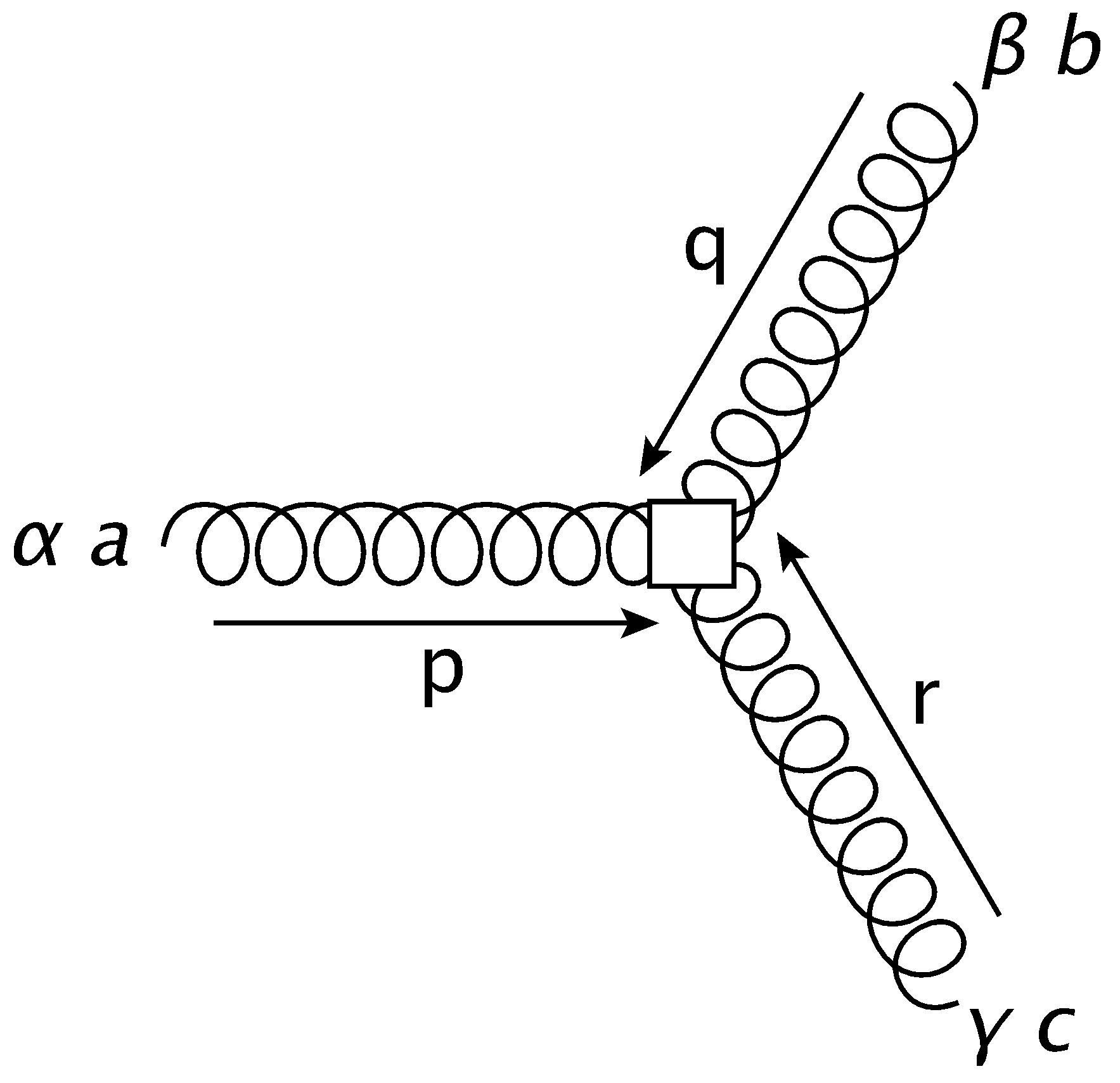}}}=\frac{3}{4}k_Wf^{abc}\text{Tr}\left\{\sigma_{\alpha\mu}\sigma_{\beta\nu}\sigma_{\gamma\rho}\gamma_5\right\}p_\mu q_\nu r_\rho
		\ee
		\be
		\vcenter{\hbox{\includegraphics[width=109pt,height=99pt]{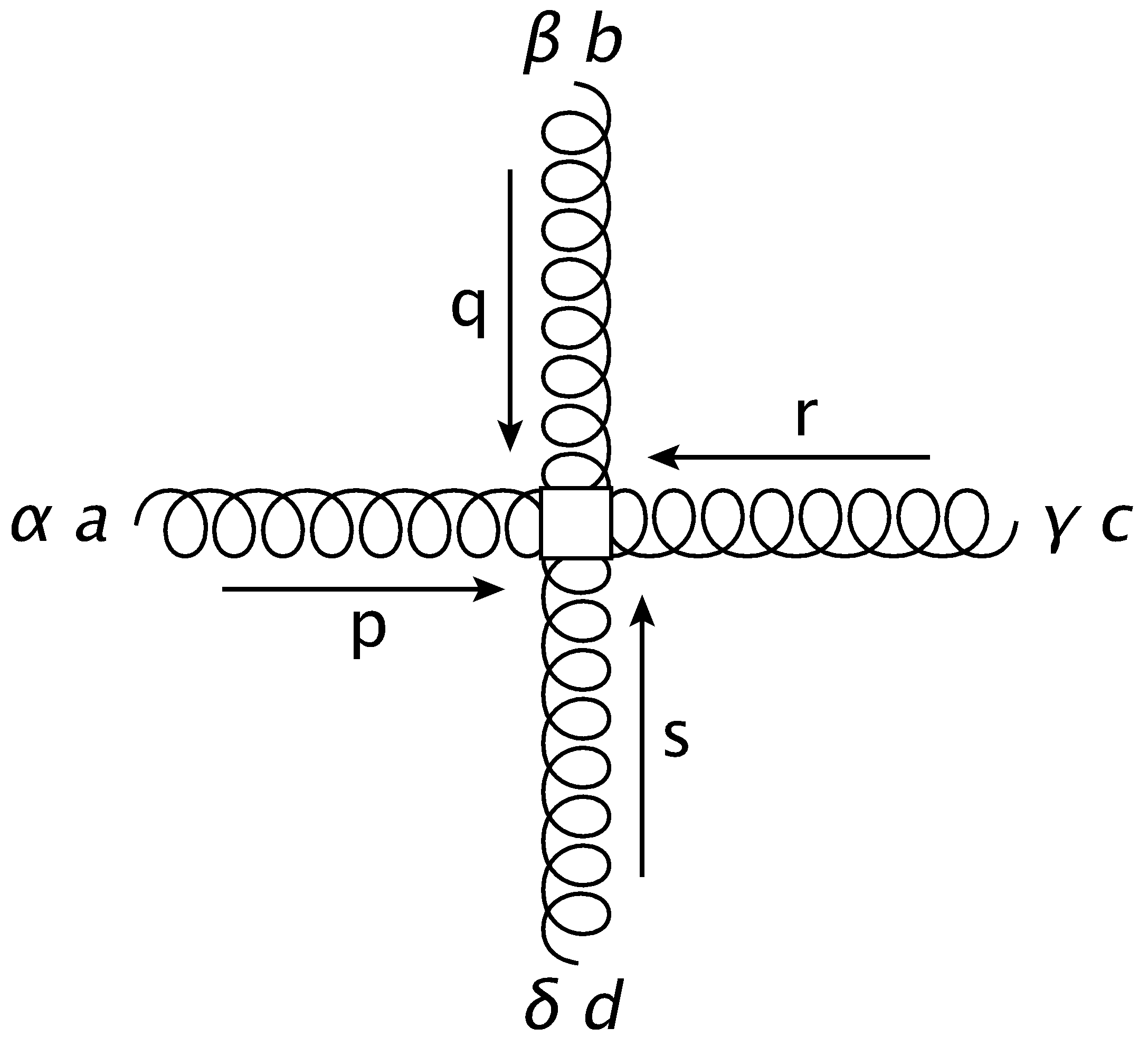}}}=
		\begin{aligned}[t]
		  -\frac{3}{4}ik_W\big[
		    &f^{abe}f^{cde}\left(p_\mu q_\nu\text{Tr}\left\{\sigma_{\mu\alpha}\sigma_{\nu\beta}\sigma_{\gamma\delta}\gamma_5\right\}+r_\mu s_\nu\text{Tr}\left\{\sigma_{\mu\gamma}\sigma_{\nu\delta}\sigma_{\alpha\beta}\gamma_5\right\}\right)\\
		    +&f^{ace}f^{bde}\left(p_\mu r_\nu\text{Tr}\left\{\sigma_{\mu\alpha}\sigma_{\nu\gamma}\sigma_{\beta\delta}\gamma_5\right\}+q_\mu s_\nu\text{Tr}\left\{\sigma_{\mu\beta}\sigma_{\nu\delta}\sigma_{\alpha\gamma}\gamma_5\right\}\right)\\
		    +&f^{ade}f^{bce}\left(p_\mu s_\nu\text{Tr}\left\{\sigma_{\mu\alpha}\sigma_{\nu\delta}\sigma_{\beta\gamma}\gamma_5\right\}+q_\mu r_\nu\text{Tr}\left\{\sigma_{\mu\beta}\sigma_{\nu\gamma}\sigma_{\alpha\delta}\gamma_5\right\}\right)
		    \big]
		\end{aligned}
		\ee

\section{Sample calculation: One-Loop Fermion Propagator}
\label{app:Zfermion}
In this appendix we discuss in some detail the one-loop calculation of the fermion propagator
for flowed fermion fields. Results for the one-loop calculation of the
flowed fermion propagator have appeared
in the literature~\cite{Luscher:2013cpa,Makino:2014taa,Monahan:2017hpu} with varying degree of detail.
We use this calculation as an example to elucidate features
of a one-loop calculation at non-vanishing flow time and to collect all the relevant tools for a perturbative calculation with flowed fermion fields.
For a more complete discussion of flowed perturbative calculations,
we refer to~\cite{Rizik:new}.

The fermion propagator
\be
S(x,y;t,s)=\langle\chi(y;s)\bar{\chi}(x;t)\rangle =
\int_pe^{ip(x-y)}\widetilde{S}(p;t,s)\,,
\label{eq:FermProp}
\ee
can be expanded in powers of the bare coupling
\be
\widetilde{S}(p;t,s) = \sum_{k=0}^{\infty} g_0^{2k} \widetilde{S}^{k}(p;t,s)\,,
\ee
with a tree-level expression
\be
\widetilde{S}^{0}(p;t,s) = \frac{e^{-p^2(t+s)}}{i\slashed{p}+m_0}\,.
\ee
The one-loop corrections can be calculated evaluating the Feynman diagrams depicted in
Eqs.~(\ref{eq:G1}-\ref{eq:G5}). There are eight nontrivial contributions to the flowed fermion propagator, of which only five are topologically distinct~\cite{Luscher:2013cpa}. The diagrams involving flow kernels present some new features compared to standard perturbative calculations in QCD. While the standard one-loop diagram in Eq.~\eqref{eq:G1} has the usual structure with tree-level propagators on the external lines, the flowed diagrams cannot truncated as easily, because they occur with one or two external kernel lines. For this reason we write the decomposition of the fermion propagator as follows
\be
\tilde{S}^{(2)}(p;t,s) = \tilde{S}^{(0)}(p;t,0)\Sigma_1^{(2)}(p)\tilde{S}^{(0)}(p;0,s) +
\sum_{i=2}^4 \left[\Gamma_{i,a}^{(2)}(p;t)\tilde{S}^{(0)}(p;0,s) + 
\tilde{S}^{(0)}(p;t,0)\Gamma_{i,b}^{(2)}(p;s) \right] + \Gamma_5^{(2)}(p;t,s)\,.
\ee
The functions $\Gamma_{i,a}^{(2)}(p;t)$ and $\Gamma_{i,b}^{(2)}(p;s)$ correspond to the first-order expansions of the external fields $\chibar(x;t)$ and $\chi(y;s)$, respectively, though they are otherwise all but formally identical. The contribution $\Gamma_5$ includes the first-order expansion of both external fields. We list the individual contributions from each Feynman diagram 
in Eqs.~(\ref{eq:G1}-\ref{eq:G5}) together with their evaluation, ignoring external propagators for brevity:
\begin{subequations}
	\begin{align}
		&\Sigma_1^{(2)}(p)=
		\vcenter{\hbox{\includegraphics[width=140pt,height=27pt]{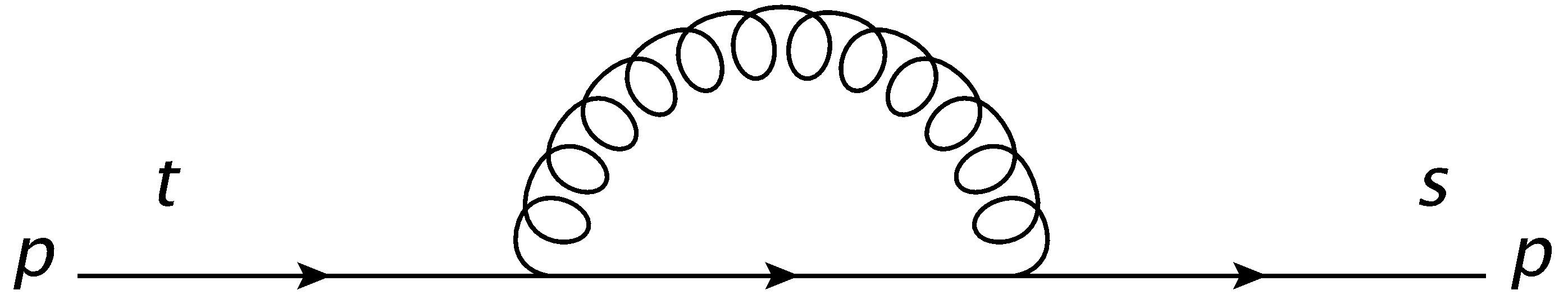}}}
                \label{eq:G1}
		\\
		\nonumber&\phantom{\Sigma_1^{(2)}(p)}=-g_0^2\frac{C_2(F)}{(4\pi)^2}\Bigg\{\left[\frac{1}{\epsilon}+\log\left(\frac{4\pi\mu^2}{p^2}\right)-\gamma_E+1\right]i\slashed{p}+4\left[\frac{1}{\epsilon}+\log\left(\frac{4\pi\mu^2}{p^2}\right)-\gamma_E+\frac{3}{2}\right]m_0+R\left(\frac{m_0^2}{p^2}\right)\Bigg\}+\mathcal{O}(\epsilon),\\
		&\Gamma_{2,a}^{(2)}(p;t)=
		\vcenter{\hbox{\includegraphics[width=140pt,height=30pt]{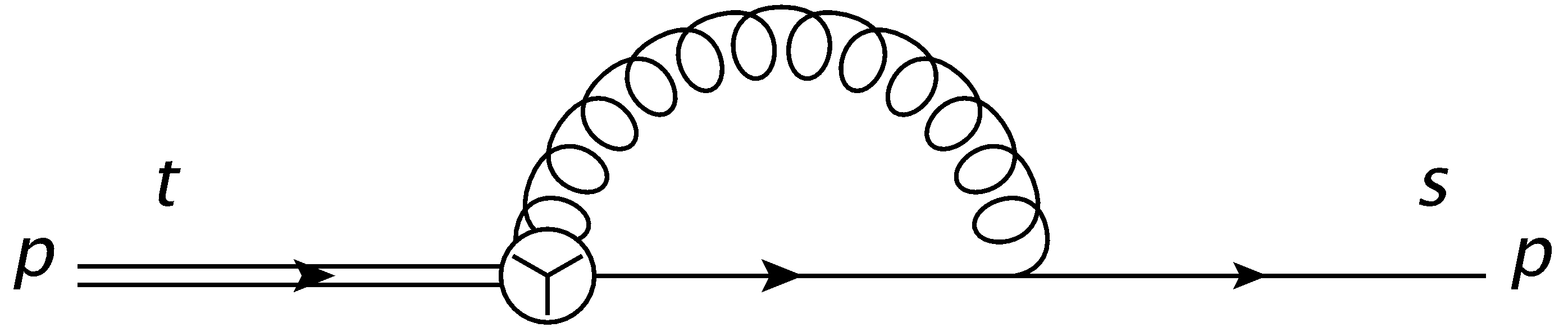}}}
		=g_0^2\frac{C_2(F)}{(4\pi)^2}\left[\frac{1}{\epsilon}+\log\left(8\pi\mu^2t\right)+1\right]+\mathcal{O}(\epsilon,t),\label{eq:G2a}\\
		&\Gamma_{2,b}^{(2)}(p;s)=
		\vcenter{\hbox{\includegraphics[width=140pt,height=30pt]{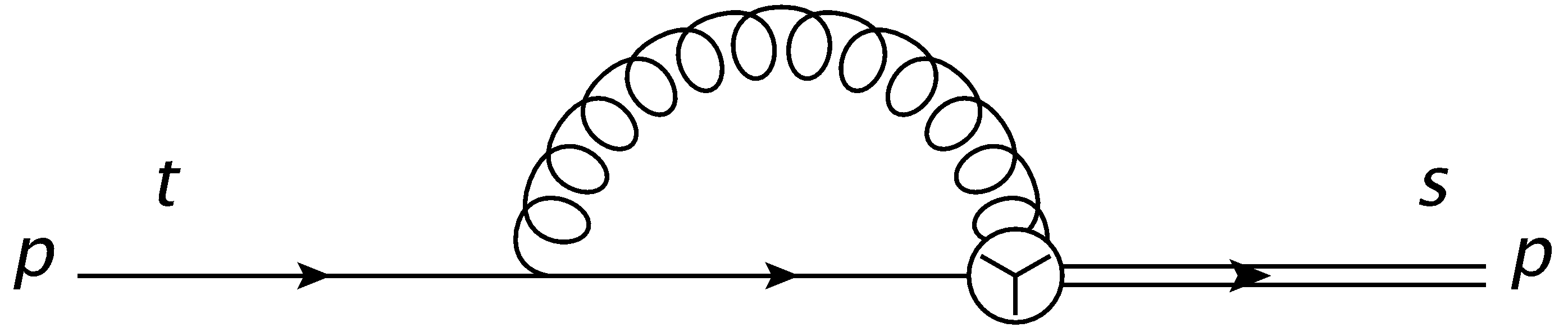}}}
		=g_0^2\frac{C_2(F)}{(4\pi)^2}\left[\frac{1}{\epsilon}+\log\left(8\pi\mu^2s\right)+1\right]+\mathcal{O}(\epsilon,s),\label{eq:G2b}\\
		&\Gamma_{3,a}^{(2)}(p;t)=
		\vcenter{\hbox{\includegraphics[width=140pt,height=30pt]{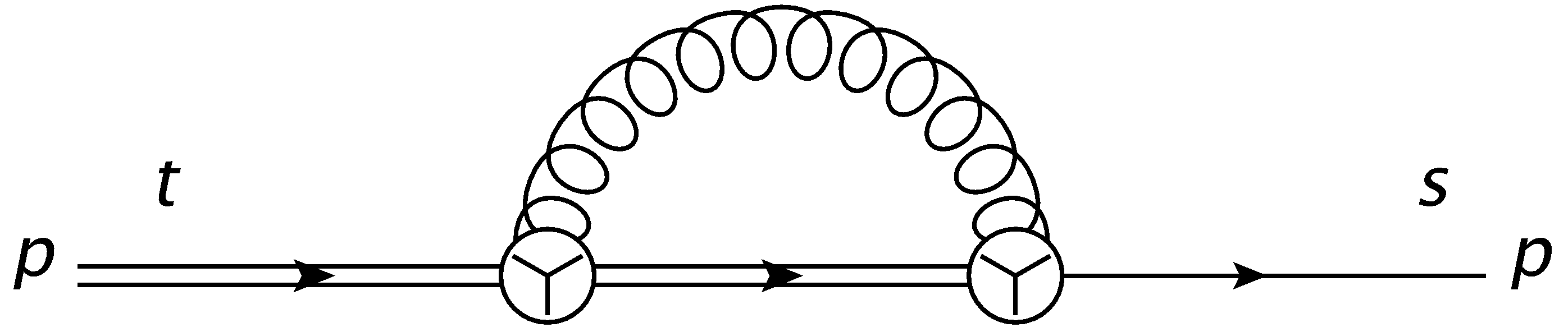}}}
		=0+\mathcal{O}(t),\label{eq:G3a}\\
		&\Gamma_{3,b}^{(2)}(p;s)=
		\vcenter{\hbox{\includegraphics[width=140pt,height=30pt]{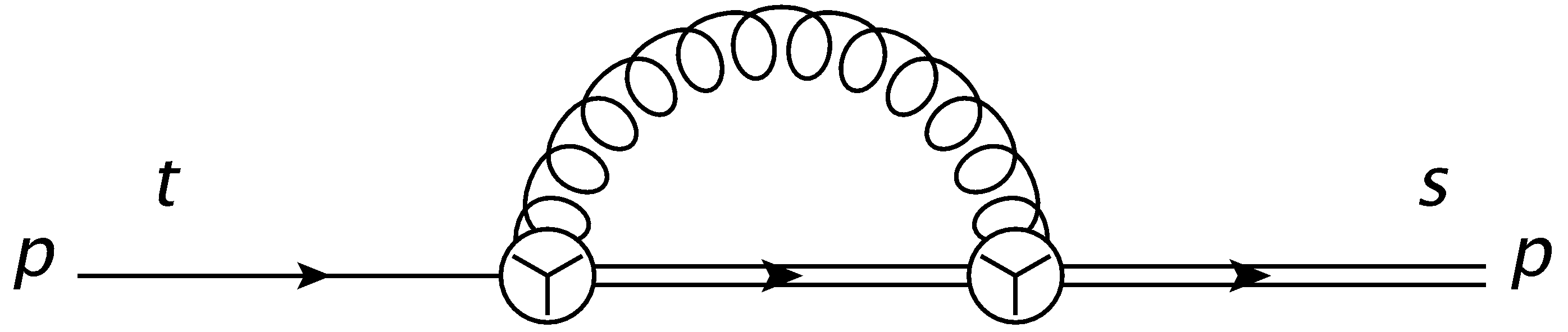}}}
		=0+\mathcal{O}(s),\label{eq:G3b}\\
		&\Gamma_{4,a}^{(2)}(p;t)=
		\vcenter{\hbox{\includegraphics[width=140pt,height=50pt]{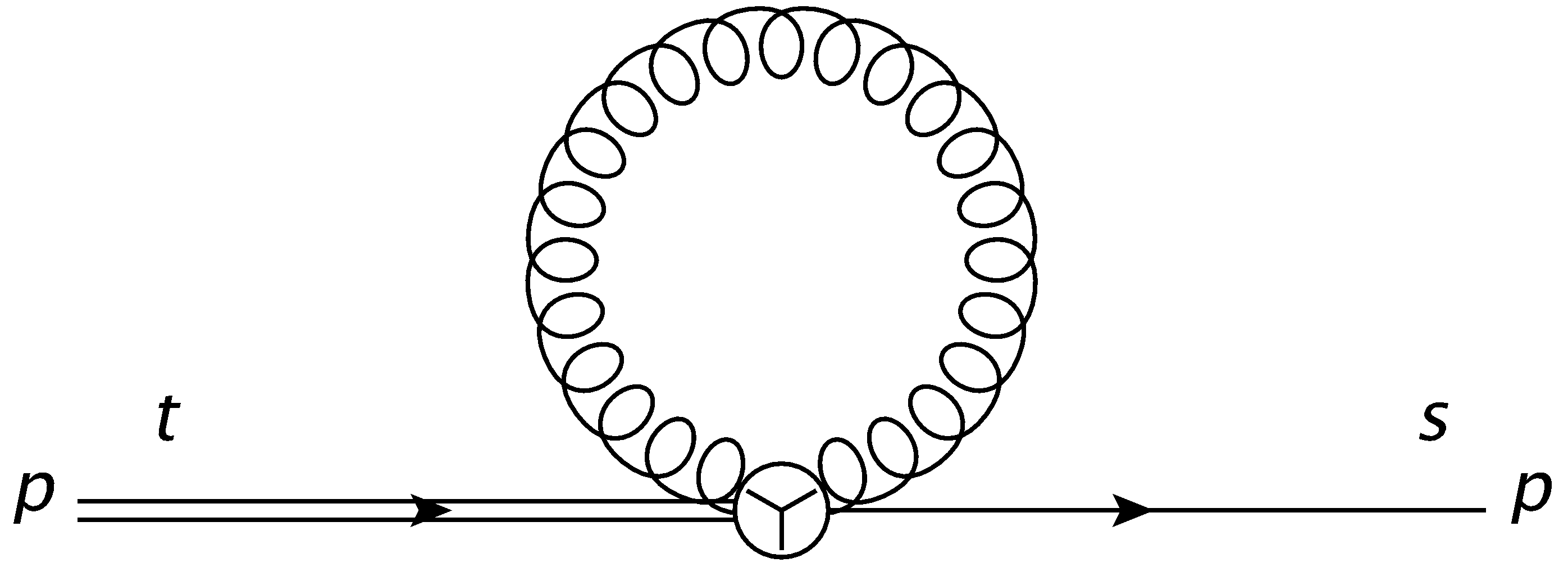}}}
		=-2g_0^2\frac{C_2(F)}{(4\pi)^2}\left[\frac{1}{\epsilon}+\log\left(8\pi\mu^2t\right)+\frac{1}{2}\right]+\mathcal{O}(\epsilon,t),\label{eq:G4a}\\
		&\Gamma_{4,b}^{(2)}(p;s)=
		\vcenter{\hbox{\includegraphics[width=140pt,height=50pt]{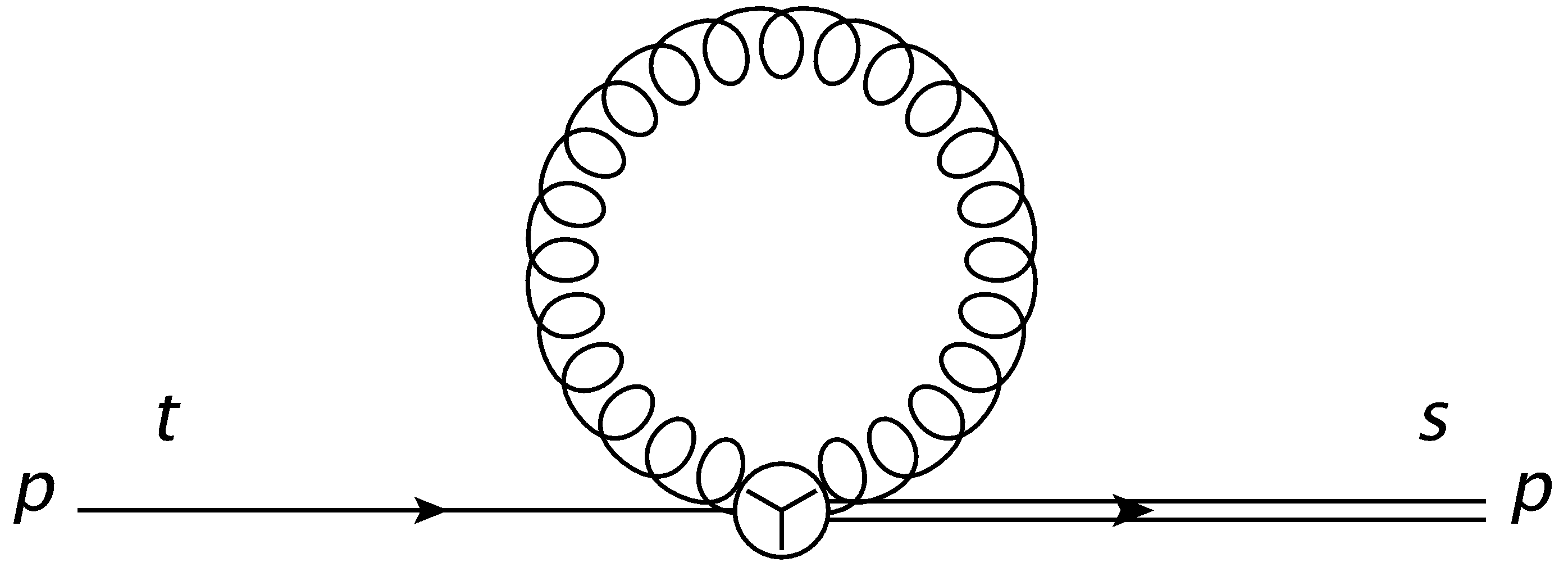}}}
		=-2g_0^2\frac{C_2(F)}{(4\pi)^2}\left[\frac{1}{\epsilon}+\log\left(8\pi\mu^2s\right)+\frac{1}{2}\right]+\mathcal{O}(\epsilon,s),\label{eq:G4b}\\
		&\Gamma_5^{(2)}(p;t,s)=
		\vcenter{\hbox{\includegraphics[width=140pt,height=30pt]{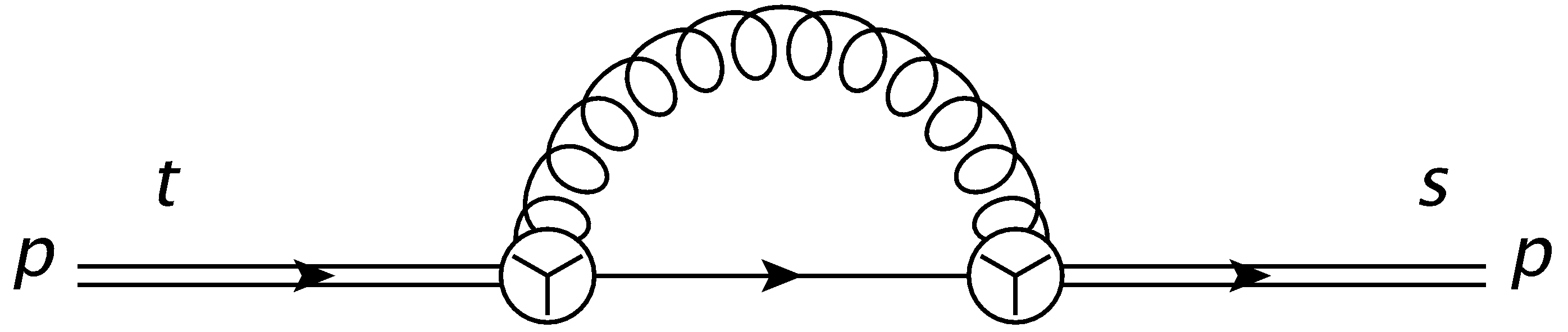}}}
		=0+\mathcal{O}(s,t),\label{eq:G5}
	\end{align}
\end{subequations}
where $R(m_0^2/p^2)$ is a remainder that vanishes for $m_0^2\ll p^2$.
The calculation of the first diagram $\Sigma_1^{(2)}(p)$ is identical
to the standard QCD quark self-energy with tree-level external quark propagators
carrying the flow-time dependence. We regulate the divergent integral with dimensional regularization
with $d=4-2\epsilon$ and $\epsilon>0$.

The next contribution, proportional to $\Gamma_{2,a}^{(2)}(p,t)$, contains a flow kernel and vertex.
Following the Feynman rules we outline in Appendix~\ref{app:rules} it is straightforward to write
\be
\Gamma_{2,a}^{(2)}(p;t) = -2ig_0^2C_2(F)\int_0^tdu\ e^{-p^2(t-u)}\int_q\frac{e^{-q^2u}}{q^2+m_0^2}\frac{e^{-(p+q)^2u}}{(p+q)^2}(iq^2+m_0\slashed{q}).
\ee
In standard perturbation theory, the integrand would next be recast with Feynman parameterization,
shifted, decomposed into scalar integrals, and brought to a spherically-symmetric form for integration in $d$ dimensions.
Specifically, the integrand must be isotropic, so that the $(d-1)$-dimensional
surface may be integrated separately from the radial portion.
This luxury is not afforded to us, however, as in this case,
the gluon propagator introduces an exponential factor, $e^{-(2p\cdot q)t}$,
which is only linear in the momentum $q$.
No Feynman parameterization and corresponding shift in the integration variable will fix this;
the exponential is neither even nor odd.
Our solution is to reparameterize the propagator \textit{\`a la} Schwinger
and to study the MacLaurin series of the cross-term:
\be
\frac{e^{-(p+q)^2u}}{(p+q)^2}=\int_0^\infty dz\ e^{-(p+q)^2(u+z)}=
\int_0^\infty dz\ e^{-(p^2+q^2)(u+z)}\sum_{n=0}^\infty\frac{(-2(u+z))^n}{n!}p_{\mu_1}\cdots p_{\mu_n}q_{\mu_1}\cdots q_{\mu_n},
\ee
where the sum over all $\mu_n$ is implied. The symmetry of this structure is now manifest;
that is, terms of even $n$ are even, and terms of odd $n$ are odd.
We now let $m_0\rightarrow0$, so that 
\be
\Gamma_{2,a}^{(2)}(p;t)=2g_0^2C_2(F)\sum_{n=0}^\infty\frac{4^n}{(2n)!}p_{I_{2n}}\int_0^tdu\int_0^\infty dz\ e^{-p^2(t+z)}(u+z)^{2n}\int_qe^{-q^2(2u+z)}q_{I_{2n}}+\mathcal{O}(m_0)\,.
\label{eq:G2Massless}
\ee
        
Indeed, in the complete calculation of the flowed diagrams of Eqs.~(\ref{eq:G2a}-\ref{eq:G5}), the mass only contributes at $\mathcal{O}(t)$. 
This allows for a concise demonstration of the techniques used in this article.
In general the kernel diagrams do not contribute to all orders in the same way as the standard QCD diagrams
~\eqref{eq:G1}. The full renormalization requires a coalescence of four semi-independent resummations.
For this reason we only consider the leading ${\cal O}g_0^2)$ corrections and how they
affect the wave function renormalization of the flowed fields.
The above integral employs the multi-index $I_n=(\mu_1,\mu_2,\dots,\mu_n)$.
Note that the multi-index above is a $2n$-tuple, because we neglect the mass and therefore 
the only term remaining outside of the gluon propagator, $iq^2$, is even,
and we may drop all odd $n$ through the reindexation $n\rightarrow2n$.
We have also rearranged the order of integration.
In order to justify this, we invoke Tonelli:
if the four integrals (including the sum, an integral with respect to the counting measure)
in \eqref{eq:G2Massless} converge in some order, then we are free to choose any order,
since the full integrand is strictly non-negative, and all domains of integration are clearly measure spaces with $\sigma$-finite measures.
With this in mind, we freely reorder the integrals,
and impose \textit{a posteriori} restrictions on the integrals as we discover them.
The momentum integral may now be calculated. Due to Lorentz invariance,
the only available structure with the total indicial symmetry of the $q_{I_{2n}}$
is the appropriately normalized sum over all $(2n-1)!!$
products of $n$ metric tensors,
where the indices are distributed according to all possible pairings.
For example, for $n=2$, we find
\be
	\int_qf(q^2)q_{I_{2n}}=\int_qf(q^2)q_{I_{4}}=\frac{\delta_{\mu_1\mu_2}\delta_{\mu_3\mu_4}+\delta_{\mu_1\mu_3}\delta_{\mu_2\mu_4}+\delta_{\mu_1\mu_4}\delta_{\mu_2\mu_3}}{d(d+2)}\int_qf(q^2)(q^2)^2,
	\label{eq:Symmetry2}
\ee
for some smooth function $f$. In general, we have 
\be
	\int_qf(q^2)q_{I_{2n}}=\int_qf(q^2)q_{I_{4}}=\frac{1}{(d)_{n,2}}S_{I_{2n}}^{(2n)}\int_qf(q^2)(q^2)^n,
\ee
where $(d)_{n,2}=\frac{2^n\Gamma(d/2+n)}{\Gamma(d/2)}$ is a Pochhammer $k$-symbol, and the tensor
\be
	S_{I_{2n}}^{(2n)}=\sum_{i=1}^{(2n-1)!!}\prod_{j=1}^n\delta_{\mu_{\sigma_i(2j-1)}\mu_{\sigma_i(2j)}}
\ee
is the generalization of the structure in \eqref{eq:Symmetry2}.
Each $\sigma_i$ is a permutation of the set $[2n]\subset\mathbb{N}$
corresponding to one of the $(2n-1)!!$ partitions without ordering of $[2n]$ into $n$ two-element subsets.
For clarity, inspect the indices in \eqref{eq:Symmetry2};
each term splits the set $\{1,2,3,4\}$ into two unordered pairs,
but the pairings are never the same.
Indeed, any permutation of the indices simply permutes the summands.
Thus the commutativity under addition of the terms in $S_{I_{2n}}^{(2n)}$
reproduces exact symmetry of the product of vectors $q_{I_{2n}}$.
Further, we integrate over the $(d-1)$-sphere to isolate the radial integral:
\be
	\begin{aligned}
		\Gamma_{2,a}^{(2)}(p;t)=&4g_0^2\frac{C_2(F)}{(4\pi)^2}\frac{(4\pi\mu^2)^{2-d/2}}{\Gamma(d/2)}\sum_{n=0}^\infty\frac{4^n}{(d)_{n,2}(2n)!}p_{I_{2n}}S^{(2n)}_{I_{2n}}\int_0^tdu\int_0^\infty dz\ e^{-p^2(t+z)}(u+z)^{2n}\\
		&\times\int_0^\infty q^{d-1}dq\ e^{-q^2(2u+z)}(q^2)^n+\mathcal{O}(m_0).
	\end{aligned}
\ee
The radial part is a simple gamma function,
and the momenta $p_{I_{2n}}$ saturate $S^{(2n)}_{I_{2n}}$,
so that after some simplification, we have
\be
	\Gamma_{2,a}^{(2)}(p;t)=2g_0^2\frac{C_2(F)}{(4\pi)^2}\left(\frac{4\pi\mu^2}{p^2}\right)^{2-d/2}\sum_{n=0}^\infty\frac{\tau}{n!}\int_0^1d\alpha\int_0^\infty d\zeta\ e^{-(\tau+\zeta)}\frac{(\alpha\tau+\zeta)^{2n}}{(2\alpha\tau+\zeta)^{d/2+n}}+\mathcal{O}(m_0).
\ee
where $\tau=p^2t$ and $\zeta = p^2 z$.
For $n\geq1$, every term is at least $\mathcal{O}(\tau)$,
since the numerator then dominates near $d=4$. Retaining only the $n=0$ term,
\be
	\begin{aligned}
		\Gamma_{2,a}^{(2)}(p;t)&=2g_0^2\frac{C_2(F)}{(4\pi)^2}\left(\frac{4\pi\mu^2}{p^2}\right)^{2-d/2}\tau\int_0^1d\alpha\int_0^\infty d\zeta\ e^{-(\tau+\zeta)}(2\alpha\tau+\zeta)^{-d/2}+\mathcal{O}(m_0,\tau)\\
		&=g_0^2\frac{C_2(F)}{(4\pi)^2}\left(\frac{4\pi\mu^2}{p^2}\right)^{\epsilon}\frac{e^{-\tau}\gamma(\epsilon,2\tau)}{1-\epsilon}+\mathcal{O}(m_0,\tau)\\
		&=g_0^2\frac{C_2(F)}{(4\pi)^2}\left[\frac{1}{\epsilon}+\log(8\pi\mu^2t)+1\right]+\mathcal{O}(m_0,t,\epsilon)\\
	\end{aligned}
\ee
as in \eqref{eq:G2a}. The error of $\mcO(m_0)$ is added here as a formality; it is absorbed into the $\mcO(t)$ term in the complete calculation.
The other graphs are calculated by similar means,
and we arrive at the one-loop self-energy for flowed fermions:
\be
S^{(2)}(x,y;t,s)=\int_p\frac{e^{ip(x-y)}}{i\slashed{p}}
\left\{1-g_0^2\frac{C_2(F)}{(4\pi)^2}\left[\frac{3}{\epsilon}+\log\left[(8\pi\mu^2)^2st\right]+
  \log\left(\frac{4\pi\mu^2}{p^2}\right)-\gamma_E+1\right]\right\}+\mathcal{O}(m_0,s,t,g_0^4)\,.
\ee
To renormalize the propagator, following ref.~\cite{Luscher:2013cpa},
we define the renormalized flowed fermion fields as
\be
\chi_{\textrm{R}}(x,t) = Z_\chi^{1/2} \chi(x,t)\,, \quad
\chibar_{\textrm{R}}(x,t) =  \chibar(x,t) Z_\chi^{1/2}\,,
\ee
so that the renormalized propagator reads
\be
S_R(x,t;y,s) = Z_\chi S(x,t;y,s)\,.
\ee
If we impose the family of conditions
\be
S_R\big|_{p^2=\mu^2=1/(8\pi\sqrt{st})}=S^{(0)},
\ee
we obtain
\be
Z_\chi \cdot
\left\{1-g_0^2\frac{C_2(F)}{(4\pi)^2}\left[\frac{3}{\epsilon}+\log\left[(8\pi\mu^2)^2st\right]
  +\log\left(\frac{4\pi\mu^2}{p^2}\right)-\gamma_E+1\right]\right\}\Bigg|_{s=t,p^2=\mu^2=1/(8\pi\sqrt{st})}=1+\mathcal{O}(g_0^4).
\ee
Expanding $Z_\chi$ in powers of the bare coupling
\be
Z_\chi = 1 + \sum_{k=1}^\infty g_0^{2k}Z_\chi^{(k)}\,,
\ee
we find
\be
Z_\chi=1+g_0^2\frac{C_2(F)}{(4\pi)^2}\left\{\frac{3}{\epsilon}+\log(4\pi)-\gamma_E+1\right\}+
\mathcal{O}(g_0^4).
\ee
We note that if we choose the MS scheme we obtain the same result already obtained in Ref.~\cite{Luscher:2013cpa}, and that pole contribution matches the results of \cite{Makino:2014taa,Monahan:2017hpu}. The finite terms, which depend on the choice of renormalization condition, have not, to our knowledge, appeared in the literature.

\bibliography{cpodd_pt.bib}

\begin{thebibliography}{90}%
\makeatletter
\providecommand \@ifxundefined [1]{%
 \@ifx{#1\undefined}
}%
\providecommand \@ifnum [1]{%
 \ifnum #1\expandafter \@firstoftwo
 \else \expandafter \@secondoftwo
 \fi
}%
\providecommand \@ifx [1]{%
 \ifx #1\expandafter \@firstoftwo
 \else \expandafter \@secondoftwo
 \fi
}%
\providecommand \natexlab [1]{#1}%
\providecommand \enquote  [1]{``#1''}%
\providecommand \bibnamefont  [1]{#1}%
\providecommand \bibfnamefont [1]{#1}%
\providecommand \citenamefont [1]{#1}%
\providecommand \href@noop [0]{\@secondoftwo}%
\providecommand \href [0]{\begingroup \@sanitize@url \@href}%
\providecommand \@href[1]{\@@startlink{#1}\@@href}%
\providecommand \@@href[1]{\endgroup#1\@@endlink}%
\providecommand \@sanitize@url [0]{\catcode `\\12\catcode `\$12\catcode
  `\&12\catcode `\#12\catcode `\^12\catcode `\_12\catcode `\%12\relax}%
\providecommand \@@startlink[1]{}%
\providecommand \@@endlink[0]{}%
\providecommand \url  [0]{\begingroup\@sanitize@url \@url }%
\providecommand \@url [1]{\endgroup\@href {#1}{\urlprefix }}%
\providecommand \urlprefix  [0]{URL }%
\providecommand \Eprint [0]{\href }%
\providecommand \doibase [0]{https://doi.org/}%
\providecommand \selectlanguage [0]{\@gobble}%
\providecommand \bibinfo  [0]{\@secondoftwo}%
\providecommand \bibfield  [0]{\@secondoftwo}%
\providecommand \translation [1]{[#1]}%
\providecommand \BibitemOpen [0]{}%
\providecommand \bibitemStop [0]{}%
\providecommand \bibitemNoStop [0]{.\EOS\space}%
\providecommand \EOS [0]{\spacefactor3000\relax}%
\providecommand \BibitemShut  [1]{\csname bibitem#1\endcsname}%
\let\auto@bib@innerbib\@empty
\bibitem [{\citenamefont {Abel}\ \emph {et~al.}(2020)\citenamefont {Abel} \emph
  {et~al.}}]{Abel:2020gbr}%
  \BibitemOpen
  \bibfield  {author} {\bibinfo {author} {\bibfnamefont {C.}~\bibnamefont
  {Abel}} \emph {et~al.} (\bibinfo {collaboration} {nEDM}),\ }\bibfield
  {title} {\bibinfo {title} {{Measurement of the permanent electric dipole
  moment of the neutron}},\ }\href
  {https://doi.org/10.1103/PhysRevLett.124.081803} {\bibfield  {journal}
  {\bibinfo  {journal} {Phys. Rev. Lett.}\ }\textbf {\bibinfo {volume} {124}},\
  \bibinfo {pages} {081803} (\bibinfo {year} {2020})},\ \Eprint
  {https://arxiv.org/abs/2001.11966} {arXiv:2001.11966 [hep-ex]} \BibitemShut
  {NoStop}%
\bibitem [{\citenamefont {Pendlebury}\ \emph {et~al.}(2015)\citenamefont
  {Pendlebury} \emph {et~al.}}]{Afach:2015sja}%
  \BibitemOpen
  \bibfield  {author} {\bibinfo {author} {\bibfnamefont {J.~M.}\ \bibnamefont
  {Pendlebury}} \emph {et~al.},\ }\bibfield  {title} {\bibinfo {title}
  {{Revised experimental upper limit on the electric dipole moment of the
  neutron}},\ }\href {https://doi.org/10.1103/PhysRevD.92.092003} {\bibfield
  {journal} {\bibinfo  {journal} {Phys. Rev.}\ }\textbf {\bibinfo {volume}
  {D92}},\ \bibinfo {pages} {092003} (\bibinfo {year} {2015})},\ \Eprint
  {https://arxiv.org/abs/1509.04411} {arXiv:1509.04411 [hep-ex]} \BibitemShut
  {NoStop}%
\bibitem [{\citenamefont {Chupp}\ \emph {et~al.}(2019)\citenamefont {Chupp},
  \citenamefont {Fierlinger}, \citenamefont {Ramsey-Musolf},\ and\
  \citenamefont {Singh}}]{Chupp:2017rkp}%
  \BibitemOpen
  \bibfield  {author} {\bibinfo {author} {\bibfnamefont {T.}~\bibnamefont
  {Chupp}}, \bibinfo {author} {\bibfnamefont {P.}~\bibnamefont {Fierlinger}},
  \bibinfo {author} {\bibfnamefont {M.}~\bibnamefont {Ramsey-Musolf}},\ and\
  \bibinfo {author} {\bibfnamefont {J.}~\bibnamefont {Singh}},\ }\bibfield
  {title} {\bibinfo {title} {{Electric dipole moments of atoms, molecules,
  nuclei, and particles}},\ }\href
  {https://doi.org/10.1103/RevModPhys.91.015001} {\bibfield  {journal}
  {\bibinfo  {journal} {Rev. Mod. Phys.}\ }\textbf {\bibinfo {volume} {91}},\
  \bibinfo {pages} {015001} (\bibinfo {year} {2019})},\ \Eprint
  {https://arxiv.org/abs/1710.02504} {arXiv:1710.02504 [physics.atom-ph]}
  \BibitemShut {NoStop}%
\bibitem [{\citenamefont {Seng}(2015)}]{Seng:2014lea}%
  \BibitemOpen
  \bibfield  {author} {\bibinfo {author} {\bibfnamefont {C.-Y.}\ \bibnamefont
  {Seng}},\ }\bibfield  {title} {\bibinfo {title} {{Reexamination of The
  Standard Model Nucleon Electric Dipole Moment}},\ }\href
  {https://doi.org/10.1103/PhysRevC.91.025502} {\bibfield  {journal} {\bibinfo
  {journal} {Phys. Rev.}\ }\textbf {\bibinfo {volume} {C91}},\ \bibinfo {pages}
  {025502} (\bibinfo {year} {2015})},\ \Eprint
  {https://arxiv.org/abs/1411.1476} {arXiv:1411.1476 [hep-ph]} \BibitemShut
  {NoStop}%
\bibitem [{\citenamefont {Dragos}\ \emph {et~al.}(2019)\citenamefont {Dragos},
  \citenamefont {Luu}, \citenamefont {Shindler}, \citenamefont {de~Vries},\
  and\ \citenamefont {Yousif}}]{Dragos:2019oxn}%
  \BibitemOpen
  \bibfield  {author} {\bibinfo {author} {\bibfnamefont {J.}~\bibnamefont
  {Dragos}}, \bibinfo {author} {\bibfnamefont {T.}~\bibnamefont {Luu}},
  \bibinfo {author} {\bibfnamefont {A.}~\bibnamefont {Shindler}}, \bibinfo
  {author} {\bibfnamefont {J.}~\bibnamefont {de~Vries}},\ and\ \bibinfo
  {author} {\bibfnamefont {A.}~\bibnamefont {Yousif}},\ }\href@noop {}
  {\bibinfo {title} {{Confirming the Existence of the strong CP Problem in
  Lattice QCD with the Gradient Flow}}} (\bibinfo {year} {2019}),\ \Eprint
  {https://arxiv.org/abs/1902.03254} {arXiv:1902.03254 [hep-lat]} \BibitemShut
  {NoStop}%
\bibitem [{\citenamefont {Demir}\ \emph {et~al.}(2003)\citenamefont {Demir},
  \citenamefont {Pospelov},\ and\ \citenamefont {Ritz}}]{Demir:2002gg}%
  \BibitemOpen
  \bibfield  {author} {\bibinfo {author} {\bibfnamefont {D.~A.}\ \bibnamefont
  {Demir}}, \bibinfo {author} {\bibfnamefont {M.}~\bibnamefont {Pospelov}},\
  and\ \bibinfo {author} {\bibfnamefont {A.}~\bibnamefont {Ritz}},\ }\bibfield
  {title} {\bibinfo {title} {{Hadronic EDMs, the Weinberg operator, and light
  gluinos}},\ }\href {https://doi.org/10.1103/PhysRevD.67.015007} {\bibfield
  {journal} {\bibinfo  {journal} {Phys. Rev. D}\ }\textbf {\bibinfo {volume}
  {67}},\ \bibinfo {pages} {015007} (\bibinfo {year} {2003})},\ \Eprint
  {https://arxiv.org/abs/hep-ph/0208257} {arXiv:hep-ph/0208257} \BibitemShut
  {NoStop}%
\bibitem [{\citenamefont {Haisch}\ and\ \citenamefont
  {Hala}(2019)}]{Haisch:2019bml}%
  \BibitemOpen
  \bibfield  {author} {\bibinfo {author} {\bibfnamefont {U.}~\bibnamefont
  {Haisch}}\ and\ \bibinfo {author} {\bibfnamefont {A.}~\bibnamefont {Hala}},\
  }\bibfield  {title} {\bibinfo {title} {{Sum rules for CP-violating operators
  of Weinberg type}},\ }\href {https://doi.org/10.1007/JHEP11(2019)154}
  {\bibfield  {journal} {\bibinfo  {journal} {JHEP}\ }\textbf {\bibinfo
  {volume} {11}},\ \bibinfo {pages} {154}},\ \Eprint
  {https://arxiv.org/abs/1909.08955} {arXiv:1909.08955 [hep-ph]} \BibitemShut
  {NoStop}%
\bibitem [{\citenamefont {de~Vries}\ \emph {et~al.}(2011)\citenamefont
  {de~Vries}, \citenamefont {Timmermans}, \citenamefont {Mereghetti},\ and\
  \citenamefont {van Kolck}}]{deVries:2010ah}%
  \BibitemOpen
  \bibfield  {author} {\bibinfo {author} {\bibfnamefont {J.}~\bibnamefont
  {de~Vries}}, \bibinfo {author} {\bibfnamefont {R.}~\bibnamefont
  {Timmermans}}, \bibinfo {author} {\bibfnamefont {E.}~\bibnamefont
  {Mereghetti}},\ and\ \bibinfo {author} {\bibfnamefont {U.}~\bibnamefont {van
  Kolck}},\ }\bibfield  {title} {\bibinfo {title} {{The Nucleon Electric Dipole
  Form Factor From Dimension-Six Time-Reversal Violation}},\ }\href
  {https://doi.org/10.1016/j.physletb.2010.11.042} {\bibfield  {journal}
  {\bibinfo  {journal} {Phys. Lett. B}\ }\textbf {\bibinfo {volume} {695}},\
  \bibinfo {pages} {268} (\bibinfo {year} {2011})},\ \Eprint
  {https://arxiv.org/abs/1006.2304} {arXiv:1006.2304 [hep-ph]} \BibitemShut
  {NoStop}%
\bibitem [{\citenamefont {Mereghetti}\ \emph {et~al.}(2011)\citenamefont
  {Mereghetti}, \citenamefont {de~Vries}, \citenamefont {Hockings},
  \citenamefont {Maekawa},\ and\ \citenamefont {van
  Kolck}}]{Mereghetti:2010kp}%
  \BibitemOpen
  \bibfield  {author} {\bibinfo {author} {\bibfnamefont {E.}~\bibnamefont
  {Mereghetti}}, \bibinfo {author} {\bibfnamefont {J.}~\bibnamefont
  {de~Vries}}, \bibinfo {author} {\bibfnamefont {W.}~\bibnamefont {Hockings}},
  \bibinfo {author} {\bibfnamefont {C.}~\bibnamefont {Maekawa}},\ and\ \bibinfo
  {author} {\bibfnamefont {U.}~\bibnamefont {van Kolck}},\ }\bibfield  {title}
  {\bibinfo {title} {{The Electric Dipole Form Factor of the Nucleon in Chiral
  Perturbation Theory to Sub-leading Order}},\ }\href
  {https://doi.org/10.1016/j.physletb.2010.12.018} {\bibfield  {journal}
  {\bibinfo  {journal} {Phys.Lett.}\ }\textbf {\bibinfo {volume} {B696}},\
  \bibinfo {pages} {97} (\bibinfo {year} {2011})},\ \Eprint
  {https://arxiv.org/abs/1010.4078} {arXiv:1010.4078 [hep-ph]} \BibitemShut
  {NoStop}%
\bibitem [{\citenamefont {Shintani}\ \emph {et~al.}(2005)\citenamefont
  {Shintani}, \citenamefont {Aoki}, \citenamefont {Ishizuka}, \citenamefont
  {Kanaya}, \citenamefont {Kikukawa}, \citenamefont {Kuramashi}, \citenamefont
  {Okawa}, \citenamefont {Tanigchi}, \citenamefont {Ukawa},\ and\ \citenamefont
  {Yoshie}}]{Shintani:2005xg}%
  \BibitemOpen
  \bibfield  {author} {\bibinfo {author} {\bibfnamefont {E.}~\bibnamefont
  {Shintani}}, \bibinfo {author} {\bibfnamefont {S.}~\bibnamefont {Aoki}},
  \bibinfo {author} {\bibfnamefont {N.}~\bibnamefont {Ishizuka}}, \bibinfo
  {author} {\bibfnamefont {K.}~\bibnamefont {Kanaya}}, \bibinfo {author}
  {\bibfnamefont {Y.}~\bibnamefont {Kikukawa}}, \bibinfo {author}
  {\bibfnamefont {Y.}~\bibnamefont {Kuramashi}}, \bibinfo {author}
  {\bibfnamefont {M.}~\bibnamefont {Okawa}}, \bibinfo {author} {\bibfnamefont
  {Y.}~\bibnamefont {Tanigchi}}, \bibinfo {author} {\bibfnamefont
  {A.}~\bibnamefont {Ukawa}},\ and\ \bibinfo {author} {\bibfnamefont
  {T.}~\bibnamefont {Yoshie}},\ }\bibfield  {title} {\bibinfo {title} {{Neutron
  electric dipole moment from lattice QCD}},\ }\href
  {https://doi.org/10.1103/PhysRevD.72.014504} {\bibfield  {journal} {\bibinfo
  {journal} {Phys. Rev.}\ }\textbf {\bibinfo {volume} {D72}},\ \bibinfo {pages}
  {014504} (\bibinfo {year} {2005})},\ \Eprint
  {https://arxiv.org/abs/hep-lat/0505022} {arXiv:hep-lat/0505022 [hep-lat]}
  \BibitemShut {NoStop}%
\bibitem [{\citenamefont {Berruto}\ \emph {et~al.}(2006)\citenamefont
  {Berruto}, \citenamefont {Blum}, \citenamefont {Orginos},\ and\ \citenamefont
  {Soni}}]{Berruto:2005hg}%
  \BibitemOpen
  \bibfield  {author} {\bibinfo {author} {\bibfnamefont {F.}~\bibnamefont
  {Berruto}}, \bibinfo {author} {\bibfnamefont {T.}~\bibnamefont {Blum}},
  \bibinfo {author} {\bibfnamefont {K.}~\bibnamefont {Orginos}},\ and\ \bibinfo
  {author} {\bibfnamefont {A.}~\bibnamefont {Soni}},\ }\bibfield  {title}
  {\bibinfo {title} {{Calculation of the neutron electric dipole moment with
  two dynamical flavors of domain wall fermions}},\ }\href
  {https://doi.org/10.1103/PhysRevD.73.054509} {\bibfield  {journal} {\bibinfo
  {journal} {Phys. Rev.}\ }\textbf {\bibinfo {volume} {D73}},\ \bibinfo {pages}
  {054509} (\bibinfo {year} {2006})},\ \Eprint
  {https://arxiv.org/abs/hep-lat/0512004} {arXiv:hep-lat/0512004 [hep-lat]}
  \BibitemShut {NoStop}%
\bibitem [{\citenamefont {Guo}\ \emph {et~al.}(2015)\citenamefont {Guo},
  \citenamefont {Horsley}, \citenamefont {Mei{\ss}ner}, \citenamefont
  {Nakamura}, \citenamefont {Perlt}, \citenamefont {Rakow}, \citenamefont
  {Schierholz}, \citenamefont {Schiller},\ and\ \citenamefont
  {Zanotti}}]{Guo:2015tla}%
  \BibitemOpen
  \bibfield  {author} {\bibinfo {author} {\bibfnamefont {F.~K.}\ \bibnamefont
  {Guo}}, \bibinfo {author} {\bibfnamefont {R.}~\bibnamefont {Horsley}},
  \bibinfo {author} {\bibfnamefont {U.-G.}\ \bibnamefont {Mei{\ss}ner}},
  \bibinfo {author} {\bibfnamefont {Y.}~\bibnamefont {Nakamura}}, \bibinfo
  {author} {\bibfnamefont {H.}~\bibnamefont {Perlt}}, \bibinfo {author}
  {\bibfnamefont {P.~E.~L.}\ \bibnamefont {Rakow}}, \bibinfo {author}
  {\bibfnamefont {G.}~\bibnamefont {Schierholz}}, \bibinfo {author}
  {\bibfnamefont {A.}~\bibnamefont {Schiller}},\ and\ \bibinfo {author}
  {\bibfnamefont {J.~M.}\ \bibnamefont {Zanotti}},\ }\bibfield  {title}
  {\bibinfo {title} {{The electric dipole moment of the neutron from 2+1 flavor
  lattice QCD}},\ }\href {https://doi.org/10.1103/PhysRevLett.115.062001}
  {\bibfield  {journal} {\bibinfo  {journal} {Phys. Rev. Lett.}\ }\textbf
  {\bibinfo {volume} {115}},\ \bibinfo {pages} {062001} (\bibinfo {year}
  {2015})},\ \Eprint {https://arxiv.org/abs/1502.02295} {arXiv:1502.02295
  [hep-lat]} \BibitemShut {NoStop}%
\bibitem [{\citenamefont {Shindler}\ \emph {et~al.}(2015)\citenamefont
  {Shindler}, \citenamefont {Luu},\ and\ \citenamefont
  {de~Vries}}]{Shindler:2015aqa}%
  \BibitemOpen
  \bibfield  {author} {\bibinfo {author} {\bibfnamefont {A.}~\bibnamefont
  {Shindler}}, \bibinfo {author} {\bibfnamefont {T.}~\bibnamefont {Luu}},\ and\
  \bibinfo {author} {\bibfnamefont {J.}~\bibnamefont {de~Vries}},\ }\bibfield
  {title} {\bibinfo {title} {{Nucleon electric dipole moment with the gradient
  flow: The {$\theta$}-term contribution}},\ }\href
  {https://doi.org/10.1103/PhysRevD.92.094518} {\bibfield  {journal} {\bibinfo
  {journal} {Phys. Rev.}\ }\textbf {\bibinfo {volume} {D92}},\ \bibinfo {pages}
  {094518} (\bibinfo {year} {2015})},\ \Eprint
  {https://arxiv.org/abs/1507.02343} {arXiv:1507.02343 [hep-lat]} \BibitemShut
  {NoStop}%
\bibitem [{\citenamefont {Alexandrou}\ \emph {et~al.}(2016)\citenamefont
  {Alexandrou}, \citenamefont {Athenodorou}, \citenamefont {Constantinou},
  \citenamefont {Hadjiyiannakou}, \citenamefont {Jansen}, \citenamefont
  {Koutsou}, \citenamefont {Ottnad},\ and\ \citenamefont
  {Petschlies}}]{Alexandrou:2015spa}%
  \BibitemOpen
  \bibfield  {author} {\bibinfo {author} {\bibfnamefont {C.}~\bibnamefont
  {Alexandrou}}, \bibinfo {author} {\bibfnamefont {A.}~\bibnamefont
  {Athenodorou}}, \bibinfo {author} {\bibfnamefont {M.}~\bibnamefont
  {Constantinou}}, \bibinfo {author} {\bibfnamefont {K.}~\bibnamefont
  {Hadjiyiannakou}}, \bibinfo {author} {\bibfnamefont {K.}~\bibnamefont
  {Jansen}}, \bibinfo {author} {\bibfnamefont {G.}~\bibnamefont {Koutsou}},
  \bibinfo {author} {\bibfnamefont {K.}~\bibnamefont {Ottnad}},\ and\ \bibinfo
  {author} {\bibfnamefont {M.}~\bibnamefont {Petschlies}},\ }\bibfield  {title}
  {\bibinfo {title} {{Neutron electric dipole moment using $N_{f} =2+1+1$
  twisted mass fermions}},\ }\href {https://doi.org/10.1103/PhysRevD.93.074503}
  {\bibfield  {journal} {\bibinfo  {journal} {Phys. Rev.}\ }\textbf {\bibinfo
  {volume} {D93}},\ \bibinfo {pages} {074503} (\bibinfo {year} {2016})},\
  \Eprint {https://arxiv.org/abs/1510.05823} {arXiv:1510.05823 [hep-lat]}
  \BibitemShut {NoStop}%
\bibitem [{\citenamefont {Shintani}\ \emph {et~al.}(2016)\citenamefont
  {Shintani}, \citenamefont {Blum}, \citenamefont {Izubuchi},\ and\
  \citenamefont {Soni}}]{Shintani:2015vsx}%
  \BibitemOpen
  \bibfield  {author} {\bibinfo {author} {\bibfnamefont {E.}~\bibnamefont
  {Shintani}}, \bibinfo {author} {\bibfnamefont {T.}~\bibnamefont {Blum}},
  \bibinfo {author} {\bibfnamefont {T.}~\bibnamefont {Izubuchi}},\ and\
  \bibinfo {author} {\bibfnamefont {A.}~\bibnamefont {Soni}},\ }\bibfield
  {title} {\bibinfo {title} {{Neutron and proton electric dipole moments from
  $N_f=2+1$ domain-wall fermion lattice QCD}},\ }\href
  {https://doi.org/10.1103/PhysRevD.93.094503} {\bibfield  {journal} {\bibinfo
  {journal} {Phys. Rev.}\ }\textbf {\bibinfo {volume} {D93}},\ \bibinfo {pages}
  {094503} (\bibinfo {year} {2016})},\ \Eprint
  {https://arxiv.org/abs/1512.00566} {arXiv:1512.00566 [hep-lat]} \BibitemShut
  {NoStop}%
\bibitem [{\citenamefont {Abramczyk}\ \emph {et~al.}(2017)\citenamefont
  {Abramczyk}, \citenamefont {Aoki}, \citenamefont {Blum}, \citenamefont
  {Izubuchi}, \citenamefont {Ohki},\ and\ \citenamefont
  {Syritsyn}}]{Abramczyk:2017oxr}%
  \BibitemOpen
  \bibfield  {author} {\bibinfo {author} {\bibfnamefont {M.}~\bibnamefont
  {Abramczyk}}, \bibinfo {author} {\bibfnamefont {S.}~\bibnamefont {Aoki}},
  \bibinfo {author} {\bibfnamefont {T.}~\bibnamefont {Blum}}, \bibinfo {author}
  {\bibfnamefont {T.}~\bibnamefont {Izubuchi}}, \bibinfo {author}
  {\bibfnamefont {H.}~\bibnamefont {Ohki}},\ and\ \bibinfo {author}
  {\bibfnamefont {S.}~\bibnamefont {Syritsyn}},\ }\bibfield  {title} {\bibinfo
  {title} {{Lattice calculation of electric dipole moments and form factors of
  the nucleon}},\ }\href {https://doi.org/10.1103/PhysRevD.96.014501}
  {\bibfield  {journal} {\bibinfo  {journal} {Phys. Rev.}\ }\textbf {\bibinfo
  {volume} {D96}},\ \bibinfo {pages} {014501} (\bibinfo {year} {2017})},\
  \Eprint {https://arxiv.org/abs/1701.07792} {arXiv:1701.07792 [hep-lat]}
  \BibitemShut {NoStop}%
\bibitem [{\citenamefont {Yoon}\ \emph {et~al.}(2018)\citenamefont {Yoon},
  \citenamefont {Bhattacharya},\ and\ \citenamefont {Gupta}}]{Yoon:2017tag}%
  \BibitemOpen
  \bibfield  {author} {\bibinfo {author} {\bibfnamefont {B.}~\bibnamefont
  {Yoon}}, \bibinfo {author} {\bibfnamefont {T.}~\bibnamefont {Bhattacharya}},\
  and\ \bibinfo {author} {\bibfnamefont {R.}~\bibnamefont {Gupta}},\ }\bibfield
   {title} {\bibinfo {title} {{Neutron Electric Dipole Moment on the
  Lattice}},\ }\href {https://doi.org/10.1051/epjconf/201817501014} {\bibfield
  {journal} {\bibinfo  {journal} {EPJ Web Conf.}\ }\textbf {\bibinfo {volume}
  {175}},\ \bibinfo {pages} {01014} (\bibinfo {year} {2018})},\ \Eprint
  {https://arxiv.org/abs/1712.08557} {arXiv:1712.08557 [hep-lat]} \BibitemShut
  {NoStop}%
\bibitem [{\citenamefont {Izubuchi}\ \emph {et~al.}(2020)\citenamefont
  {Izubuchi}, \citenamefont {Ohki},\ and\ \citenamefont
  {Syritsyn}}]{Izubuchi:2020ngl}%
  \BibitemOpen
  \bibfield  {author} {\bibinfo {author} {\bibfnamefont {T.}~\bibnamefont
  {Izubuchi}}, \bibinfo {author} {\bibfnamefont {H.}~\bibnamefont {Ohki}},\
  and\ \bibinfo {author} {\bibfnamefont {S.}~\bibnamefont {Syritsyn}},\
  }\bibfield  {title} {\bibinfo {title} {{Computing Nucleon Electric Dipole
  Moment from lattice QCD}},\ }in\ \href@noop {} {\emph {\bibinfo {booktitle}
  {{37th International Symposium on Lattice Field Theory}}}}\ (\bibinfo {year}
  {2020})\ \Eprint {https://arxiv.org/abs/2004.10449} {arXiv:2004.10449
  [hep-lat]} \BibitemShut {NoStop}%
\bibitem [{\citenamefont {Dragos}\ \emph
  {et~al.}(2018{\natexlab{a}})\citenamefont {Dragos}, \citenamefont {Luu},
  \citenamefont {Shindler}, \citenamefont {de~Vries},\ and\ \citenamefont
  {Yousif}}]{Dragos:2018uzd}%
  \BibitemOpen
  \bibfield  {author} {\bibinfo {author} {\bibfnamefont {J.}~\bibnamefont
  {Dragos}}, \bibinfo {author} {\bibfnamefont {T.}~\bibnamefont {Luu}},
  \bibinfo {author} {\bibfnamefont {A.}~\bibnamefont {Shindler}}, \bibinfo
  {author} {\bibfnamefont {J.}~\bibnamefont {de~Vries}},\ and\ \bibinfo
  {author} {\bibfnamefont {A.}~\bibnamefont {Yousif}},\ }\bibfield  {title}
  {\bibinfo {title} {{Improvements to Nucleon Matrix Elements within a $\theta$
  Vacuum from Lattice QCD}}\ }(\bibinfo {year} {2018})\ \Eprint
  {https://arxiv.org/abs/1809.03487} {arXiv:1809.03487 [hep-lat]} \BibitemShut
  {NoStop}%
\bibitem [{\citenamefont {Shindler}\ \emph {et~al.}(2014)\citenamefont
  {Shindler}, \citenamefont {de~Vries},\ and\ \citenamefont
  {Luu}}]{Shindler:2014oha}%
  \BibitemOpen
  \bibfield  {author} {\bibinfo {author} {\bibfnamefont {A.}~\bibnamefont
  {Shindler}}, \bibinfo {author} {\bibfnamefont {J.}~\bibnamefont {de~Vries}},\
  and\ \bibinfo {author} {\bibfnamefont {T.}~\bibnamefont {Luu}},\ }\bibfield
  {title} {\bibinfo {title} {{Beyond-the-Standard-Model matrix elements with
  the gradient flow}},\ }\bibfield  {booktitle} {\emph {\bibinfo {booktitle}
  {{Proceedings, 32nd International Symposium on Lattice Field Theory (Lattice
  2014): Brookhaven, NY, USA, June 23-28, 2014}}},\ }\href
  {https://doi.org/10.22323/1.214.0251} {\bibfield  {journal} {\bibinfo
  {journal} {PoS}\ }\textbf {\bibinfo {volume} {LATTICE2014}},\ \bibinfo
  {pages} {251} (\bibinfo {year} {2014})},\ \Eprint
  {https://arxiv.org/abs/1409.2735} {arXiv:1409.2735 [hep-lat]} \BibitemShut
  {NoStop}%
\bibitem [{\citenamefont {Narayanan}\ and\ \citenamefont
  {Neuberger}(2006)}]{Narayanan:2006rf}%
  \BibitemOpen
  \bibfield  {author} {\bibinfo {author} {\bibfnamefont {R.}~\bibnamefont
  {Narayanan}}\ and\ \bibinfo {author} {\bibfnamefont {H.}~\bibnamefont
  {Neuberger}},\ }\bibfield  {title} {\bibinfo {title} {{Infinite N phase
  transitions in continuum Wilson loop operators}},\ }\href
  {https://doi.org/10.1088/1126-6708/2006/03/064} {\bibfield  {journal}
  {\bibinfo  {journal} {JHEP}\ }\textbf {\bibinfo {volume} {0603}},\ \bibinfo
  {pages} {064}},\ \Eprint {https://arxiv.org/abs/hep-th/0601210}
  {arXiv:hep-th/0601210 [hep-th]} \BibitemShut {NoStop}%
\bibitem [{\citenamefont {{L\"uscher, M.}}(2010)}]{Luscher:2010iy}%
  \BibitemOpen
  \bibfield  {author} {\bibinfo {author} {\bibnamefont {{L\"uscher, M.}}},\
  }\bibfield  {title} {\bibinfo {title} {{Properties and uses of the Wilson
  flow in lattice QCD}},\ }\href {https://doi.org/10.1007/JHEP08(2010)071}
  {\bibfield  {journal} {\bibinfo  {journal} {JHEP}\ }\textbf {\bibinfo
  {volume} {1008}},\ \bibinfo {pages} {071}},\ \Eprint
  {https://arxiv.org/abs/1006.4518} {arXiv:1006.4518 [hep-lat]} \BibitemShut
  {NoStop}%
\bibitem [{\citenamefont {{L\"uscher, M.}}\ and\ \citenamefont
  {Weisz}(2011)}]{Luscher:2011bx}%
  \BibitemOpen
  \bibfield  {author} {\bibinfo {author} {\bibnamefont {{L\"uscher, M.}}}\ and\
  \bibinfo {author} {\bibfnamefont {P.}~\bibnamefont {Weisz}},\ }\bibfield
  {title} {\bibinfo {title} {{Perturbative analysis of the gradient flow in
  non-abelian gauge theories}},\ }\href
  {https://doi.org/10.1007/JHEP02(2011)051} {\bibfield  {journal} {\bibinfo
  {journal} {JHEP}\ }\textbf {\bibinfo {volume} {1102}},\ \bibinfo {pages}
  {051}},\ \Eprint {https://arxiv.org/abs/1101.0963} {arXiv:1101.0963 [hep-th]}
  \BibitemShut {NoStop}%
\bibitem [{\citenamefont {{L\"uscher, M.}}(2013)}]{Luscher:2013cpa}%
  \BibitemOpen
  \bibfield  {author} {\bibinfo {author} {\bibnamefont {{L\"uscher, M.}}},\
  }\bibfield  {title} {\bibinfo {title} {{Chiral symmetry and the Yang--Mills
  gradient flow}},\ }\href {https://doi.org/10.1007/JHEP04(2013)123} {\bibfield
   {journal} {\bibinfo  {journal} {JHEP}\ }\textbf {\bibinfo {volume} {1304}},\
  \bibinfo {pages} {123}},\ \Eprint {https://arxiv.org/abs/1302.5246}
  {arXiv:1302.5246 [hep-lat]} \BibitemShut {NoStop}%
\bibitem [{\citenamefont {Dragos}\ \emph
  {et~al.}(2018{\natexlab{b}})\citenamefont {Dragos}, \citenamefont {Luu},
  \citenamefont {Shindler},\ and\ \citenamefont {de~Vries}}]{Dragos:2017wms}%
  \BibitemOpen
  \bibfield  {author} {\bibinfo {author} {\bibfnamefont {J.}~\bibnamefont
  {Dragos}}, \bibinfo {author} {\bibfnamefont {T.}~\bibnamefont {Luu}},
  \bibinfo {author} {\bibfnamefont {A.}~\bibnamefont {Shindler}},\ and\
  \bibinfo {author} {\bibfnamefont {J.}~\bibnamefont {de~Vries}},\ }\bibfield
  {title} {\bibinfo {title} {{Electric Dipole Moment Results from lattice
  QCD}},\ }\bibfield  {booktitle} {\emph {\bibinfo {booktitle} {{Proceedings,
  35th International Symposium on Lattice Field Theory (Lattice 2017): Granada,
  Spain, June 18-24, 2017}}},\ }\href
  {https://doi.org/10.1051/epjconf/201817506018} {\bibfield  {journal}
  {\bibinfo  {journal} {EPJ Web Conf.}\ }\textbf {\bibinfo {volume} {175}},\
  \bibinfo {pages} {06018} (\bibinfo {year} {2018}{\natexlab{b}})},\ \Eprint
  {https://arxiv.org/abs/1711.04730} {arXiv:1711.04730 [hep-lat]} \BibitemShut
  {NoStop}%
\bibitem [{\citenamefont {Fodor}\ \emph {et~al.}(2012)\citenamefont {Fodor},
  \citenamefont {Holland}, \citenamefont {Kuti}, \citenamefont {Nogradi},\ and\
  \citenamefont {Wong}}]{Fodor:2012td}%
  \BibitemOpen
  \bibfield  {author} {\bibinfo {author} {\bibfnamefont {Z.}~\bibnamefont
  {Fodor}}, \bibinfo {author} {\bibfnamefont {K.}~\bibnamefont {Holland}},
  \bibinfo {author} {\bibfnamefont {J.}~\bibnamefont {Kuti}}, \bibinfo {author}
  {\bibfnamefont {D.}~\bibnamefont {Nogradi}},\ and\ \bibinfo {author}
  {\bibfnamefont {C.~H.}\ \bibnamefont {Wong}},\ }\bibfield  {title} {\bibinfo
  {title} {{The Yang-Mills gradient flow in finite volume}},\ }\href
  {https://doi.org/10.1007/JHEP11(2012)007} {\bibfield  {journal} {\bibinfo
  {journal} {JHEP}\ }\textbf {\bibinfo {volume} {1211}},\ \bibinfo {pages}
  {007}},\ \Eprint {https://arxiv.org/abs/1208.1051} {arXiv:1208.1051
  [hep-lat]} \BibitemShut {NoStop}%
\bibitem [{\citenamefont {Fodor}\ \emph {et~al.}(2014)\citenamefont {Fodor},
  \citenamefont {Holland}, \citenamefont {Kuti}, \citenamefont {Mondal},
  \citenamefont {Nogradi},\ and\ \citenamefont {Wong}}]{Fodor:2014cpa}%
  \BibitemOpen
  \bibfield  {author} {\bibinfo {author} {\bibfnamefont {Z.}~\bibnamefont
  {Fodor}}, \bibinfo {author} {\bibfnamefont {K.}~\bibnamefont {Holland}},
  \bibinfo {author} {\bibfnamefont {J.}~\bibnamefont {Kuti}}, \bibinfo {author}
  {\bibfnamefont {S.}~\bibnamefont {Mondal}}, \bibinfo {author} {\bibfnamefont
  {D.}~\bibnamefont {Nogradi}},\ and\ \bibinfo {author} {\bibfnamefont {C.~H.}\
  \bibnamefont {Wong}},\ }\bibfield  {title} {\bibinfo {title} {{The lattice
  gradient flow at tree-level and its improvement}},\ }\href
  {https://doi.org/10.1007/JHEP09(2014)018} {\bibfield  {journal} {\bibinfo
  {journal} {JHEP}\ }\textbf {\bibinfo {volume} {09}},\ \bibinfo {pages}
  {018}},\ \Eprint {https://arxiv.org/abs/1406.0827} {arXiv:1406.0827
  [hep-lat]} \BibitemShut {NoStop}%
\bibitem [{\citenamefont {Fritzsch}\ and\ \citenamefont
  {Ramos}(2013)}]{Fritzsch:2013je}%
  \BibitemOpen
  \bibfield  {author} {\bibinfo {author} {\bibfnamefont {P.}~\bibnamefont
  {Fritzsch}}\ and\ \bibinfo {author} {\bibfnamefont {A.}~\bibnamefont
  {Ramos}},\ }\bibfield  {title} {\bibinfo {title} {{The gradient flow coupling
  in the Schr{\"o}dinger Functional}},\ }\href
  {https://doi.org/10.1007/JHEP10(2013)008} {\bibfield  {journal} {\bibinfo
  {journal} {JHEP}\ }\textbf {\bibinfo {volume} {10}},\ \bibinfo {pages}
  {008}},\ \Eprint {https://arxiv.org/abs/1301.4388} {arXiv:1301.4388
  [hep-lat]} \BibitemShut {NoStop}%
\bibitem [{\citenamefont {Ramos}(2014)}]{Ramos:2014kla}%
  \BibitemOpen
  \bibfield  {author} {\bibinfo {author} {\bibfnamefont {A.}~\bibnamefont
  {Ramos}},\ }\bibfield  {title} {\bibinfo {title} {{The gradient flow running
  coupling with twisted boundary conditions}},\ }\href
  {https://doi.org/10.1007/JHEP11(2014)101} {\bibfield  {journal} {\bibinfo
  {journal} {JHEP}\ }\textbf {\bibinfo {volume} {11}},\ \bibinfo {pages}
  {101}},\ \Eprint {https://arxiv.org/abs/1409.1445} {arXiv:1409.1445
  [hep-lat]} \BibitemShut {NoStop}%
\bibitem [{\citenamefont {Ramos}\ and\ \citenamefont
  {Sint}(2016)}]{Ramos:2015baa}%
  \BibitemOpen
  \bibfield  {author} {\bibinfo {author} {\bibfnamefont {A.}~\bibnamefont
  {Ramos}}\ and\ \bibinfo {author} {\bibfnamefont {S.}~\bibnamefont {Sint}},\
  }\bibfield  {title} {\bibinfo {title} {{Symanzik improvement of the gradient
  flow in lattice gauge theories}},\ }\href
  {https://doi.org/10.1140/epjc/s10052-015-3831-9} {\bibfield  {journal}
  {\bibinfo  {journal} {Eur. Phys. J.}\ }\textbf {\bibinfo {volume} {C76}},\
  \bibinfo {pages} {15} (\bibinfo {year} {2016})},\ \Eprint
  {https://arxiv.org/abs/1508.05552} {arXiv:1508.05552 [hep-lat]} \BibitemShut
  {NoStop}%
\bibitem [{\citenamefont {Dalla~Brida}\ \emph
  {et~al.}(2017{\natexlab{a}})\citenamefont {Dalla~Brida}, \citenamefont
  {Fritzsch}, \citenamefont {Korzec}, \citenamefont {Ramos}, \citenamefont
  {Sint},\ and\ \citenamefont {Sommer}}]{DallaBrida:2016kgh}%
  \BibitemOpen
  \bibfield  {author} {\bibinfo {author} {\bibfnamefont {M.}~\bibnamefont
  {Dalla~Brida}}, \bibinfo {author} {\bibfnamefont {P.}~\bibnamefont
  {Fritzsch}}, \bibinfo {author} {\bibfnamefont {T.}~\bibnamefont {Korzec}},
  \bibinfo {author} {\bibfnamefont {A.}~\bibnamefont {Ramos}}, \bibinfo
  {author} {\bibfnamefont {S.}~\bibnamefont {Sint}},\ and\ \bibinfo {author}
  {\bibfnamefont {R.}~\bibnamefont {Sommer}} (\bibinfo {collaboration}
  {ALPHA}),\ }\bibfield  {title} {\bibinfo {title} {{Slow running of the
  Gradient Flow coupling from 200 MeV to 4 GeV in $N_{\rm f}=3$ QCD}},\ }\href
  {https://doi.org/10.1103/PhysRevD.95.014507} {\bibfield  {journal} {\bibinfo
  {journal} {Phys. Rev.}\ }\textbf {\bibinfo {volume} {D95}},\ \bibinfo {pages}
  {014507} (\bibinfo {year} {2017}{\natexlab{a}})},\ \Eprint
  {https://arxiv.org/abs/1607.06423} {arXiv:1607.06423 [hep-lat]} \BibitemShut
  {NoStop}%
\bibitem [{\citenamefont {Dalla~Brida}\ and\ \citenamefont
  {Ramos}(2019)}]{DallaBrida:2019wur}%
  \BibitemOpen
  \bibfield  {author} {\bibinfo {author} {\bibfnamefont {M.}~\bibnamefont
  {Dalla~Brida}}\ and\ \bibinfo {author} {\bibfnamefont {A.}~\bibnamefont
  {Ramos}},\ }\bibfield  {title} {\bibinfo {title} {{The gradient flow coupling
  at high-energy and the scale of SU(3) Yang-Mills theory}},\ }\href
  {https://doi.org/10.1140/epjc/s10052-019-7228-z} {\bibfield  {journal}
  {\bibinfo  {journal} {Eur. Phys. J.}\ }\textbf {\bibinfo {volume} {C79}},\
  \bibinfo {pages} {720} (\bibinfo {year} {2019})},\ \Eprint
  {https://arxiv.org/abs/1905.05147} {arXiv:1905.05147 [hep-lat]} \BibitemShut
  {NoStop}%
\bibitem [{\citenamefont {Ishikawa}\ \emph {et~al.}(2017)\citenamefont
  {Ishikawa}, \citenamefont {Kanamori}, \citenamefont {Murakami}, \citenamefont
  {Nakamura}, \citenamefont {Okawa},\ and\ \citenamefont
  {Ueno}}]{Ishikawa:2017xam}%
  \BibitemOpen
  \bibfield  {author} {\bibinfo {author} {\bibfnamefont {K.-I.}\ \bibnamefont
  {Ishikawa}}, \bibinfo {author} {\bibfnamefont {I.}~\bibnamefont {Kanamori}},
  \bibinfo {author} {\bibfnamefont {Y.}~\bibnamefont {Murakami}}, \bibinfo
  {author} {\bibfnamefont {A.}~\bibnamefont {Nakamura}}, \bibinfo {author}
  {\bibfnamefont {M.}~\bibnamefont {Okawa}},\ and\ \bibinfo {author}
  {\bibfnamefont {R.}~\bibnamefont {Ueno}},\ }\bibfield  {title} {\bibinfo
  {title} {{Non-perturbative determination of the $\Lambda$-parameter in the
  pure SU(3) gauge theory from the twisted gradient flow coupling}},\ }\href
  {https://doi.org/10.1007/JHEP12(2017)067} {\bibfield  {journal} {\bibinfo
  {journal} {JHEP}\ }\textbf {\bibinfo {volume} {12}},\ \bibinfo {pages}
  {067}},\ \Eprint {https://arxiv.org/abs/1702.06289} {arXiv:1702.06289
  [hep-lat]} \BibitemShut {NoStop}%
\bibitem [{\citenamefont {Asakawa}\ \emph {et~al.}(2014)\citenamefont
  {Asakawa}, \citenamefont {Hatsuda}, \citenamefont {Itou}, \citenamefont
  {Kitazawa},\ and\ \citenamefont {Suzuki}}]{Asakawa:2013laa}%
  \BibitemOpen
  \bibfield  {author} {\bibinfo {author} {\bibfnamefont {M.}~\bibnamefont
  {Asakawa}}, \bibinfo {author} {\bibfnamefont {T.}~\bibnamefont {Hatsuda}},
  \bibinfo {author} {\bibfnamefont {E.}~\bibnamefont {Itou}}, \bibinfo {author}
  {\bibfnamefont {M.}~\bibnamefont {Kitazawa}},\ and\ \bibinfo {author}
  {\bibfnamefont {H.}~\bibnamefont {Suzuki}} (\bibinfo {collaboration}
  {FlowQCD}),\ }\bibfield  {title} {\bibinfo {title} {{Thermodynamics of SU(3)
  gauge theory from gradient flow on the lattice}},\ }\href
  {https://doi.org/10.1103/PhysRevD.90.011501, 10.1103/PhysRevD.92.059902}
  {\bibfield  {journal} {\bibinfo  {journal} {Phys. Rev.}\ }\textbf {\bibinfo
  {volume} {D90}},\ \bibinfo {pages} {011501} (\bibinfo {year} {2014})},\
  \bibinfo {note} {[Erratum: Phys. Rev.D92,no.5,059902(2015)]},\ \Eprint
  {https://arxiv.org/abs/1312.7492} {arXiv:1312.7492 [hep-lat]} \BibitemShut
  {NoStop}%
\bibitem [{\citenamefont {Kitazawa}\ \emph {et~al.}(2016)\citenamefont
  {Kitazawa}, \citenamefont {Iritani}, \citenamefont {Asakawa}, \citenamefont
  {Hatsuda},\ and\ \citenamefont {Suzuki}}]{Kitazawa:2016dsl}%
  \BibitemOpen
  \bibfield  {author} {\bibinfo {author} {\bibfnamefont {M.}~\bibnamefont
  {Kitazawa}}, \bibinfo {author} {\bibfnamefont {T.}~\bibnamefont {Iritani}},
  \bibinfo {author} {\bibfnamefont {M.}~\bibnamefont {Asakawa}}, \bibinfo
  {author} {\bibfnamefont {T.}~\bibnamefont {Hatsuda}},\ and\ \bibinfo {author}
  {\bibfnamefont {H.}~\bibnamefont {Suzuki}},\ }\bibfield  {title} {\bibinfo
  {title} {{Equation of State for SU(3) Gauge Theory via the Energy-Momentum
  Tensor under Gradient Flow}},\ }\href
  {https://doi.org/10.1103/PhysRevD.94.114512} {\bibfield  {journal} {\bibinfo
  {journal} {Phys. Rev.}\ }\textbf {\bibinfo {volume} {D94}},\ \bibinfo {pages}
  {114512} (\bibinfo {year} {2016})},\ \Eprint
  {https://arxiv.org/abs/1610.07810} {arXiv:1610.07810 [hep-lat]} \BibitemShut
  {NoStop}%
\bibitem [{\citenamefont {Taniguchi}\ \emph {et~al.}(2017)\citenamefont
  {Taniguchi}, \citenamefont {Ejiri}, \citenamefont {Iwami}, \citenamefont
  {Kanaya}, \citenamefont {Kitazawa}, \citenamefont {Suzuki}, \citenamefont
  {Umeda},\ and\ \citenamefont {Wakabayashi}}]{Taniguchi:2016ofw}%
  \BibitemOpen
  \bibfield  {author} {\bibinfo {author} {\bibfnamefont {Y.}~\bibnamefont
  {Taniguchi}}, \bibinfo {author} {\bibfnamefont {S.}~\bibnamefont {Ejiri}},
  \bibinfo {author} {\bibfnamefont {R.}~\bibnamefont {Iwami}}, \bibinfo
  {author} {\bibfnamefont {K.}~\bibnamefont {Kanaya}}, \bibinfo {author}
  {\bibfnamefont {M.}~\bibnamefont {Kitazawa}}, \bibinfo {author}
  {\bibfnamefont {H.}~\bibnamefont {Suzuki}}, \bibinfo {author} {\bibfnamefont
  {T.}~\bibnamefont {Umeda}},\ and\ \bibinfo {author} {\bibfnamefont
  {N.}~\bibnamefont {Wakabayashi}},\ }\bibfield  {title} {\bibinfo {title}
  {{Exploring $N_{f}$ = 2+1 QCD thermodynamics from the gradient flow}},\
  }\href {https://doi.org/10.1103/PhysRevD.96.014509,
  10.1103/PhysRevD.99.059904} {\bibfield  {journal} {\bibinfo  {journal} {Phys.
  Rev.}\ }\textbf {\bibinfo {volume} {D96}},\ \bibinfo {pages} {014509}
  (\bibinfo {year} {2017})},\ \bibinfo {note} {[Erratum: Phys.
  Rev.D99,no.5,059904(2019)]},\ \Eprint {https://arxiv.org/abs/1609.01417}
  {arXiv:1609.01417 [hep-lat]} \BibitemShut {NoStop}%
\bibitem [{\citenamefont {Kitazawa}\ \emph {et~al.}(2017)\citenamefont
  {Kitazawa}, \citenamefont {Iritani}, \citenamefont {Asakawa},\ and\
  \citenamefont {Hatsuda}}]{Kitazawa:2017qab}%
  \BibitemOpen
  \bibfield  {author} {\bibinfo {author} {\bibfnamefont {M.}~\bibnamefont
  {Kitazawa}}, \bibinfo {author} {\bibfnamefont {T.}~\bibnamefont {Iritani}},
  \bibinfo {author} {\bibfnamefont {M.}~\bibnamefont {Asakawa}},\ and\ \bibinfo
  {author} {\bibfnamefont {T.}~\bibnamefont {Hatsuda}},\ }\bibfield  {title}
  {\bibinfo {title} {{Correlations of the energy-momentum tensor via gradient
  flow in SU(3) Yang-Mills theory at finite temperature}},\ }\href
  {https://doi.org/10.1103/PhysRevD.96.111502} {\bibfield  {journal} {\bibinfo
  {journal} {Phys. Rev.}\ }\textbf {\bibinfo {volume} {D96}},\ \bibinfo {pages}
  {111502} (\bibinfo {year} {2017})},\ \Eprint
  {https://arxiv.org/abs/1708.01415} {arXiv:1708.01415 [hep-lat]} \BibitemShut
  {NoStop}%
\bibitem [{\citenamefont {Iritani}\ \emph {et~al.}(2019)\citenamefont
  {Iritani}, \citenamefont {Kitazawa}, \citenamefont {Suzuki},\ and\
  \citenamefont {Takaura}}]{Iritani:2018idk}%
  \BibitemOpen
  \bibfield  {author} {\bibinfo {author} {\bibfnamefont {T.}~\bibnamefont
  {Iritani}}, \bibinfo {author} {\bibfnamefont {M.}~\bibnamefont {Kitazawa}},
  \bibinfo {author} {\bibfnamefont {H.}~\bibnamefont {Suzuki}},\ and\ \bibinfo
  {author} {\bibfnamefont {H.}~\bibnamefont {Takaura}},\ }\bibfield  {title}
  {\bibinfo {title} {{Thermodynamics in quenched QCD: energy--momentum tensor
  with two-loop order coefficients in the gradient-flow formalism}},\ }\href
  {https://doi.org/10.1093/ptep/ptz001} {\bibfield  {journal} {\bibinfo
  {journal} {PTEP}\ }\textbf {\bibinfo {volume} {2019}},\ \bibinfo {pages}
  {023B02} (\bibinfo {year} {2019})},\ \Eprint
  {https://arxiv.org/abs/1812.06444} {arXiv:1812.06444 [hep-lat]} \BibitemShut
  {NoStop}%
\bibitem [{\citenamefont {Eller}\ and\ \citenamefont
  {Moore}(2018)}]{Eller:2018yje}%
  \BibitemOpen
  \bibfield  {author} {\bibinfo {author} {\bibfnamefont {A.~M.}\ \bibnamefont
  {Eller}}\ and\ \bibinfo {author} {\bibfnamefont {G.~D.}\ \bibnamefont
  {Moore}},\ }\bibfield  {title} {\bibinfo {title} {{Gradient-flowed thermal
  correlators: how much flow is too much?}},\ }\href
  {https://doi.org/10.1103/PhysRevD.97.114507} {\bibfield  {journal} {\bibinfo
  {journal} {Phys.\ Rev.\ D}\ }\textbf {\bibinfo {volume} {97}},\ \bibinfo
  {pages} {114507} (\bibinfo {year} {2018})},\ \Eprint
  {https://arxiv.org/abs/1802.04562} {arXiv:1802.04562 [hep-lat]} \BibitemShut
  {NoStop}%
\bibitem [{\citenamefont {Suzuki}(2013)}]{Suzuki:2013gza}%
  \BibitemOpen
  \bibfield  {author} {\bibinfo {author} {\bibfnamefont {H.}~\bibnamefont
  {Suzuki}},\ }\bibfield  {title} {\bibinfo {title} {{Energy–momentum tensor
  from the Yang–Mills gradient flow}},\ }\href
  {https://doi.org/10.1093/ptep/ptt059} {\bibfield  {journal} {\bibinfo
  {journal} {PTEP}\ }\textbf {\bibinfo {volume} {2013}},\ \bibinfo {pages}
  {083B03} (\bibinfo {year} {2013})},\ \Eprint
  {https://arxiv.org/abs/1304.0533} {arXiv:1304.0533 [hep-lat]} \BibitemShut
  {NoStop}%
\bibitem [{\citenamefont {Del~Debbio}\ \emph {et~al.}(2013)\citenamefont
  {Del~Debbio}, \citenamefont {Patella},\ and\ \citenamefont
  {Rago}}]{DelDebbio:2013zaa}%
  \BibitemOpen
  \bibfield  {author} {\bibinfo {author} {\bibfnamefont {L.}~\bibnamefont
  {Del~Debbio}}, \bibinfo {author} {\bibfnamefont {A.}~\bibnamefont
  {Patella}},\ and\ \bibinfo {author} {\bibfnamefont {A.}~\bibnamefont
  {Rago}},\ }\bibfield  {title} {\bibinfo {title} {{Space-time symmetries and
  the Yang-Mills gradient flow}},\ }\href
  {https://doi.org/10.1007/JHEP11(2013)212} {\bibfield  {journal} {\bibinfo
  {journal} {JHEP}\ }\textbf {\bibinfo {volume} {11}},\ \bibinfo {pages}
  {212}},\ \Eprint {https://arxiv.org/abs/1306.1173} {arXiv:1306.1173 [hep-th]}
  \BibitemShut {NoStop}%
\bibitem [{\citenamefont {Makino}\ and\ \citenamefont
  {Suzuki}(2014)}]{Makino:2014taa}%
  \BibitemOpen
  \bibfield  {author} {\bibinfo {author} {\bibfnamefont {H.}~\bibnamefont
  {Makino}}\ and\ \bibinfo {author} {\bibfnamefont {H.}~\bibnamefont
  {Suzuki}},\ }\bibfield  {title} {\bibinfo {title} {{Lattice energy--momentum
  tensor from the Yang-Mills gradient flow--inclusion of fermion fields}},\
  }\href {https://doi.org/10.1093/ptep/ptu070} {\bibfield  {journal} {\bibinfo
  {journal} {PTEP}\ }\textbf {\bibinfo {volume} {2014}},\ \bibinfo {pages}
  {063B02} (\bibinfo {year} {2014})},\ \Eprint
  {https://arxiv.org/abs/1403.4772} {arXiv:1403.4772 [hep-lat]} \BibitemShut
  {NoStop}%
\bibitem [{\citenamefont {Harlander}\ \emph {et~al.}(2018)\citenamefont
  {Harlander}, \citenamefont {Kluth},\ and\ \citenamefont
  {Lange}}]{Harlander:2018zpi}%
  \BibitemOpen
  \bibfield  {author} {\bibinfo {author} {\bibfnamefont {R.~V.}\ \bibnamefont
  {Harlander}}, \bibinfo {author} {\bibfnamefont {Y.}~\bibnamefont {Kluth}},\
  and\ \bibinfo {author} {\bibfnamefont {F.}~\bibnamefont {Lange}},\ }\bibfield
   {title} {\bibinfo {title} {{The two-loop energy–momentum tensor within the
  gradient-flow formalism}},\ }\href
  {https://doi.org/10.1140/epjc/s10052-019-7327-x,
  10.1140/epjc/s10052-018-6415-7} {\bibfield  {journal} {\bibinfo  {journal}
  {Eur. Phys. J.}\ }\textbf {\bibinfo {volume} {C78}},\ \bibinfo {pages} {944}
  (\bibinfo {year} {2018})},\ \bibinfo {note} {[Erratum: Eur. Phys.
  J.C79,no.10,858(2019)]},\ \Eprint {https://arxiv.org/abs/1808.09837}
  {arXiv:1808.09837 [hep-lat]} \BibitemShut {NoStop}%
\bibitem [{\citenamefont {Borsanyi}\ \emph {et~al.}(2012)\citenamefont
  {Borsanyi}, \citenamefont {Durr}, \citenamefont {Fodor}, \citenamefont
  {Hoelbling}, \citenamefont {Katz} \emph {et~al.}}]{Borsanyi:2012zs}%
  \BibitemOpen
  \bibfield  {author} {\bibinfo {author} {\bibfnamefont {S.}~\bibnamefont
  {Borsanyi}}, \bibinfo {author} {\bibfnamefont {S.}~\bibnamefont {Durr}},
  \bibinfo {author} {\bibfnamefont {Z.}~\bibnamefont {Fodor}}, \bibinfo
  {author} {\bibfnamefont {C.}~\bibnamefont {Hoelbling}}, \bibinfo {author}
  {\bibfnamefont {S.~D.}\ \bibnamefont {Katz}}, \emph {et~al.},\ }\bibfield
  {title} {\bibinfo {title} {{High-precision scale setting in lattice QCD}},\
  }\href {https://doi.org/10.1007/JHEP09(2012)010} {\bibfield  {journal}
  {\bibinfo  {journal} {JHEP}\ }\textbf {\bibinfo {volume} {1209}},\ \bibinfo
  {pages} {010}},\ \Eprint {https://arxiv.org/abs/1203.4469} {arXiv:1203.4469
  [hep-lat]} \BibitemShut {NoStop}%
\bibitem [{\citenamefont {H{\"o}llwieser}\ \emph {et~al.}(2020)\citenamefont
  {H{\"o}llwieser}, \citenamefont {Knechtli},\ and\ \citenamefont
  {Korzec}}]{Hollwieser:2020qri}%
  \BibitemOpen
  \bibfield  {author} {\bibinfo {author} {\bibfnamefont {R.}~\bibnamefont
  {H{\"o}llwieser}}, \bibinfo {author} {\bibfnamefont {F.}~\bibnamefont
  {Knechtli}},\ and\ \bibinfo {author} {\bibfnamefont {T.}~\bibnamefont
  {Korzec}} (\bibinfo {collaboration} {ALPHA}),\ }\href@noop {} {\bibinfo
  {title} {{Scale setting for $N_f=3+1$ QCD}}} (\bibinfo {year} {2020}),\
  \Eprint {https://arxiv.org/abs/2002.02866} {arXiv:2002.02866 [hep-lat]}
  \BibitemShut {NoStop}%
\bibitem [{\citenamefont {Lin}\ \emph {et~al.}(2015)\citenamefont {Lin},
  \citenamefont {Ogawa},\ and\ \citenamefont {Ramos}}]{Lin:2015zpa}%
  \BibitemOpen
  \bibfield  {author} {\bibinfo {author} {\bibfnamefont {C.~J.~D.}\
  \bibnamefont {Lin}}, \bibinfo {author} {\bibfnamefont {K.}~\bibnamefont
  {Ogawa}},\ and\ \bibinfo {author} {\bibfnamefont {A.}~\bibnamefont {Ramos}},\
  }\bibfield  {title} {\bibinfo {title} {{The Yang-Mills gradient flow and
  SU(3) gauge theory with 12 massless fundamental fermions in a colour-twisted
  box}},\ }\href {https://doi.org/10.1007/JHEP12(2015)103} {\bibfield
  {journal} {\bibinfo  {journal} {JHEP}\ }\textbf {\bibinfo {volume} {12}},\
  \bibinfo {pages} {103}},\ \Eprint {https://arxiv.org/abs/1510.05755}
  {arXiv:1510.05755 [hep-lat]} \BibitemShut {NoStop}%
\bibitem [{\citenamefont {Hasenfratz}\ and\ \citenamefont
  {Schaich}(2018)}]{Hasenfratz:2016dou}%
  \BibitemOpen
  \bibfield  {author} {\bibinfo {author} {\bibfnamefont {A.}~\bibnamefont
  {Hasenfratz}}\ and\ \bibinfo {author} {\bibfnamefont {D.}~\bibnamefont
  {Schaich}},\ }\bibfield  {title} {\bibinfo {title} {{Nonperturbative $\beta$
  function of twelve-flavor SU(3) gauge theory}},\ }\href
  {https://doi.org/10.1007/JHEP02(2018)132} {\bibfield  {journal} {\bibinfo
  {journal} {JHEP}\ }\textbf {\bibinfo {volume} {02}},\ \bibinfo {pages}
  {132}},\ \Eprint {https://arxiv.org/abs/1610.10004} {arXiv:1610.10004
  [hep-lat]} \BibitemShut {NoStop}%
\bibitem [{\citenamefont {DeGrand}(2017)}]{DeGrand:2017gbi}%
  \BibitemOpen
  \bibfield  {author} {\bibinfo {author} {\bibfnamefont {T.}~\bibnamefont
  {DeGrand}},\ }\bibfield  {title} {\bibinfo {title} {{Simple chromatic
  properties of gradient flow}},\ }\href
  {https://doi.org/10.1103/PhysRevD.95.114512} {\bibfield  {journal} {\bibinfo
  {journal} {Phys. Rev. D}\ }\textbf {\bibinfo {volume} {95}},\ \bibinfo
  {pages} {114512} (\bibinfo {year} {2017})},\ \Eprint
  {https://arxiv.org/abs/1701.00793} {arXiv:1701.00793 [hep-lat]} \BibitemShut
  {NoStop}%
\bibitem [{\citenamefont {Bribian}\ and\ \citenamefont
  {Garcia~Perez}(2019)}]{Bribian:2019ybc}%
  \BibitemOpen
  \bibfield  {author} {\bibinfo {author} {\bibfnamefont {E.~I.}\ \bibnamefont
  {Bribian}}\ and\ \bibinfo {author} {\bibfnamefont {M.}~\bibnamefont
  {Garcia~Perez}},\ }\bibfield  {title} {\bibinfo {title} {{The twisted
  gradient flow coupling at one loop}},\ }\href
  {https://doi.org/10.1007/JHEP03(2019)200} {\bibfield  {journal} {\bibinfo
  {journal} {JHEP}\ }\textbf {\bibinfo {volume} {03}},\ \bibinfo {pages}
  {200}},\ \Eprint {https://arxiv.org/abs/1903.08029} {arXiv:1903.08029
  [hep-lat]} \BibitemShut {NoStop}%
\bibitem [{\citenamefont {Hirakida}\ \emph {et~al.}(2019)\citenamefont
  {Hirakida}, \citenamefont {Itou},\ and\ \citenamefont
  {Kouno}}]{Hirakida:2018uoy}%
  \BibitemOpen
  \bibfield  {author} {\bibinfo {author} {\bibfnamefont {T.}~\bibnamefont
  {Hirakida}}, \bibinfo {author} {\bibfnamefont {E.}~\bibnamefont {Itou}},\
  and\ \bibinfo {author} {\bibfnamefont {H.}~\bibnamefont {Kouno}},\ }\bibfield
   {title} {\bibinfo {title} {{Thermodynamics for pure SU(2) gauge theory using
  gradient flow}},\ }\href {https://doi.org/10.1093/ptep/ptz003} {\bibfield
  {journal} {\bibinfo  {journal} {PTEP}\ }\textbf {\bibinfo {volume} {2019}},\
  \bibinfo {pages} {033B01} (\bibinfo {year} {2019})},\ \Eprint
  {https://arxiv.org/abs/1805.07106} {arXiv:1805.07106 [hep-lat]} \BibitemShut
  {NoStop}%
\bibitem [{\citenamefont {Carosso}\ \emph {et~al.}(2018)\citenamefont
  {Carosso}, \citenamefont {Hasenfratz},\ and\ \citenamefont
  {Neil}}]{Carosso:2018bmz}%
  \BibitemOpen
  \bibfield  {author} {\bibinfo {author} {\bibfnamefont {A.}~\bibnamefont
  {Carosso}}, \bibinfo {author} {\bibfnamefont {A.}~\bibnamefont
  {Hasenfratz}},\ and\ \bibinfo {author} {\bibfnamefont {E.~T.}\ \bibnamefont
  {Neil}},\ }\bibfield  {title} {\bibinfo {title} {{Nonperturbative
  Renormalization of Operators in Near-Conformal Systems Using Gradient
  Flows}},\ }\href {https://doi.org/10.1103/PhysRevLett.121.201601} {\bibfield
  {journal} {\bibinfo  {journal} {Phys. Rev. Lett.}\ }\textbf {\bibinfo
  {volume} {121}},\ \bibinfo {pages} {201601} (\bibinfo {year} {2018})},\
  \Eprint {https://arxiv.org/abs/1806.01385} {arXiv:1806.01385 [hep-lat]}
  \BibitemShut {NoStop}%
\bibitem [{\citenamefont {Hasenfratz}\ \emph {et~al.}(2019)\citenamefont
  {Hasenfratz}, \citenamefont {Rebbi},\ and\ \citenamefont
  {Witzel}}]{Hasenfratz:2019dpr}%
  \BibitemOpen
  \bibfield  {author} {\bibinfo {author} {\bibfnamefont {A.}~\bibnamefont
  {Hasenfratz}}, \bibinfo {author} {\bibfnamefont {C.}~\bibnamefont {Rebbi}},\
  and\ \bibinfo {author} {\bibfnamefont {O.}~\bibnamefont {Witzel}},\
  }\bibfield  {title} {\bibinfo {title} {{Gradient flow step-scaling function
  for SU(3) with twelve flavors}},\ }\href
  {https://doi.org/10.1103/PhysRevD.100.114508} {\bibfield  {journal} {\bibinfo
   {journal} {Phys. Rev. D}\ }\textbf {\bibinfo {volume} {100}},\ \bibinfo
  {pages} {114508} (\bibinfo {year} {2019})},\ \Eprint
  {https://arxiv.org/abs/1909.05842} {arXiv:1909.05842 [hep-lat]} \BibitemShut
  {NoStop}%
\bibitem [{\citenamefont {Bennett}\ \emph {et~al.}(2019)\citenamefont
  {Bennett}, \citenamefont {Hong}, \citenamefont {Lee}, \citenamefont {Lin},
  \citenamefont {Lucini}, \citenamefont {Piai},\ and\ \citenamefont
  {Vadacchino}}]{Bennett:2019jzz}%
  \BibitemOpen
  \bibfield  {author} {\bibinfo {author} {\bibfnamefont {E.}~\bibnamefont
  {Bennett}}, \bibinfo {author} {\bibfnamefont {D.~K.}\ \bibnamefont {Hong}},
  \bibinfo {author} {\bibfnamefont {J.-W.}\ \bibnamefont {Lee}}, \bibinfo
  {author} {\bibfnamefont {C.-J.~D.}\ \bibnamefont {Lin}}, \bibinfo {author}
  {\bibfnamefont {B.}~\bibnamefont {Lucini}}, \bibinfo {author} {\bibfnamefont
  {M.}~\bibnamefont {Piai}},\ and\ \bibinfo {author} {\bibfnamefont
  {D.}~\bibnamefont {Vadacchino}},\ }\bibfield  {title} {\bibinfo {title}
  {{Sp(4) gauge theories on the lattice: $N_f=2$ dynamical fundamental
  fermions}},\ }\href {https://doi.org/10.1007/JHEP12(2019)053} {\bibfield
  {journal} {\bibinfo  {journal} {JHEP}\ }\textbf {\bibinfo {volume} {12}},\
  \bibinfo {pages} {053}},\ \Eprint {https://arxiv.org/abs/1909.12662}
  {arXiv:1909.12662 [hep-lat]} \BibitemShut {NoStop}%
\bibitem [{\citenamefont {DeGrand}(2020)}]{DeGrand:2020utq}%
  \BibitemOpen
  \bibfield  {author} {\bibinfo {author} {\bibfnamefont {T.}~\bibnamefont
  {DeGrand}},\ }\href@noop {} {\bibinfo {title} {{Topological susceptibility in
  QCD with two flavors and 3-5 colors -- a pilot study}}} (\bibinfo {year}
  {2020}),\ \Eprint {https://arxiv.org/abs/2004.09649} {arXiv:2004.09649
  [hep-lat]} \BibitemShut {NoStop}%
\bibitem [{\citenamefont {L{\"u}scher}(2014)}]{Luscher:2014kea}%
  \BibitemOpen
  \bibfield  {author} {\bibinfo {author} {\bibfnamefont {M.}~\bibnamefont
  {L{\"u}scher}},\ }\bibfield  {title} {\bibinfo {title} {{Step scaling and the
  Yang-Mills gradient flow}},\ }\href {https://doi.org/10.1007/JHEP06(2014)105}
  {\bibfield  {journal} {\bibinfo  {journal} {JHEP}\ }\textbf {\bibinfo
  {volume} {06}},\ \bibinfo {pages} {105}},\ \Eprint
  {https://arxiv.org/abs/1404.5930} {arXiv:1404.5930 [hep-lat]} \BibitemShut
  {NoStop}%
\bibitem [{\citenamefont {Monahan}\ and\ \citenamefont
  {Orginos}(2014)}]{Monahan:2013lwa}%
  \BibitemOpen
  \bibfield  {author} {\bibinfo {author} {\bibfnamefont {C.}~\bibnamefont
  {Monahan}}\ and\ \bibinfo {author} {\bibfnamefont {K.}~\bibnamefont
  {Orginos}},\ }\bibfield  {title} {\bibinfo {title} {{Finite volume
  renormalization scheme for fermionic operators}},\ }\bibfield  {booktitle}
  {\emph {\bibinfo {booktitle} {{Proceedings, 31st International Symposium on
  Lattice Field Theory (Lattice 2013): Mainz, Germany, July 29-August 3,
  2013}}},\ }\href {https://doi.org/10.22323/1.187.0443} {\bibfield  {journal}
  {\bibinfo  {journal} {PoS}\ }\textbf {\bibinfo {volume} {Lattice2013}},\
  \bibinfo {pages} {443} (\bibinfo {year} {2014})},\ \Eprint
  {https://arxiv.org/abs/1311.2310} {arXiv:1311.2310 [hep-lat]} \BibitemShut
  {NoStop}%
\bibitem [{\citenamefont {Monahan}\ and\ \citenamefont
  {Orginos}(2017)}]{Monahan:2016bvm}%
  \BibitemOpen
  \bibfield  {author} {\bibinfo {author} {\bibfnamefont {C.}~\bibnamefont
  {Monahan}}\ and\ \bibinfo {author} {\bibfnamefont {K.}~\bibnamefont
  {Orginos}},\ }\bibfield  {title} {\bibinfo {title} {{Quasi parton
  distributions and the gradient flow}},\ }\href
  {https://doi.org/10.1007/JHEP03(2017)116} {\bibfield  {journal} {\bibinfo
  {journal} {JHEP}\ }\textbf {\bibinfo {volume} {03}},\ \bibinfo {pages}
  {116}},\ \Eprint {https://arxiv.org/abs/1612.01584} {arXiv:1612.01584
  [hep-lat]} \BibitemShut {NoStop}%
\bibitem [{\citenamefont {Monahan}(2018)}]{Monahan:2017hpu}%
  \BibitemOpen
  \bibfield  {author} {\bibinfo {author} {\bibfnamefont {C.}~\bibnamefont
  {Monahan}},\ }\bibfield  {title} {\bibinfo {title} {{Smeared
  quasidistributions in perturbation theory}},\ }\href
  {https://doi.org/10.1103/PhysRevD.97.054507} {\bibfield  {journal} {\bibinfo
  {journal} {Phys. Rev.}\ }\textbf {\bibinfo {volume} {D97}},\ \bibinfo {pages}
  {054507} (\bibinfo {year} {2018})},\ \Eprint
  {https://arxiv.org/abs/1710.04607} {arXiv:1710.04607 [hep-lat]} \BibitemShut
  {NoStop}%
\bibitem [{\citenamefont {Endo}\ \emph {et~al.}(2015)\citenamefont {Endo},
  \citenamefont {Hieda}, \citenamefont {Miura},\ and\ \citenamefont
  {Suzuki}}]{Endo:2015iea}%
  \BibitemOpen
  \bibfield  {author} {\bibinfo {author} {\bibfnamefont {T.}~\bibnamefont
  {Endo}}, \bibinfo {author} {\bibfnamefont {K.}~\bibnamefont {Hieda}},
  \bibinfo {author} {\bibfnamefont {D.}~\bibnamefont {Miura}},\ and\ \bibinfo
  {author} {\bibfnamefont {H.}~\bibnamefont {Suzuki}},\ }\bibfield  {title}
  {\bibinfo {title} {{Universal formula for the flavor non-singlet axial-vector
  current from the gradient flow}},\ }\href
  {https://doi.org/10.1093/ptep/ptv058} {\bibfield  {journal} {\bibinfo
  {journal} {PTEP}\ }\textbf {\bibinfo {volume} {2015}},\ \bibinfo {pages}
  {053B03} (\bibinfo {year} {2015})},\ \Eprint
  {https://arxiv.org/abs/1502.01809} {arXiv:1502.01809 [hep-lat]} \BibitemShut
  {NoStop}%
\bibitem [{\citenamefont {Hieda}\ and\ \citenamefont
  {Suzuki}(2016)}]{Hieda:2016lly}%
  \BibitemOpen
  \bibfield  {author} {\bibinfo {author} {\bibfnamefont {K.}~\bibnamefont
  {Hieda}}\ and\ \bibinfo {author} {\bibfnamefont {H.}~\bibnamefont {Suzuki}},\
  }\bibfield  {title} {\bibinfo {title} {{Small flow-time representation of
  fermion bilinear operators}},\ }\href
  {https://doi.org/10.1142/S021773231650214X} {\bibfield  {journal} {\bibinfo
  {journal} {Mod. Phys. Lett.}\ }\textbf {\bibinfo {volume} {A31}},\ \bibinfo
  {pages} {1650214} (\bibinfo {year} {2016})},\ \Eprint
  {https://arxiv.org/abs/1606.04193} {arXiv:1606.04193 [hep-lat]} \BibitemShut
  {NoStop}%
\bibitem [{\citenamefont {Harlander}\ and\ \citenamefont
  {Neumann}(2016)}]{Harlander:2016vzb}%
  \BibitemOpen
  \bibfield  {author} {\bibinfo {author} {\bibfnamefont {R.~V.}\ \bibnamefont
  {Harlander}}\ and\ \bibinfo {author} {\bibfnamefont {T.}~\bibnamefont
  {Neumann}},\ }\bibfield  {title} {\bibinfo {title} {{The perturbative QCD
  gradient flow to three loops}},\ }\href
  {https://doi.org/10.1007/JHEP06(2016)161} {\bibfield  {journal} {\bibinfo
  {journal} {JHEP}\ }\textbf {\bibinfo {volume} {06}},\ \bibinfo {pages}
  {161}},\ \Eprint {https://arxiv.org/abs/1606.03756} {arXiv:1606.03756
  [hep-ph]} \BibitemShut {NoStop}%
\bibitem [{\citenamefont {Artz}\ \emph {et~al.}(2019)\citenamefont {Artz},
  \citenamefont {Harlander}, \citenamefont {Lange}, \citenamefont {Neumann},\
  and\ \citenamefont {Prausa}}]{Artz:2019bpr}%
  \BibitemOpen
  \bibfield  {author} {\bibinfo {author} {\bibfnamefont {J.}~\bibnamefont
  {Artz}}, \bibinfo {author} {\bibfnamefont {R.~V.}\ \bibnamefont {Harlander}},
  \bibinfo {author} {\bibfnamefont {F.}~\bibnamefont {Lange}}, \bibinfo
  {author} {\bibfnamefont {T.}~\bibnamefont {Neumann}},\ and\ \bibinfo {author}
  {\bibfnamefont {M.}~\bibnamefont {Prausa}},\ }\bibfield  {title} {\bibinfo
  {title} {{Results and techniques for higher order calculations within the
  gradient-flow formalism}},\ }\href {https://doi.org/10.1007/JHEP06(2019)121,
  10.1007/JHEP10(2019)032} {\bibfield  {journal} {\bibinfo  {journal} {JHEP}\
  }\textbf {\bibinfo {volume} {06}},\ \bibinfo {pages} {121}},\ \bibinfo {note}
  {[erratum: JHEP10,032(2019)]},\ \Eprint {https://arxiv.org/abs/1905.00882}
  {arXiv:1905.00882 [hep-lat]} \BibitemShut {NoStop}%
\bibitem [{\citenamefont {Dalla~Brida}\ \emph
  {et~al.}(2017{\natexlab{b}})\citenamefont {Dalla~Brida}, \citenamefont
  {Garofalo},\ and\ \citenamefont {Kennedy}}]{DallaBrida:2017pex}%
  \BibitemOpen
  \bibfield  {author} {\bibinfo {author} {\bibfnamefont {M.}~\bibnamefont
  {Dalla~Brida}}, \bibinfo {author} {\bibfnamefont {M.}~\bibnamefont
  {Garofalo}},\ and\ \bibinfo {author} {\bibfnamefont {A.}~\bibnamefont
  {Kennedy}},\ }\bibfield  {title} {\bibinfo {title} {{Investigation of New
  Methods for Numerical Stochastic Perturbation Theory in $\varphi^4$
  Theory}},\ }\href {https://doi.org/10.1103/PhysRevD.96.054502} {\bibfield
  {journal} {\bibinfo  {journal} {Phys.\ Rev.\ D}\ }\textbf {\bibinfo {volume}
  {96}},\ \bibinfo {pages} {054502} (\bibinfo {year} {2017}{\natexlab{b}})},\
  \Eprint {https://arxiv.org/abs/1703.04406} {arXiv:1703.04406 [hep-lat]}
  \BibitemShut {NoStop}%
\bibitem [{\citenamefont {Dalla~Brida}\ and\ \citenamefont
  {L{\"u}scher}(2017)}]{DallaBrida:2017tru}%
  \BibitemOpen
  \bibfield  {author} {\bibinfo {author} {\bibfnamefont {M.}~\bibnamefont
  {Dalla~Brida}}\ and\ \bibinfo {author} {\bibfnamefont {M.}~\bibnamefont
  {L{\"u}scher}},\ }\bibfield  {title} {\bibinfo {title} {{SMD-based numerical
  stochastic perturbation theory}},\ }\href
  {https://doi.org/10.1140/epjc/s10052-017-4839-0} {\bibfield  {journal}
  {\bibinfo  {journal} {Eur.\ Phys.\ J.\ C}\ }\textbf {\bibinfo {volume}
  {77}},\ \bibinfo {pages} {308} (\bibinfo {year} {2017})},\ \Eprint
  {https://arxiv.org/abs/1703.04396} {arXiv:1703.04396 [hep-lat]} \BibitemShut
  {NoStop}%
\bibitem [{\citenamefont {Rizik}\ \emph {et~al.}(2018)\citenamefont {Rizik},
  \citenamefont {Monahan},\ and\ \citenamefont {Shindler}}]{Rizik:2018lrz}%
  \BibitemOpen
  \bibfield  {author} {\bibinfo {author} {\bibfnamefont {M.}~\bibnamefont
  {Rizik}}, \bibinfo {author} {\bibfnamefont {C.}~\bibnamefont {Monahan}},\
  and\ \bibinfo {author} {\bibfnamefont {A.}~\bibnamefont {Shindler}},\
  }\bibfield  {title} {\bibinfo {title} {{Renormalization of CP-Violating Pure
  Gauge Operators in Perturbative QCD Using the Gradient Flow}},\ }in\
  \href@noop {} {\emph {\bibinfo {booktitle} {{36th International Symposium on
  Lattice Field Theory (Lattice 2018) East Lansing, MI, United States, July
  22-28, 2018}}}}\ (\bibinfo {year} {2018})\ \Eprint
  {https://arxiv.org/abs/1810.05637} {arXiv:1810.05637 [hep-lat]} \BibitemShut
  {NoStop}%
\bibitem [{\citenamefont {Kim}\ \emph {et~al.}(2018)\citenamefont {Kim},
  \citenamefont {Dragos}, \citenamefont {Shindler}, \citenamefont {Luu},\ and\
  \citenamefont {de~Vries}}]{Kim:2018rce}%
  \BibitemOpen
  \bibfield  {author} {\bibinfo {author} {\bibfnamefont {J.}~\bibnamefont
  {Kim}}, \bibinfo {author} {\bibfnamefont {J.}~\bibnamefont {Dragos}},
  \bibinfo {author} {\bibfnamefont {A.}~\bibnamefont {Shindler}}, \bibinfo
  {author} {\bibfnamefont {T.}~\bibnamefont {Luu}},\ and\ \bibinfo {author}
  {\bibfnamefont {J.}~\bibnamefont {de~Vries}},\ }\bibfield  {title} {\bibinfo
  {title} {{Towards a determination of the nucleon EDM from the quark
  chromo-EDM operator with the gradient flow}},\ }in\ \href@noop {} {\emph
  {\bibinfo {booktitle} {{36th International Symposium on Lattice Field Theory
  (Lattice 2018) East Lansing, MI, United States, July 22-28, 2018}}}}\
  (\bibinfo {year} {2018})\ \Eprint {https://arxiv.org/abs/1810.10301}
  {arXiv:1810.10301 [hep-lat]} \BibitemShut {NoStop}%
\bibitem [{\citenamefont {Reyes}\ \emph {et~al.}(2018)\citenamefont {Reyes},
  \citenamefont {Dragos}, \citenamefont {Kim}, \citenamefont {Shindler},\ and\
  \citenamefont {Luu}}]{Reyes:2018ucu}%
  \BibitemOpen
  \bibfield  {author} {\bibinfo {author} {\bibfnamefont {J.~G.}\ \bibnamefont
  {Reyes}}, \bibinfo {author} {\bibfnamefont {J.}~\bibnamefont {Dragos}},
  \bibinfo {author} {\bibfnamefont {J.}~\bibnamefont {Kim}}, \bibinfo {author}
  {\bibfnamefont {A.}~\bibnamefont {Shindler}},\ and\ \bibinfo {author}
  {\bibfnamefont {T.}~\bibnamefont {Luu}},\ }\bibfield  {title} {\bibinfo
  {title} {{Expansion coefficient of the pseudo-scalar density using the
  gradient flow in lattice QCD}}\ }(\bibinfo {year} {2018})\ \Eprint
  {https://arxiv.org/abs/1811.11798} {arXiv:1811.11798 [hep-lat]} \BibitemShut
  {NoStop}%
\bibitem [{\citenamefont {Weinberg}(1989)}]{Weinberg:1989dx}%
  \BibitemOpen
  \bibfield  {author} {\bibinfo {author} {\bibfnamefont {S.}~\bibnamefont
  {Weinberg}},\ }\bibfield  {title} {\bibinfo {title} {{Larger Higgs Exchange
  Terms in the Neutron Electric Dipole Moment}},\ }\href
  {https://doi.org/10.1103/PhysRevLett.63.2333} {\bibfield  {journal} {\bibinfo
   {journal} {Phys. Rev. Lett.}\ }\textbf {\bibinfo {volume} {63}},\ \bibinfo
  {pages} {2333} (\bibinfo {year} {1989})}\BibitemShut {NoStop}%
\bibitem [{\citenamefont {Braaten}\ \emph
  {et~al.}(1990{\natexlab{a}})\citenamefont {Braaten}, \citenamefont {Li},\
  and\ \citenamefont {Yuan}}]{Braaten:1990zt}%
  \BibitemOpen
  \bibfield  {author} {\bibinfo {author} {\bibfnamefont {E.}~\bibnamefont
  {Braaten}}, \bibinfo {author} {\bibfnamefont {C.~S.}\ \bibnamefont {Li}},\
  and\ \bibinfo {author} {\bibfnamefont {T.~C.}\ \bibnamefont {Yuan}},\
  }\bibfield  {title} {\bibinfo {title} {{The Gluon Color - Electric Dipole
  Moment and Its Anomalous Dimension}},\ }\href
  {https://doi.org/10.1103/PhysRevD.42.276} {\bibfield  {journal} {\bibinfo
  {journal} {Phys. Rev. D}\ }\textbf {\bibinfo {volume} {42}},\ \bibinfo
  {pages} {276} (\bibinfo {year} {1990}{\natexlab{a}})}\BibitemShut {NoStop}%
\bibitem [{\citenamefont {Braaten}\ \emph
  {et~al.}(1990{\natexlab{b}})\citenamefont {Braaten}, \citenamefont {Li},\
  and\ \citenamefont {Yuan}}]{Braaten:1990gq}%
  \BibitemOpen
  \bibfield  {author} {\bibinfo {author} {\bibfnamefont {E.}~\bibnamefont
  {Braaten}}, \bibinfo {author} {\bibfnamefont {C.-S.}\ \bibnamefont {Li}},\
  and\ \bibinfo {author} {\bibfnamefont {T.-C.}\ \bibnamefont {Yuan}},\
  }\bibfield  {title} {\bibinfo {title} {{The Evolution of Weinberg's Gluonic
  {CP} Violation Operator}},\ }\href
  {https://doi.org/10.1103/PhysRevLett.64.1709} {\bibfield  {journal} {\bibinfo
   {journal} {Phys. Rev. Lett.}\ }\textbf {\bibinfo {volume} {64}},\ \bibinfo
  {pages} {1709} (\bibinfo {year} {1990}{\natexlab{b}})}\BibitemShut {NoStop}%
\bibitem [{\citenamefont {Degrassi}\ \emph {et~al.}(2005)\citenamefont
  {Degrassi}, \citenamefont {Franco}, \citenamefont {Marchetti},\ and\
  \citenamefont {Silvestrini}}]{Degrassi:2005zd}%
  \BibitemOpen
  \bibfield  {author} {\bibinfo {author} {\bibfnamefont {G.}~\bibnamefont
  {Degrassi}}, \bibinfo {author} {\bibfnamefont {E.}~\bibnamefont {Franco}},
  \bibinfo {author} {\bibfnamefont {S.}~\bibnamefont {Marchetti}},\ and\
  \bibinfo {author} {\bibfnamefont {L.}~\bibnamefont {Silvestrini}},\
  }\bibfield  {title} {\bibinfo {title} {{QCD corrections to the electric
  dipole moment of the neutron in the MSSM}},\ }\href
  {https://doi.org/10.1088/1126-6708/2005/11/044} {\bibfield  {journal}
  {\bibinfo  {journal} {JHEP}\ }\textbf {\bibinfo {volume} {11}},\ \bibinfo
  {pages} {044}},\ \Eprint {https://arxiv.org/abs/hep-ph/0510137}
  {arXiv:hep-ph/0510137} \BibitemShut {NoStop}%
\bibitem [{\citenamefont {de~Vries}\ \emph {et~al.}(2019)\citenamefont
  {de~Vries}, \citenamefont {Falcioni}, \citenamefont {Herzog},\ and\
  \citenamefont {Ruijl}}]{deVries:2019nsu}%
  \BibitemOpen
  \bibfield  {author} {\bibinfo {author} {\bibfnamefont {J.}~\bibnamefont
  {de~Vries}}, \bibinfo {author} {\bibfnamefont {G.}~\bibnamefont {Falcioni}},
  \bibinfo {author} {\bibfnamefont {F.}~\bibnamefont {Herzog}},\ and\ \bibinfo
  {author} {\bibfnamefont {B.}~\bibnamefont {Ruijl}},\ }\href@noop {} {\bibinfo
  {title} {{Two- and three-loop anomalous dimensions of Weinberg's
  dimension-six CP-odd gluonic operator}}} (\bibinfo {year} {2019}),\ \Eprint
  {https://arxiv.org/abs/1907.04923} {arXiv:1907.04923 [hep-ph]} \BibitemShut
  {NoStop}%
\bibitem [{\citenamefont {Lüscher}(2014)}]{Luscher:2013vga}%
  \BibitemOpen
  \bibfield  {author} {\bibinfo {author} {\bibfnamefont {M.}~\bibnamefont
  {Lüscher}},\ }\bibfield  {title} {\bibinfo {title} {{Future applications of
  the Yang-Mills gradient flow in lattice QCD}},\ }\href
  {https://doi.org/10.22323/1.187.0016} {\bibfield  {journal} {\bibinfo
  {journal} {PoS}\ }\textbf {\bibinfo {volume} {LATTICE2013}},\ \bibinfo
  {pages} {016} (\bibinfo {year} {2014})},\ \Eprint
  {https://arxiv.org/abs/1308.5598} {arXiv:1308.5598 [hep-lat]} \BibitemShut
  {NoStop}%
\bibitem [{\citenamefont {Martinelli}\ \emph {et~al.}(1993)\citenamefont
  {Martinelli}, \citenamefont {Petrarca}, \citenamefont {Sachrajda},\ and\
  \citenamefont {Vladikas}}]{Martinelli:1993dq}%
  \BibitemOpen
  \bibfield  {author} {\bibinfo {author} {\bibfnamefont {G.}~\bibnamefont
  {Martinelli}}, \bibinfo {author} {\bibfnamefont {S.}~\bibnamefont
  {Petrarca}}, \bibinfo {author} {\bibfnamefont {C.~T.}\ \bibnamefont
  {Sachrajda}},\ and\ \bibinfo {author} {\bibfnamefont {A.}~\bibnamefont
  {Vladikas}},\ }\bibfield  {title} {\bibinfo {title} {{Nonperturbative
  renormalization of two quark operators with an improved lattice fermion
  action}},\ }\href {https://doi.org/10.1016/0370-2693(93)90562-V} {\bibfield
  {journal} {\bibinfo  {journal} {Phys. Lett. B}\ }\textbf {\bibinfo {volume}
  {311}},\ \bibinfo {pages} {241} (\bibinfo {year} {1993})},\ \bibinfo {note}
  {[Erratum: Phys.Lett.B 317, 660 (1993)]}\BibitemShut {NoStop}%
\bibitem [{\citenamefont {Martinelli}\ \emph {et~al.}(1995)\citenamefont
  {Martinelli}, \citenamefont {Pittori}, \citenamefont {Sachrajda},
  \citenamefont {Testa},\ and\ \citenamefont {Vladikas}}]{Martinelli:1994ty}%
  \BibitemOpen
  \bibfield  {author} {\bibinfo {author} {\bibfnamefont {G.}~\bibnamefont
  {Martinelli}}, \bibinfo {author} {\bibfnamefont {C.}~\bibnamefont {Pittori}},
  \bibinfo {author} {\bibfnamefont {C.~T.}\ \bibnamefont {Sachrajda}}, \bibinfo
  {author} {\bibfnamefont {M.}~\bibnamefont {Testa}},\ and\ \bibinfo {author}
  {\bibfnamefont {A.}~\bibnamefont {Vladikas}},\ }\bibfield  {title} {\bibinfo
  {title} {{A General method for nonperturbative renormalization of lattice
  operators}},\ }\href {https://doi.org/10.1016/0550-3213(95)00126-D}
  {\bibfield  {journal} {\bibinfo  {journal} {Nucl. Phys. B}\ }\textbf
  {\bibinfo {volume} {445}},\ \bibinfo {pages} {81} (\bibinfo {year} {1995})},\
  \Eprint {https://arxiv.org/abs/hep-lat/9411010} {arXiv:hep-lat/9411010}
  \BibitemShut {NoStop}%
\bibitem [{\citenamefont {Donini}\ \emph {et~al.}(1999)\citenamefont {Donini},
  \citenamefont {Gimenez}, \citenamefont {Martinelli}, \citenamefont {Talevi},\
  and\ \citenamefont {Vladikas}}]{Donini:1999sf}%
  \BibitemOpen
  \bibfield  {author} {\bibinfo {author} {\bibfnamefont {A.}~\bibnamefont
  {Donini}}, \bibinfo {author} {\bibfnamefont {V.}~\bibnamefont {Gimenez}},
  \bibinfo {author} {\bibfnamefont {G.}~\bibnamefont {Martinelli}}, \bibinfo
  {author} {\bibfnamefont {M.}~\bibnamefont {Talevi}},\ and\ \bibinfo {author}
  {\bibfnamefont {A.}~\bibnamefont {Vladikas}},\ }\bibfield  {title} {\bibinfo
  {title} {{Nonperturbative renormalization of lattice four fermion operators
  without power subtractions}},\ }\href {https://doi.org/10.1007/s100529900097}
  {\bibfield  {journal} {\bibinfo  {journal} {Eur. Phys. J. C}\ }\textbf
  {\bibinfo {volume} {10}},\ \bibinfo {pages} {121} (\bibinfo {year} {1999})},\
  \Eprint {https://arxiv.org/abs/hep-lat/9902030} {arXiv:hep-lat/9902030}
  \BibitemShut {NoStop}%
\bibitem [{\citenamefont {Bhattacharya}\ \emph {et~al.}(2015)\citenamefont
  {Bhattacharya}, \citenamefont {Cirigliano}, \citenamefont {Gupta},
  \citenamefont {Mereghetti},\ and\ \citenamefont
  {Yoon}}]{Bhattacharya:2015rsa}%
  \BibitemOpen
  \bibfield  {author} {\bibinfo {author} {\bibfnamefont {T.}~\bibnamefont
  {Bhattacharya}}, \bibinfo {author} {\bibfnamefont {V.}~\bibnamefont
  {Cirigliano}}, \bibinfo {author} {\bibfnamefont {R.}~\bibnamefont {Gupta}},
  \bibinfo {author} {\bibfnamefont {E.}~\bibnamefont {Mereghetti}},\ and\
  \bibinfo {author} {\bibfnamefont {B.}~\bibnamefont {Yoon}},\ }\bibfield
  {title} {\bibinfo {title} {{Dimension-5 CP-odd operators: QCD mixing and
  renormalization}},\ }\href {https://doi.org/10.1103/PhysRevD.92.114026}
  {\bibfield  {journal} {\bibinfo  {journal} {Phys. Rev.}\ }\textbf {\bibinfo
  {volume} {D92}},\ \bibinfo {pages} {114026} (\bibinfo {year} {2015})},\
  \Eprint {https://arxiv.org/abs/1502.07325} {arXiv:1502.07325 [hep-ph]}
  \BibitemShut {NoStop}%
\bibitem [{\citenamefont {Cirigliano}\ \emph {et~al.}(2020)\citenamefont
  {Cirigliano}, \citenamefont {Mereghetti},\ and\ \citenamefont
  {Stoffer}}]{Cirigliano:2020msr}%
  \BibitemOpen
  \bibfield  {author} {\bibinfo {author} {\bibfnamefont {V.}~\bibnamefont
  {Cirigliano}}, \bibinfo {author} {\bibfnamefont {E.}~\bibnamefont
  {Mereghetti}},\ and\ \bibinfo {author} {\bibfnamefont {P.}~\bibnamefont
  {Stoffer}},\ }\href@noop {} {\bibinfo {title} {{Non-perturbative
  renormalization scheme for the CP-odd three-gluon operator}}} (\bibinfo
  {year} {2020}),\ \Eprint {https://arxiv.org/abs/2004.03576} {arXiv:2004.03576
  [hep-ph]} \BibitemShut {NoStop}%
\bibitem [{\citenamefont {Gimenez}\ \emph {et~al.}(2004)\citenamefont
  {Gimenez}, \citenamefont {Giusti}, \citenamefont {Guerriero}, \citenamefont
  {Lubicz}, \citenamefont {Martinelli}, \citenamefont {Petrarca}, \citenamefont
  {Reyes}, \citenamefont {Taglienti},\ and\ \citenamefont
  {Trevigne}}]{Gimenez:2004me}%
  \BibitemOpen
  \bibfield  {author} {\bibinfo {author} {\bibfnamefont {V.}~\bibnamefont
  {Gimenez}}, \bibinfo {author} {\bibfnamefont {L.}~\bibnamefont {Giusti}},
  \bibinfo {author} {\bibfnamefont {S.}~\bibnamefont {Guerriero}}, \bibinfo
  {author} {\bibfnamefont {V.}~\bibnamefont {Lubicz}}, \bibinfo {author}
  {\bibfnamefont {G.}~\bibnamefont {Martinelli}}, \bibinfo {author}
  {\bibfnamefont {S.}~\bibnamefont {Petrarca}}, \bibinfo {author}
  {\bibfnamefont {J.}~\bibnamefont {Reyes}}, \bibinfo {author} {\bibfnamefont
  {B.}~\bibnamefont {Taglienti}},\ and\ \bibinfo {author} {\bibfnamefont
  {E.}~\bibnamefont {Trevigne}},\ }\bibfield  {title} {\bibinfo {title}
  {{Non-perturbative renormalization of lattice operators in coordinate
  space}},\ }\href {https://doi.org/10.1016/j.physletb.2004.07.053} {\bibfield
  {journal} {\bibinfo  {journal} {Phys. Lett. B}\ }\textbf {\bibinfo {volume}
  {598}},\ \bibinfo {pages} {227} (\bibinfo {year} {2004})},\ \Eprint
  {https://arxiv.org/abs/hep-lat/0406019} {arXiv:hep-lat/0406019} \BibitemShut
  {NoStop}%
\bibitem [{\citenamefont {Chetyrkin}\ and\ \citenamefont
  {Maier}(2011)}]{Chetyrkin:2010dx}%
  \BibitemOpen
  \bibfield  {author} {\bibinfo {author} {\bibfnamefont {K.}~\bibnamefont
  {Chetyrkin}}\ and\ \bibinfo {author} {\bibfnamefont {A.}~\bibnamefont
  {Maier}},\ }\bibfield  {title} {\bibinfo {title} {{Massless correlators of
  vector, scalar and tensor currents in position space at orders $\alpha_s^3$
  and $\alpha_s^4$: Explicit analytical results}},\ }\href
  {https://doi.org/10.1016/j.nuclphysb.2010.11.007} {\bibfield  {journal}
  {\bibinfo  {journal} {Nucl. Phys. B}\ }\textbf {\bibinfo {volume} {844}},\
  \bibinfo {pages} {266} (\bibinfo {year} {2011})},\ \Eprint
  {https://arxiv.org/abs/1010.1145} {arXiv:1010.1145 [hep-ph]} \BibitemShut
  {NoStop}%
\bibitem [{\citenamefont {Tomii}\ and\ \citenamefont
  {Christ}(2019)}]{Tomii:2018zix}%
  \BibitemOpen
  \bibfield  {author} {\bibinfo {author} {\bibfnamefont {M.}~\bibnamefont
  {Tomii}}\ and\ \bibinfo {author} {\bibfnamefont {N.~H.}\ \bibnamefont
  {Christ}},\ }\bibfield  {title} {\bibinfo {title} {{$O(4)$-symmetric
  position-space renormalization of lattice operators}},\ }\href
  {https://doi.org/10.1103/PhysRevD.99.014515} {\bibfield  {journal} {\bibinfo
  {journal} {Phys. Rev. D}\ }\textbf {\bibinfo {volume} {99}},\ \bibinfo
  {pages} {014515} (\bibinfo {year} {2019})},\ \Eprint
  {https://arxiv.org/abs/1811.11238} {arXiv:1811.11238 [hep-lat]} \BibitemShut
  {NoStop}%
\bibitem [{\citenamefont {Engel}\ \emph {et~al.}(2013)\citenamefont {Engel},
  \citenamefont {Ramsey-Musolf},\ and\ \citenamefont {van
  Kolck}}]{Engel:2013lsa}%
  \BibitemOpen
  \bibfield  {author} {\bibinfo {author} {\bibfnamefont {J.}~\bibnamefont
  {Engel}}, \bibinfo {author} {\bibfnamefont {M.~J.}\ \bibnamefont
  {Ramsey-Musolf}},\ and\ \bibinfo {author} {\bibfnamefont {U.}~\bibnamefont
  {van Kolck}},\ }\bibfield  {title} {\bibinfo {title} {{Electric Dipole
  Moments of Nucleons, Nuclei, and Atoms: The Standard Model and Beyond}},\
  }\href {https://doi.org/10.1016/j.ppnp.2013.03.003} {\bibfield  {journal}
  {\bibinfo  {journal} {Prog.Part.Nucl.Phys.}\ }\textbf {\bibinfo {volume}
  {71}},\ \bibinfo {pages} {21} (\bibinfo {year} {2013})},\ \Eprint
  {https://arxiv.org/abs/1303.2371} {arXiv:1303.2371 [nucl-th]} \BibitemShut
  {NoStop}%
\bibitem [{\citenamefont {Panagopoulos}\ and\ \citenamefont
  {Vicari}(1990)}]{Panagopoulos:1989zn}%
  \BibitemOpen
  \bibfield  {author} {\bibinfo {author} {\bibfnamefont {H.}~\bibnamefont
  {Panagopoulos}}\ and\ \bibinfo {author} {\bibfnamefont {E.}~\bibnamefont
  {Vicari}},\ }\bibfield  {title} {\bibinfo {title} {{The Trilinear Gluon
  Condensate on the Lattice}},\ }\href
  {https://doi.org/10.1016/0550-3213(90)90039-G} {\bibfield  {journal}
  {\bibinfo  {journal} {Nucl. Phys. B}\ }\textbf {\bibinfo {volume} {332}},\
  \bibinfo {pages} {261} (\bibinfo {year} {1990})}\BibitemShut {NoStop}%
\bibitem [{Note1()}]{Note1}%
  \BibitemOpen
  \bibinfo {note} {We will not consider contributions from totally disconnected
  diagrams.}\BibitemShut {Stop}%
\bibitem [{\citenamefont {Rizik}()}]{Rizik:new}%
  \BibitemOpen
  \bibfield  {author} {\bibinfo {author} {\bibfnamefont {M.}~\bibnamefont
  {Rizik}},\ }\bibinfo {note} {in preparation (unpublished)}\BibitemShut
  {NoStop}%
\bibitem [{\citenamefont {'t~Hooft}\ and\ \citenamefont
  {Veltman}(1972)}]{tHooft:1972tcz}%
  \BibitemOpen
  \bibfield  {author} {\bibinfo {author} {\bibfnamefont {G.}~\bibnamefont
  {'t~Hooft}}\ and\ \bibinfo {author} {\bibfnamefont {M.}~\bibnamefont
  {Veltman}},\ }\bibfield  {title} {\bibinfo {title} {{Regularization and
  Renormalization of Gauge Fields}},\ }\href
  {https://doi.org/10.1016/0550-3213(72)90279-9} {\bibfield  {journal}
  {\bibinfo  {journal} {Nucl. Phys. B}\ }\textbf {\bibinfo {volume} {44}},\
  \bibinfo {pages} {189} (\bibinfo {year} {1972})}\BibitemShut {NoStop}%
\bibitem [{\citenamefont {Breitenlohner}\ and\ \citenamefont
  {Maison}(1977)}]{Breitenlohner:1977hr}%
  \BibitemOpen
  \bibfield  {author} {\bibinfo {author} {\bibfnamefont {P.}~\bibnamefont
  {Breitenlohner}}\ and\ \bibinfo {author} {\bibfnamefont {D.}~\bibnamefont
  {Maison}},\ }\bibfield  {title} {\bibinfo {title} {{Dimensional
  Renormalization and the Action Principle}},\ }\href
  {https://doi.org/10.1007/BF01609069} {\bibfield  {journal} {\bibinfo
  {journal} {Commun. Math. Phys.}\ }\textbf {\bibinfo {volume} {52}},\ \bibinfo
  {pages} {11} (\bibinfo {year} {1977})}\BibitemShut {NoStop}%
\bibitem [{\citenamefont {Buras}\ and\ \citenamefont
  {Weisz}(1990)}]{Buras:1989xd}%
  \BibitemOpen
  \bibfield  {author} {\bibinfo {author} {\bibfnamefont {A.~J.}\ \bibnamefont
  {Buras}}\ and\ \bibinfo {author} {\bibfnamefont {P.~H.}\ \bibnamefont
  {Weisz}},\ }\bibfield  {title} {\bibinfo {title} {{QCD Nonleading Corrections
  to Weak Decays in Dimensional Regularization and 't Hooft-Veltman Schemes}},\
  }\href {https://doi.org/10.1016/0550-3213(90)90223-Z} {\bibfield  {journal}
  {\bibinfo  {journal} {Nucl. Phys. B}\ }\textbf {\bibinfo {volume} {333}},\
  \bibinfo {pages} {66} (\bibinfo {year} {1990})}\BibitemShut {NoStop}%
\bibitem [{\citenamefont {SymLat}()}]{SymLat:2020}%
  \BibitemOpen
  \bibfield  {author} {\bibinfo {author} {\bibnamefont {SymLat}},\ }\bibinfo
  {note} {in preparation (unpublished)}\BibitemShut {NoStop}%
\bibitem [{\citenamefont {Espriu}\ and\ \citenamefont
  {Tarrach}(1982)}]{Espriu:1982bw}%
  \BibitemOpen
  \bibfield  {author} {\bibinfo {author} {\bibfnamefont {D.}~\bibnamefont
  {Espriu}}\ and\ \bibinfo {author} {\bibfnamefont {R.}~\bibnamefont
  {Tarrach}},\ }\bibfield  {title} {\bibinfo {title} {{Renormalization of the
  Axial Anomaly Operators}},\ }\href {https://doi.org/10.1007/BF01573750}
  {\bibfield  {journal} {\bibinfo  {journal} {Z. Phys.}\ }\textbf {\bibinfo
  {volume} {C16}},\ \bibinfo {pages} {77} (\bibinfo {year} {1982})}\BibitemShut
  {NoStop}%
\end{thebibliography}%

\end{document}